\documentclass[twocolumn]{aastex631}

\usepackage[T1]{fontenc}
\usepackage{ae,aecompl}

\usepackage{graphicx}	% Including figure files
\usepackage{amsmath}	% Advanced maths commands
\usepackage{amssymb}	% Extra maths symbols

\usepackage{natbib}

\bibpunct{(}{)}{;}{a}{}{,}

\usepackage{url}

\usepackage{longtable}
\usepackage{booktabs}

\usepackage[flushleft, referable]{threeparttablex}

\usepackage{subfloat}

\DeclareGraphicsExtensions{.jpg,.pdf,.png}

\usepackage{parskip}

\usepackage{lipsum}

\usepackage{dsfont}

\usepackage{comment}

\usepackage{tablefootnote}

\def\tightitemize{\begin{itemize}\setlength{\parskip}{0.25 \parsep}}
\def\cvitemize{\begin{list}{}{\parskip=0.5\parsep\itemsep=0pt}}
\def\endcvitemize{\end{list}}
\def\be{\begin{equation}}
\def\ee{\end{equation}}
\def\bea{\begin{eqnarray*}}
\def\eea{\end{eqnarray*}}

\def\degrees{$^{\circ}$}
\def\arcsec{^{\prime\prime}}
\def\arcmin{^\prime}

\def\Lsun{L_{\odot}}
\def\Msun{M_{\odot}}
\def\Mdot{$\rm M_{\odot}~yr^{-1}$}
\def\d{$^{\dagger}$}

\def\Planck{\textit{Planck}}
\def\WISE{\textit{WISE}}
\def\Herschel{\textit{Herschel}}

\def\HST{\textit{HST}}
\def\JWST{\textit{JWST}}

\def\lenstool{\textsc{lenstool}}
\def\casa{\textsc{casa}}
\def\imfit{\textsc{imfit}}

\def\cleanlens{\texttt{cleanlens}}

\def\ss{$^\text{s}$}

\submitjournal{ApJ}

\begin{document}

\title[PASSAGES: Lens modeling of Planck DSFGs]{
PASSAGES: 
the wide-ranging, extreme intrinsic properties of $\boldsymbol{\it Planck}$-selected, lensed dusty star-forming galaxies}

\shorttitle{PASSAGES: Lens modeling of Planck DSFGs}
\shortauthors{Kamieneski et al.}

\author[0000-0001-9394-6732]{Patrick S. Kamieneski}
\affiliation{Department of Astronomy, University of Massachusetts, Amherst, MA 01003, USA} 
\affiliation{School of Earth and Space Exploration, Arizona State University, Tempe, AZ 85287-6004, USA}

\correspondingauthor{Patrick S. Kamieneski}
\email{pkamiene@asu.edu}

\author[0000-0000-0000-0000]{Min S. Yun}
\affiliation{Department of Astronomy, University of Massachusetts, Amherst, MA 01003, USA}

\author[0000-0000-0000-0000]{Kevin C. Harrington}
\affiliation{European Southern Observatory, Alonso de C\'ordova 3107, Viticura, Casilla 19001, Santiago de Chile, Chile}

\author[0000-0000-0000-0000]{James D. Lowenthal}
\affiliation{Smith College, Northampton, MA 01063, USA}

\author[0000-0000-0000-0000]{Q. Daniel Wang}
\affiliation{Department of Astronomy, University of Massachusetts, Amherst, MA 01003, USA}

\author[0000-0000-0000-0000]{Brenda L. Frye}
\affiliation{Department of Astronomy/Steward Observatory, 933 North Cherry Avenue, University of Arizona, Tucson, AZ 85721, USA}

\author[0000-0000-0000-0000]{Eric F. Jim\'enez-Andrade}
%\affiliation{National Radio Astronomy Observatory, 520 Edgemont Road, Charlottesville, VA 22903, USA}
\affiliation{Instituto de Radioastronom\'{i}a y Astrof\'{i}sica, Universidad Nacional Aut\'{o}noma de M\'{e}xico, Antigua Carretera a P\'{a}tzcuaro \# 8701, Ex-Hda. San Jos\'{e} de la Huerta, Morelia, Michoac\'{a}n, M\'{e}xico C.P. 58089}

\author[0000-0000-0000-0000]{Amit Vishwas}
\affiliation{Department of Astronomy, Cornell University, Space Sciences Building, Ithaca, NY 14853, USA}

\author[0000-0000-0000-0000]{Olivia Cooper}
\affiliation{Department of Astronomy, The University of Texas at Austin, 2515 Speedway Boulevard Stop C1400, Austin, TX 78712, USA}

\author[0000-0000-0000-0000]{Massimo Pascale}
\affiliation{Department of Astronomy, University of California, 501 Campbell Hall \#3411, Berkeley, CA 94720, USA}

\author[0000-0000-0000-0000]{Nicholas Foo}
\affiliation{Department of Astronomy/Steward Observatory, 933 North Cherry Avenue, University of Arizona, Tucson, AZ 85721, USA}

\author[0000-0000-0000-0000]{Derek Berman}
\affiliation{Department of Earth and Atmospheric Sciences, Cornell University, Ithaca, NY 14853, USA}

\author[0000-0000-0000-0000]{Anthony Englert}
\affiliation{Department of Physics, Brown University, Providence, RI 02912, USA}

\author[0000-0000-0000-0000]{Carlos Garcia Diaz}
\affiliation{Department of Astronomy, University of Massachusetts, Amherst, MA 01003, USA}

%\author{August Muench}
%\affiliation{American Astronomical Society \\
%1667 K Street NW, Suite 800 \\



%%%%%%%%%%%%%%%%%%%%%%%%%%%%%%%%%%%%%%%%%%%%%%%%%%

%%%%%%%%%%%%%%%%%%% TITLE PAGE %%%%%%%%%%%%%%%%%%%

\begin{abstract}

%This example manuscript is intended to serve as a tutorial and template for
%authors to use when writing their own AAS Journal articles. The manuscript
%includes a history of \aastex\ and includes figure and table examples to illustrate these features. Information on features not explicitly mentioned in the article can be viewed in the manuscript comments or more extensive online
%documentation. Authors are welcome replace the text, tables, figures, and
%bibliography with their own and submit the resulting manuscript to the AAS
%Journals peer review system.  The first lesson in the tutorial is to remind
%authors that the AAS Journals, the Astrophysical Journal (ApJ), the
%Astrophysical Journal Letters (ApJL), the Astronomical Journal (AJ), and
%the Planetary Science Journal (PSJ) all have a 250 word limit for the 
%abstract\footnote{Abstracts for Research Notes of the American Astronomical 
%Society (RNAAS) are limited to 150 words}.  If you exceed this length the
%Editorial office will ask you to shorten it. This abstract has 161 words.

%
%
%

The PASSAGES (\Planck\ All-Sky Survey to Analyze Gravitationally-lensed Extreme Starbursts) collaboration has recently defined a sample of 30 gravitationally-lensed dusty star-forming galaxies (DSFGs). 
These rare, submillimeter-selected objects enable high-resolution views of the most extreme sites of star formation in galaxies at Cosmic Noon.   
Here, we present 
%new lens modeling with \lenstool\ for 
the first major compilation of strong lensing analyses using \lenstool\ for PASSAGES, including
15 objects spanning $z=1.1-3.3$, using complementary information from $0.6\arcsec$-resolution 1 mm Atacama Large Millimeter/submillimeter Array (ALMA) and $0.4\arcsec$ 
%6 GHz 
5 cm Jansky Very Large Array continuum imaging, in tandem with
%high-resolution 
1.6$\micron$ {\it Hubble} and optical imaging with Gemini-S.
%Space Telescope} 
%imaging and optical imaging with Gemini-S. 
%We 
%perform lens modeling with \lenstool\ to 
%recover intrinsic, de-lensed properties of the DSFGs, parameterizing the lensing mass but not the source-plane structure.
%
%Of the overall PASSAGES sample, {\red XX of 30} objects have evidence for strong lensing ($\mu > 2$ with an image multiplicity of at least 2). 
%
Magnifications range from $\mu = {2 - 28}$ (median {$\mu=7$}), yielding
%
%lensing-corrected 
intrinsic infrared luminosities of $L_{\rm IR} = {0.2 - 5.9} \times 10^{13}~\Lsun$ (median ${1.4}\times10^{13}~\Lsun$) and inferred star formation rates of ${170-6300}~\Msun~{\rm yr}^{-1}$ (median ${1500}~\Msun~{\rm yr}^{-1}$).
These results suggest that %gravitational lensing alone cannot sufficiently explain the extreme apparent brightness of these objects.
the PASSAGES objects comprise 
%uniquely 
%serendipitous population of 
some of the most extreme known starbursts, rivaling the luminosities of even the brightest unlensed objects, 
further amplified by lensing.
The intrinsic sizes of %rest-frame 
far-infrared continuum regions
%, probing dust heated by star formation, 
are 
%on the larger end 
large ($R_{\rm e} = {1.7 - 4.3}$ kpc; median {$3.0$} kpc) but consistent with 
%previously-observed 
$L_{\rm IR}-R_{\rm e}$ scaling relations for $z>1$ DSFGs, suggesting a widespread spatial distribution of star formation.
With modestly-high angular resolution, we explore if these objects might be 
%considered 
maximal starbursts.
%approaching theoretical maximum, 
%
Instead of 
approaching Eddington-limited 
%star formation rate 
surface densities, above which radiation pressure will disrupt further star formation, they are safely sub-Eddington\textemdash at least on global, galaxy-integrated scales.
%
%
%
%
%}
%\newline{}
\end{abstract}

%\begin{keywords}
%gravitational lensing: strong -- galaxies: starburst -- submillimetre: galaxies
%\end{keywords}




%%%%%%%%%%%%%%%%%%%%%%%%%%%%%%%%%%%%%%%%%%%%%%%%%%

%%%%%%%%%%%%%%%%% BODY OF PAPER %%%%%%%%%%%%%%%%%%

\section{Introduction}
\label{sec:intro}

Dusty star-forming galaxies (DSFGs) host some of the most extreme infrared luminosities in the known Universe, often in excess of $10^{12-13}~\Lsun$, reflecting star formation rates $>100-1000$ \Mdot. %Occurring 
Observed
mainly during the peak of the cosmic star formation history at $z=1-4$, they are fundamental to our understanding of stellar mass assembly in the last 10 billion years (see reviews by \citealt{Blain:2002aa,Casey:2014aa}). 
The current galaxy evolution paradigm favors that DSFGs are progenitors of massive, quiescent elliptical galaxies (e.g., \citealt{Lilly:1999aa, Brodwin:2008aa, Tacconi:2008aa, Daddi:2009aa, Toft:2014aa}), in part because of their association with galaxy over-densities and mergers.
However, numerous questions remain regarding the rapid conversion of cold gas reservoirs into new stars, often at star formation surface densities approaching theoretical maxima. Beyond this Eddington-like limit, stellar feedback should hinder or even quench subsequent star formation \citep{Scoville:2001aa, Scoville:2003aa, Murray:2005aa, Thompson:2005aa, Andrews:2011ab, Hodge:2019aa}. Galaxies that are sub-Eddington when integrated over star-forming regions may still have Eddington-limited clumps of star formation (e.g., \citealt{Simpson:2015ab, Barcos-Munoz:2017aa}). 
%{\red more.
%}
%
Moreover, the predominant triggering mechanism for fueling this star formation has lingering uncertainties. 
There is now plentiful evidence (observational and theoretical) that major galaxy mergers play an important role (\citealt{Ivison:2002aa, Narayanan:2006aa, Narayanan:2010ab, Tacconi:2008aa}, among many others), but cold-mode accretion, in which star formation in massive galaxies is fed by smooth infall of gas, has also been put forward as a viable explanation
\citep{Keres:2005aa, Dekel:2009ab, Dekel:2009aa}.
Answering these open questions absolutely necessitates resolutions corresponding to sub-galactic physical scales.

The past decade has ushered in the discovery of a substantial number of DSFGs that are amplified in submillimeter flux by strong gravitational lensing. Due to the steep drop-off in submillimeter number counts for high-$z$ galaxies \citep{Blain:1996aa, Negrello:2007aa}, lensed DSFGs can be efficiently identified using a simple flux threshold. This was first demonstrated with large-area surveys undertaken with \Herschel\ \citep{Negrello:2010aa, Negrello:2017aa, Conley:2011aa, Wardlow:2013aa, Nayyeri:2016aa}, the South Pole Telescope (SPT; \citealt{Vieira:2010aa, Weiss:2013aa, Mocanu:2013aa, Vieira:2013aa}), the Atacama Cosmology Telescope (ACT; \citealt{Marsden:2014aa}), and more recently with {\it Planck} \citep{Planck-Collaboration:2015aa, Canameras:2015aa, Harrington:2016aa, Berman:2022aa}. 
Such lens samples are invaluable for the study of star formation at sub-kpc scales in the early Universe, thanks to the magnification in angular size afforded by strong lensing. Without lensing, the angular resolution of {\it Hubble} and \JWST\ is only sufficient to resolve physical sizes at $z>1$ of 1 kpc or greater, so the critical $10-100$pc scales where star-forming and feedback processes are most relevant remain inaccessible. These samples offer unique opportunities for statistical studies of the foreground lensing population (e.g., \citealt{Eales:2015ab, Amvrosiadis:2018aa})\textemdash with implications for the halo mass function and mass density profiles for individual halos\textemdash as the selection function depends only on intrinsic flux and magnification. This stands in contrast to the complex biases of lens detection methods based on arc morphology and foreground halo masses.

DSFGs also benefit from a strongly negative $K$-correction in the submillimeter (e.g. \citealt{Blain:1993aa}), owing to steep Rayleigh-Jeans tails in the rest-frame far-IR regime of the spectral energy distribution (SED) from dust emission. 
Increasing the distance to a galaxy of fixed luminosity would dim its observable flux at a given wavelength, were it not also the case that the increased redshift results in capturing shorter-wavelength (and therefore brighter) parts of the SED.
For example, at 850$\micron$ in the observer-frame, these effects roughly equilibrate, and the flux for an object of fixed luminosity appears uniform from $z\approx1-8$ (e.g., \citealt{Casey:2014aa}). 
It is for this reason that DSFGs can be efficiently identified in the submillimeter regime up to high redshifts (modulo their evolving number density), and this partially explains the preponderance of submillimeter-bright galaxies (SMGs; \citealt{Smail:1997aa, Hughes:1998aa, Barger:1998aa}; review by \citealt{Blain:2002aa}), a term now largely synonymous with DSFGs.

\begin{figure*}
	\includegraphics[width=\textwidth]{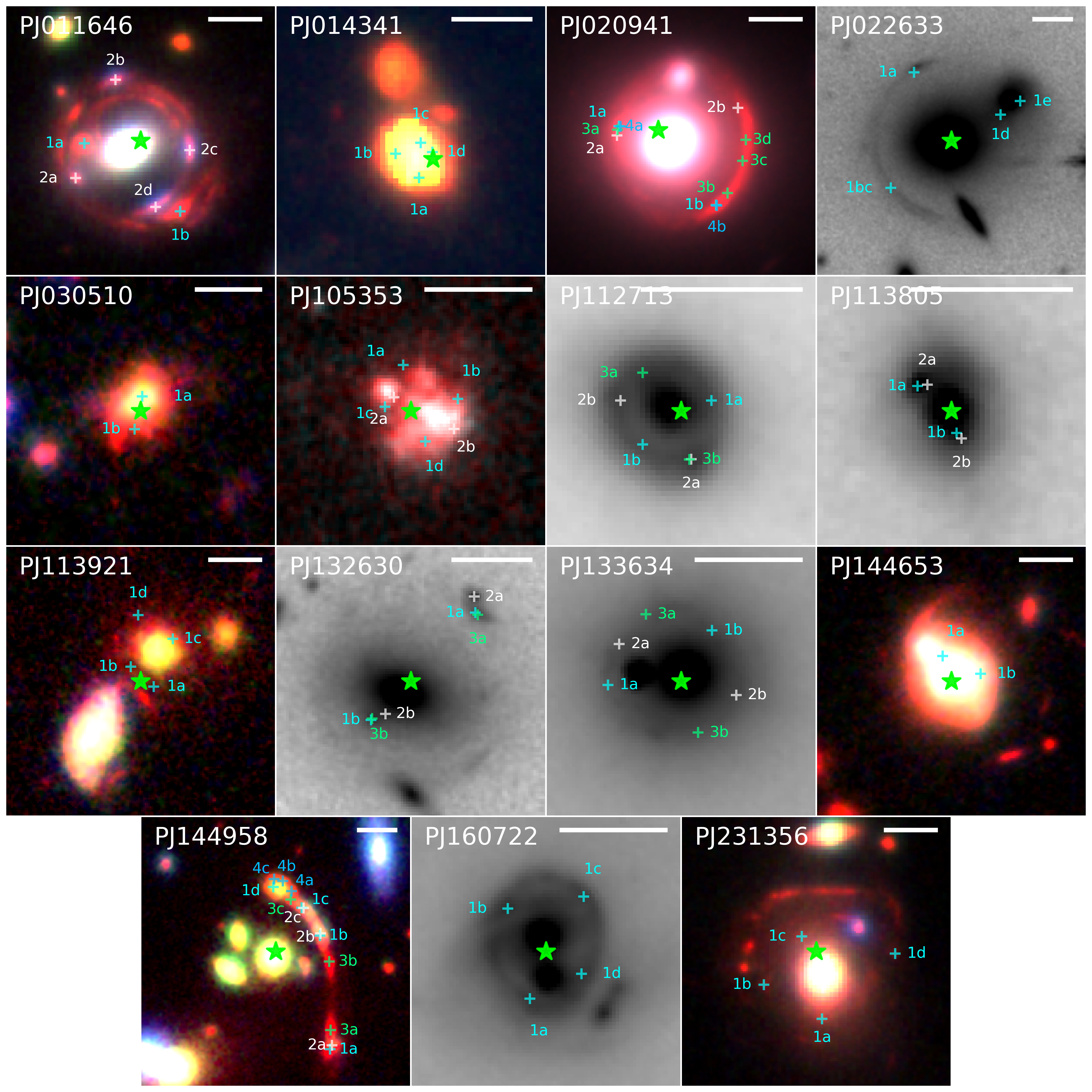}
    \caption{Postage stamp of near-IR images for the 15 PASSAGES objects modeled in this work. RGB color panels show imaging with Gemini-S $r'$ and $z'$ filters (where available) with \HST\ $H$-band (where $H$ is shown as red, $z'$ as green, and $r'$ as blue). One exception is PJ105353, which instead includes an \HST\ F110W image as the green/blue filters (\S \ref{sec:HST_obs}). Grayscale panels are those where only $H$-band is available. A white scalebar in the upper right corner of each panel represents 2$\arcsec$, and the target ID is included in the upper left corner. Each panel is north-aligned and re-centered to best showcase visible lensing features, with the WISE centroids (or target coordinates for the \HST\ observations) indicated as green stars (to facilitate easy comparison with Fig.~\ref{fig:model_SP}). Labeled cyan crosses indicate the locations of image families, summarized in Table \ref{tab:lensconstraints} (see also discussion in Section \ref{sec:family_id}).
    }
    \label{fig:postage_stamps}
\end{figure*}

The {\it Planck} All-Sky Survey to Analyze Gravitationally-lensed Extreme Starbursts (PASSAGES) sample was introduced in \citealt{Harrington:2016aa} and \citealt{Berman:2022aa}, with the scientific aims of exploring the gas fueling and induced starbursting phases in some of the most apparently IR-luminous objects yet identified ($\mu L_{\rm IR}  = 0.1 - 3.1 \times 10^{14} ~ L_\odot$, median $1.2 \times 10^{14} ~ L_\odot$; also \citealt{Harrington:2021aa}). Even assuming a fiducial magnification factor of $\mu \approx 10$, typical for galaxy-scale strong lenses (e.g. \citealt{Negrello:2017aa}), they remain intrinsically hyper-luminous, an order of magnitude higher than local ultra-luminous infrared galaxies (ULIRGs, \citealt{Sanders:1996aa, Lonsdale:2006aa}), which appear to be the closest low-$z$ analog for DSFGs in terms of observed $L_{\rm IR}$.
Such objects with IR luminosity in excess of $10^{13} L_\odot$ have typically been referred to as hyper-luminous infrared galaxies, or HyLIRGs \citep{Cutri:1994aa}.
As suggested by \citealt{Berman:2022aa}, the selection from an all-sky facility like \Planck\ can identify the rarest, most extreme 
%limit of 
star-forming galaxies. On one hand, strong gravitational lensing may boost the flux of fainter populations that would otherwise be undetected by an observation (the well-known magnification bias, e.g., \citealt{Turner:1980aa,Turner:1984aa}; \citealt{Scranton:2005aa}; \citealt{Hildebrandt:2009aa}), but it may also be the case that selection from \Planck\ reveals objects that would be detected within flux limits even without the modulation by lensing.

A detailed understanding of the PASSAGES objects requires refined and meticulously-constructed gravitational lens models. As the null geodesics that light rays from distant objects follow on their path to Earth depend on all matter\textemdash indiscriminate of luminous vs. dark matter\textemdash it is impossible to perfectly describe the exact deformation caused by lensing. While luminous, baryonic matter can offer great insight into the total underlying mass distribution, most gravitational lens modeling techniques hinge on characterizing the distortion of the background and determining what foreground mass profile is required to induce this distortion. Fortunately, lensing is achromatic, as photons of all wavelengths follow the same null geodesics. Images of different wavelengths (which presumably originate in slightly disparate spatial distributions in the source plane) offer complementary information by piercing different locations in the foreground lens. 
In this context, robust lens models of strongly-lensed sources enable investigations of $z>1$ galaxies at a (source-plane) spatial resolution that is currently possible only for local, nearby objects. 
Therefore, in this paper, we derive our first multi-wavelength-based lens models for 15 (out of 30) members of PASSAGES (all with spectroscopically-confirmed background redshifts). In so doing, we use optical and near-infrared information from {\it Hubble} and Gemini, (sub-)millimeter information from the Atacama Large Millimeter/submillimeter Array, and radio information from the {\it Karl G. Jansky} Very Large Array. 

This paper is organized as follows: in Section \ref{sec:observations}, we introduce the subset of the \Planck\ lens sample discussed in this work and the relevant near-IR, submillimeter, and radio observations from which our lens models are derived. In Section \ref{sec:lensing}, we review our approach to gravitational lens modeling with \lenstool, and discuss the identified lensing features and constraints for each object. 
%{\red Remove: In Section \ref{sec:magnifications}, we test the feasibility of a model-independent approach to predicting lensing magnification.} 
In Section \ref{sec:discussion}, we examine some of the overall properties of this PASSAGES subsample, including de-lensed luminosity and intrinsic source sizes. Finally, we summarize our conclusions in Section \ref{sec:conclusions}.

In this work, we adopt a 
%standard $\Lambda$CDM 
flat cosmological model with $H_0=70$ km s$^{-1}$ Mpc$^{-1}$, $\Omega_m = 0.3$, and $\Omega_\Lambda = 0.7$.
%{\red Double-check?}
At the median source redshift of this work, $z\approx 2.2$, $1\arcsec$ corresponds to a physical scale of 8.27 kpc; at the median lens redshift, $z \approx 0.5$, $1\arcsec = 6.10$ kpc 
\citep{Wright:2006aa}.

\section{Data}
\label{sec:observations}

\subsection{Sample selection}
\label{sec:sample}

PASSAGES consists of 30 high-$z$ objects that were identified from the {\it Planck} Catalogue of Compact Sources \citep{Planck-Collaboration:2014ab} and cross-identified with \Herschel\ and \WISE, the details of which are provided in \citet{Harrington:2016aa} and \citet{Berman:2022aa}.
Here, we select a subset of 15 members of the larger current sample, restricting our focus to primarily those objects with both a simpler lensing morphology, arising from galaxy-scale or small group-scale lens potentials (Fig.~\ref{fig:postage_stamps}), and sufficient multi-wavelength information. 
These lower-mass deflectors account for more than 75\% of the total collection of lensed objects from PASSAGES.
All of the objects in this subsample have been imaged with high angular resolution with optical/near-IR, radio, and (in most cases) submillimeter telescopes.
These data were sufficient to identify the families of lensed arcs and images that were responsible for the large submillimeter fluxes in these fields, in addition to the foreground lensing galaxies (which are predominately very faint at longer wavelengths).
%
%which has the advantage of filtering out foreground objects (which are predominately very faint at these long wavelengths) and pinpointing the lensed image families responsible for the large submillimeter flux by which the PASSAGES members are identified.
%
We summarize this subsample of objects in Table \ref{tab:sample}, and provide further details on each source in Appendix \ref{sec:object_notes}. 
Lens models for the remaining objects\textemdash most of which are cluster-scale lenses or fields lacking sub-arcsecond-resolution millimeter imaging\textemdash will be described in future works. 
Some models have already been presented publicly for PASSAGES, including: 
PJ020941.3 ({\it 9io9}; \citealt{Geach:2015aa, Geach:2018aa, Rivera:2019aa, Liu:2022ad}; and this work), 
PJ105322.6 (G145.2+50.9; \citealt{Frye:2019aa}), 
PJ105353.0 (G244.8+54.9; \citealt{Canameras:2017aa, Canameras:2017ab, Frye:2019aa}; and this work), 
PJ112714.5 (G165.7+67.0; \citealt{Frye:2019aa, Pascale:2022aa}), 
PJ132630.3 (NAv1.195; \citealt{Bussmann:2013aa}; and this work),
PJ132934.2/PJ132935.3 (the ``Cosmic Eyebrow;" \citealt{Diaz-Sanchez:2017aa}), 
and PJ142823.9 (HBootes03; \citealt{Borys:2006aa}).
For the 3 previously-modeled targets studied independently in this work (in addition to PJ113921.7, for which a magnification is given in \citealt{Canameras:2018ab}), we will examine any inconsistencies in Appendix \ref{sec:lens_model_comparison}.

In deriving the lens models in this work, we take advantage of multi-wavelength observations from the Large Millimeter Telescope, the {\it Hubble Space Telescope}, Gemini South Observatory, the {\it Karl G. Jansky} Very Large Array, and the Atacama Large Millimeter/submillimeter Array, which we outline in this section. These observations serve to spectroscopically identify source redshifts, pinpoint lensed arcs associated with the \Planck\ DSFGs, provide priors on the foreground lensing mass structure, and offer multiple sightlines through the deflecting foreground for the purpose of constraining lens mass models. Expanded details on this workflow are given in Section \ref{sec:lensing}.

\begin{table}
	\caption{Summary of PASSAGES objects included in this lens modeling study. A brief description of each object is provided in Appendix \ref{sec:object_notes}. 
	%{\red Double-checking source of some of these foreground redshifts.} 
	}
	\label{tab:sample}
	\centering
	\begin{tabular}{cccccc} % four columns, alignment for each
		\hline
		ID & RA$^a$  & Dec.$^a$ & $z_{\rm s}^b$ & $z_{\rm lens}^c$  & Ref.\\[0.5ex]
		\hline
		PJ011646 	& 01:16:46.77 &-24:37:01.9 & 2.125 & 0.555$^d$ & 1 \\[1ex]			                                
		PJ014341 	& 01:43:41.20 &-01:47:26.0 & 1.096 & 0.594 & 2 \\[1ex]
		PJ020941 	& 02:09:41.30 & 00:15:59.0 & 2.554 & 0.202 & 3, 4 \\[1ex]
		PJ022633 	& 02:26:33.98 & 23:45:28.3 & 3.120 & 0.41 & 2  \\[1ex]
		PJ030510 	& 03:05:10.62 &-30:36:30.3 & 2.263 & 0.5$^*$ & 2 \\[1ex]
		PJ105353 	& 10:53:53.15 & 05:56:18.8 & 3.005 & 1.525 & 4, 5  \\[1ex]
		PJ112713 	& 11:27:13.44 & 46:09:24.1 & 1.303 & 0.42 & 2 \\[1ex]
		PJ113805 	& 11:38:05.53 & 32:57:56.9 & 2.019 & 0.52 & 2 \\[1ex]
		PJ113921 	& 11:39:21.74 & 20:24:50.9 & 2.858 & 0.57 & 2 \\[1ex]
		PJ132630 	& 13:26:30.25 & 33:44:07.4 & 2.951 & 0.786 & 2  \\[1ex]
		PJ133634 	& 13:36:34.94 & 49:13:13.6 & 3.254 & 0.26 & 2 \\[1ex]
		PJ144653 	& 14:46:53.20 & 17:52:33.3 & 1.084 & 0.493 & 2 \\[1ex]
		PJ144958		&14:49:58.59& 22:38:36.8 & 2.153 & 0.4$^*$ & 2 \\[1ex]
		PJ160722 	& 16:07:22.77 & 73:47:02.2 & 1.482 & 0.65 & 4 \\[1ex]
		PJ231356 	& 23:13:56.64 & 01:09:17.7 & 2.217 & 0.560 & 2  \\[1ex]
		\hline\\[-0.5ex]
	\end{tabular}
	\tablecomments{
	%{\bf Notes.}   
 	$^a$ J2000 object positions are derived from \WISE\ cross-matching.
  	$^b$ Source redshifts are determined from CO detections, reported in the listed references.
   	$^c$ Foreground lens redshifts (photometric and spectroscopic) are taken from the supplied references.
	$^d$ Spectroscopic foreground redshift determined with the Multi Unit Spectroscopic Explorer (MUSE) on the Very Large Telescope (VLT).
	%, {\red ID and PI?}
	$^*$ Preliminary photometric redshift estimate (Cooper et al. in prep.). For the lensing mass estimates from Einstein radii in \S \ref{sec:Einstein_radius}, we assume a fiducial uncertainty in the lens redshift 
	%is assumed to 
	%lie in the inner 95\% range of lens redshifts for the PASSAGES sample ($\approx 0.2 - 0.9$; \citealt{Harrington:2021aa, Berman:2022aa}); 
	of $\sigma_z = \pm 0.2$ (given the inner 68\% range of lens redshifts for the PASSAGES sample, $\approx 0.3 - 0.7$; \citealt{Harrington:2021aa, Berman:2022aa}); 
	see Table \ref{tab:lensproperties}. Properties such as magnification are not impacted by this uncertainty, as parameters encapsulating the mass of the lens are largely degenerate with lens redshift.  
	{References:} (1) Kamieneski et al., in prep.; (2) \citealt{Berman:2022aa, Harrington:2021aa}; (3) \citealt{Geach:2015aa}; (4) \citealt{Harrington:2016aa}; (5) \citealt{Canameras:2015aa, Canameras:2017aa} and \citealt{Frye:2019aa}.
	}
\end{table}

\subsection{Spectroscopic source redshifts}
\label{sec:specz}

Spectroscopic redshifts of background objects are critical for deriving accurate lens models and inferring intrinsic properties in the source plane. 
An initial follow-up of candidate lensed objects was performed with 1.1 mm AzTEC imaging and 3 mm Redshift Search Receiver (RSR) spectroscopy with the Large Millimeter Telescope (LMT), described in \citet{Harrington:2016aa} and \citet{Berman:2022aa}.
The wide bandwidth of RSR ($73 - 111$ GHz) captures at least two CO transitions at $z > 3.15$, and also in a very narrow window around $z\approx 2.2$ \citep{Yun:2015aa}, which accounts for 3 members of this subsample (PJ011646, PJ133634, and PJ144958).
The photometrically-supported redshifts of all other fields were later confirmed spectroscopically to high precision with the detection of more than 160 redshifted CO and 35 [CI] emission lines by single-dish facilities\textemdash including the Green Bank Telescope (GBT), the IRAM 30 m telescope, and the Atacama Pathfinder Experiment (APEX)\textemdash as summarized by \citet{Harrington:2021aa}.

\subsection{\textrm{HST} WFC3/IR}
\label{sec:HST_obs}

All objects in this work were observed with the IR channel of the Wide Field Camera 3 (WFC3) on the {\it Hubble Space Telescope} (\HST) during Cycle 24 (Program GO-14653, PI: J. Lowenthal), except for PJ105353, which was observed in Cycle 23 (Program GO-14223, PI: B. Frye; \citealt{Frye:2019aa}). 
Full details of the observation and reduction of the \HST\ images are given in Lowenthal et al., in prep., but we provide a brief summary here.
The F160W wide-band filter ($\lambda_{\rm effective} = 1.54 \micron$, $H$-band) was used to image each member of our sample for one orbit each (with an additional orbit in F110W, $1.15 \micron$, for PJ105353). Each orbit was composed of a 5-point dithering pattern for 500 seconds each, which were combined during data reduction with {\it astrodrizzle} (\texttt{final\_pixfrac} of 0.9). A fiducial $5\sigma$ sensitivity of $m_{AB} \approx 28.7$ was reached for an unresolved point source. Due to the sub-pixel dithering setup ($0.572\arcsec$ spacing), the native pixel scale was improved by a factor of 2 to $0.065\arcsec$ (with a point spread function FWHM = $0.2\arcsec$). Before combining, the individual exposures were pipeline-reduced, flat-fielded, and calibrated with standard Space Telescope Science Institute routines (e.g., \citealt{Deustua:2016aa}).
The absolute astrometry of the drizzle-combined exposures was subsequently calibrated using the position of stars within the image frame from the {\it Gaia} Data Release 1 catalog \citep{Gaia-Collaboration:2016aa}. For fields containing at least 3 {\it Gaia} objects detected within the \HST/WFC3 field-of-view ($141\arcsec \times 125\arcsec$), we can apply a 3-dimensional astrometric solution. In fields with fewer matched {\it Gaia} detections, we apply only 2-dimensional shifts in RA/Dec. 

\subsection{ALMA 1.1 mm continuum}
\label{sec:alma_cont}

We obtained 
%Band 3 (2.6 - 3.6 mm) and 
Band 6 ($1.1 - 1.4$ mm, $211 - 275$ GHz) observations with the Atacama Large Millimeter/submillimeter Array (ALMA) during Cycle 5 (Program 2017.1.01214.S, PI: M. Yun) for eleven of these 15 targets, executed between March 22 and May 5, 2018 (with the exception of one executed on August 24, 2018). 
Additional details of the program are presented in \citet{Berman:2022aa}, but we summarize relevant details here. 
A default continuum setup was employed to target rest-frame FIR/sub-mm dust emission in the DSFGs, with a total bandwidth of 8 GHz ($250-254$ GHz and $266-270$ GHz). 
%{\red Need more on SPW setup?} 
A target angular resolution of $0.4 - 1.0\arcsec$ and largest resolvable angular scale of 6$\arcsec$ was requested, with a target sensitivity of 0.1 mJy over the full bandwidth. 
%
%\footnote{{\red To do: quote actually achieved for resolution and sensitivity.}}
%
The minimum baseline utilized was $15$ meters, and the maximum baselines ranged from $484 - 784$ meters, enabling natural-weighted synthesized beam sizes of $\sqrt{\theta_{\rm min} \theta_{\rm maj}} \approx 0.4 - 0.8\arcsec$.
Total on-source integration times ranged from 17 - 21 minutes, resulting in achieved sensitivities of $1\sigma \approx 0.07 - 0.33$ mJy.
Superb weather conditions were achieved for the observations, with the mean precipitable water vapor ranging between $0.7-1.7$ mm.

As one exception, PJ105353 was observed separately with Band 6 during Cycle 3 (Program 2015.1.01518.S, PI: N. Nesvadba)
and with Band 7 ($0.8 - 1.1$ mm, $275 - 373$ GHz) during Cycle 7 (Program 2019.1.01636.S, PI: M. Yun).
%, both with a continuum setup. 
As the Band 7 observation had an angular resolution closer to that of our Cycle 5 program than the Band 6 observations (0.35$\arcsec$ vs. 0.07$\arcsec$; \citealt{Canameras:2017aa,Canameras:2017ab}), we opt to use the Band 7 image for our analysis in this work 
%{\red (confirm?)} 
so as to mitigate the effects of highly disparate beams on derived magnifications. However, we refer to the Band 6 observation when deriving a lens model; see Appendix \ref{sec:object_notes}.

The 1 mm fluxes from ALMA recovered those measured with the LMT/AzTEC bolometer camera, suggesting that the higher-resolution interferometric images do not resolve out significant flux at large angular scales (see Figure 8 and relevant discussion in \citealt{Berman:2022aa}), ensuring a complete assessment of lensed arcs\footnote{Cluster-scale lenses, most of which are not included in this analysis, are more likely to be affected by this.}. 
This is important as, throughout this work, we use lensing magnifications at 1 mm as a proxy for the total infrared magnification, e.g. in tabulating intrinsic luminosities in Table~\ref{tab:effective_radii}.
In theory, source-plane structure (and in turn magnification) can vary continuously with infrared wavelength, but this is clearly not feasible to measure at high spectral resolution in practice.

\subsection{JVLA 6 GHz}

All members of this sub-sample were observed with the {\it Karl G. Jansky} Very Large Array (JVLA) 
%during Semester 18A 
(Program 18A-399, PI: P. Kamieneski). 
The observations were carried out between March 29 and April 30, 2018 in 13 execution blocks totaling 38.9 hours, each targeting two PASSAGES objects. 
All objects were observed at 5 cm with C-band ($4 - 8$ GHz) and full polarization for 1.5 hours each. With the WIDAR correlator configured to 3-bit sampling, the effective bandwidth of the two basebands is 4 GHz, centered at 6 GHz. 
%Each baseband is further divided into 16 spectral windows, each consisting of 64 channels. 
The targets were scheduled to be observed in pairs at similar right ascensions, alternating between the two during 3-hour tracks, to improve $uv$-coverage. The most extended A-configuration was used to provide optimal resolution ($\sqrt{\theta_{\rm min}\theta_{\rm maj}} \approx 0.3-0.7\arcsec$, median $0.4\arcsec$). With maximum and minimum baselines of 36.4 km and 0.68 km, the largest recoverable angular scale was 8.9$\arcsec$, which we assume to be larger than most PASSAGES objects. 
The 6 GHz continuum data were reduced using the Common Astronomy Software Applications, \casa\ \citep{McMullin:2007aa}. Basic flagging and calibration were performed with the VLA Calibration pipeline (version {2018.1}). Each object was first imaged using natural weighting (maximal sensitivity at the expense of slightly degraded resolution). After creating a `dirty' image with no deconvolution, the sky noise level is estimated, before using the CLEAN algorithm \citep{Hogbom:1974aa} to deconvolve down to a $2\sigma$ threshold. 
Achieved sensitivities ranged from $1\sigma = 2.6 - 11.7~\mu$Jy (median 2.9 $\mu$Jy).
%\footnote{{\red To do: give range of achieved sensitivities for context.}}
%
For fields where higher resolution is desired, we also created images with Briggs weighting (typically \texttt{robust} $ = 0.5$).

Photometry for natural-weighted 6 GHz imaging of the background DSFGs is performed using the flood-filling source extraction software \textsc{blobcat} \citep{Hales:2012aa}. Gravitational lensing renders the majority of the PASSAGES sample to be resolved, extended sources, inconsistent with simple 2-dimensional Gaussian profiles. For this reason, we use the uncorrected integrated surface brightness when performing continuum photometry (see discussion in \S 3.3 of \citeauthor{Hales:2012aa}). The total observed 
%(i.e. lensing-uncorrected) 
6 GHz flux ($S_{\rm 6 GHz}$) is reported in Table \ref{tab:effective_radii_VLA}.

\subsection{Gemini $r'$ \& $z'$}

We have additionally obtained Gemini Multi-Object Spectrograph (GMOS) $r'$ and $z'$ imaging for eight (out of 15) of the members of this sample as part of Gemini-South programs GS-2018A-Q-216 (PI: J. Lowenthal; PJ011646, PJ014341, PJ144653, PJ231356), GS-2018B-Q-123 (PI: J. Lowenthal; PJ020941, PJ030510), and GS-2020A-Q-217 (PI: J. Lowenthal, PJ113921, PJ144958).
In tandem with our $H$-band images from \HST, this three-color imaging (shown in Fig.~\ref{fig:postage_stamps}) allows us to identify multiple images, separate foreground and background objects, and identify lensing groups and clusters with techniques such as the Red Cluster Sequence, which is effective to at least $z=1.4$ \citep{Gladders:2000aa, Gladders:2005aa}. Each target was observed in the $r'$ filter for 1500s to achieve $S/N = 10$ for an object with $r'=25.0$, and in the $z'$ filter for 2100s to achieve $S/N=10$ for a $z'=23.0$ object. These limiting magnitudes are each chosen to be one magnitude fainter than a typical early-type L$^\star$ cluster member at $z\sim 0.5$ \citep{Gladders:2005aa}, in order to capture the foreground objects most likely to be responsible for lensing. Each field was observed with dithered 300-second exposures to cover chip gaps within the GMOS $5.5\arcmin \times5.5\arcmin$ field-of-view. The observations have a pixel scale of 0.080$\arcsec$. 
Observations were taken on July 13-14, 2018 for semester 2018A, on October 7, 2018 for semester 2018B, and on February 20, 2020 for semester 2020A.
Absolute image astrometry was corrected in the same manner as the \HST\ images using Gaia catalog DR1 before creating RGB images (see Fig.~\ref{fig:postage_stamps}).
Images were bias-subtracted and flat-fielded using the Data Reduction for Astronomy from Gemini Observatory North and South (\textsc{dragons}\footnote{\url{https://github.com/GeminiDRSoftware/DRAGONS}}) software, as will be described in Cooper et al., in prep.

\section{Gravitational lens modeling with LENSTOOL}
\label{sec:lensing}

% More basic intro:

All gravitational lens modeling in this work was performed with the publicly-available software, \lenstool\footnote{\url{https://projets.lam.fr/projects/lenstool/wiki}} \citep{Kneib:1993aa, Kneib:1996aa,Jullo:2007aa, Jullo:2009aa}.
\lenstool\ uses parametric forms to model the foreground lens mass distribution that contributes sufficiently to deflect light from background objects. 
This deflection is described by the simple lens equation,
\begin{equation}
\boldsymbol{\beta} = \boldsymbol{\theta} - \frac{D_{LS}}{D_S} \boldsymbol{\hat{\alpha}(\xi)},
\end{equation}
where $\boldsymbol{\beta}$, $\boldsymbol{\theta}$, and $\boldsymbol{\hat{\alpha}(\xi)}$ are the vector quantities describing (respectively) the intrinsic angular position of the source, the observed angular position after deflection, and the deflection angle at impact parameter $\boldsymbol{\xi}$. $D_S$ and $D_{LS}$ are the angular diameter distances to the source plane and from the lens plane to the source plane, respectively. 
The deflection angle itself is the integral of surface mass density in the lens plane:
\begin{equation}
\boldsymbol{\hat{\alpha}(\xi)} = \frac{4G}{c^2} \int_{\mathds{R}^2} d^2 \xi' ~\Sigma(\boldsymbol{\xi'})\boldsymbol{\frac{\xi - \xi'}{|\xi - \xi'^2|}}
\end{equation}
where $G$ is the Newtonian gravitational constant and $c$ is the speed of light. $\Sigma$ is surface mass density evaluated at the location of the impact parameter $\xi$, after projecting the volumetric density distribution on to the lens plane \citep{Schneider:1992aa}.
Thus, provided the locations $\boldsymbol{\theta}$ (and redshifts) of multiple images identified visually by the user, one can use this formalism to model the distribution of mass in the lens plane.
%The model parameters are constrained by the locations and redshifts of multiple images identified visually by the user. 
The process of isolating image families is usually iterative, where the model may predict unidentified image locations. When confirmed, these added images can be incorporated into the model for further refinement. However, for many of the lensed systems in this sample, which are predominantly galaxy-scale lenses, identifying image families can be straightforward. We describe this step in more detail in the following section, and label the image systems in Fig.~\ref{fig:postage_stamps}.

The only information included in the derived lens models for \lenstool\ are the multiple image catalogs supplied by the user, and the parametric mass models are likewise defined by the user. The input image catalogs we constructed are provided in Table~\ref{tab:lensconstraints}. For some lenses, this can grossly oversimplify the information provided by imaging itself, such as the extended ring-like features seen in ALMA and \HST\ observations for many PASSAGES objects (e.g., Fig.~\ref{fig:postage_stamps}). When these arcs and rings are clumpy, it may be possible to identify sub-images, or smaller-scale features that appear inside the broader multiple images. For interferometric imaging, this may be challenging, as certain spatial scales are filtered out, and correlated noise can introduce spurious point-like sources. As lens models for the PASSAGES sample are continually refined in the future, we will ideally incorporate information about extended rings, which often closely trace the critical curve.

Lens models for most cluster-scale members of the PASSAGES sample will be formulated using both parametric modeling (e.g. \lenstool) and light-traces-mass models, in an approach similar to \citet{Zitrin:2009aa,Zitrin:2015aa} and \citet{Frye:2019aa}\footnote{See also \citealt{Broadhurst:2005aa}, \citealt{Acebron:2018aa}, and \citealt{Cibirka:2018aa}.
}, the latter of which covers several members of the full PASSAGES sample that were identified independently by \citealt{Canameras:2015aa}. Here, we narrow our focus primarily to the galaxy-scale and group-scale lenses, which are more likely to be adequately described by simpler parametric mass models than larger, cluster-scale lenses.

\begin{figure*}
	\centering
	{
	\setlength\tabcolsep{2 pt}
	\begin{tabular}{@{}l l}
	\includegraphics[width=0.495\textwidth]{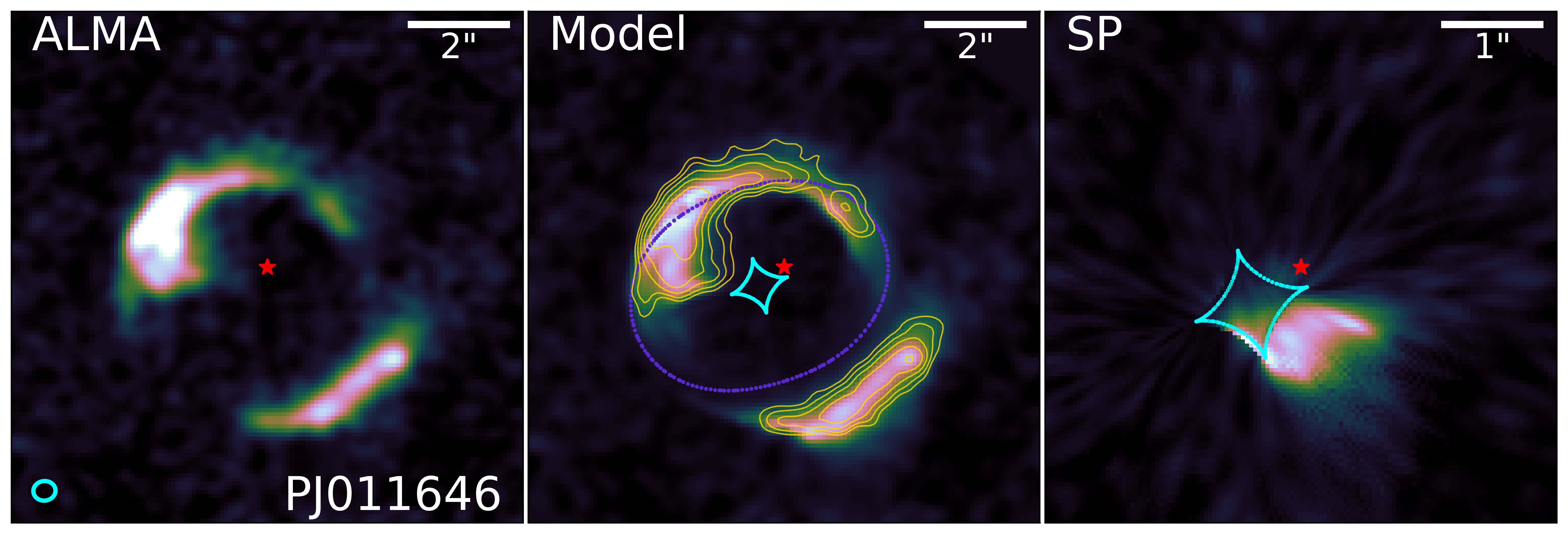}
	&\includegraphics[width=0.495\textwidth]{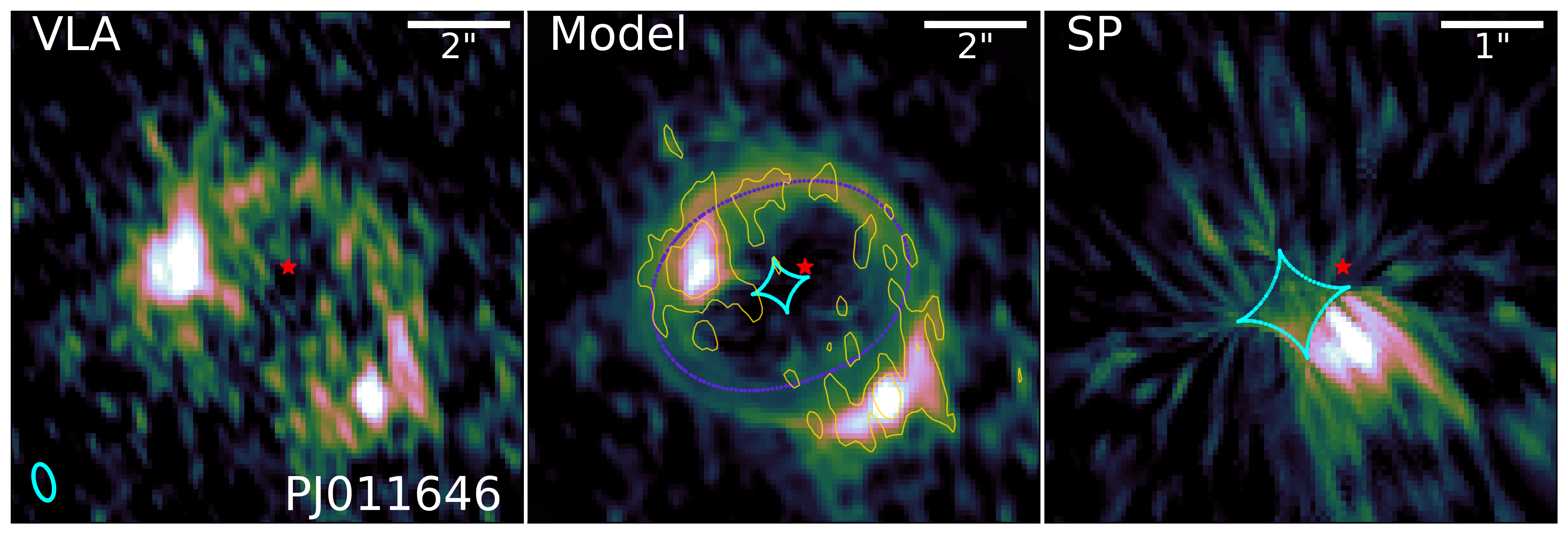}\\
	\includegraphics[width=0.495\textwidth]{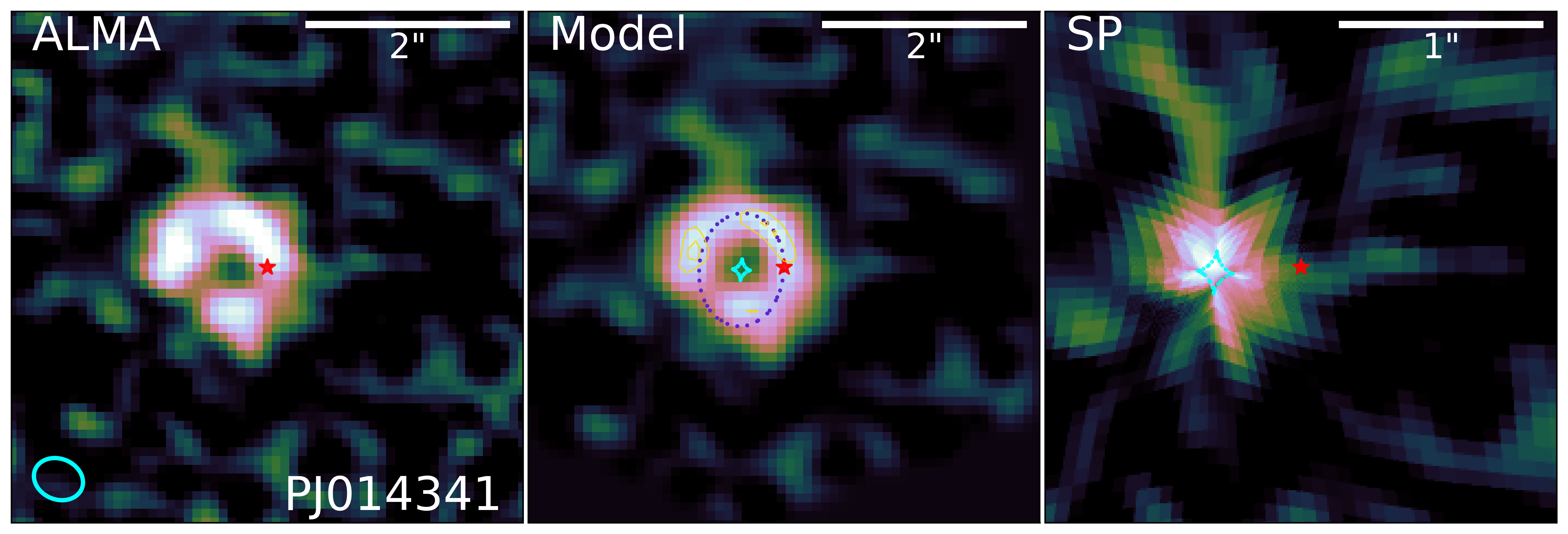}
	&\includegraphics[width=0.495\textwidth]{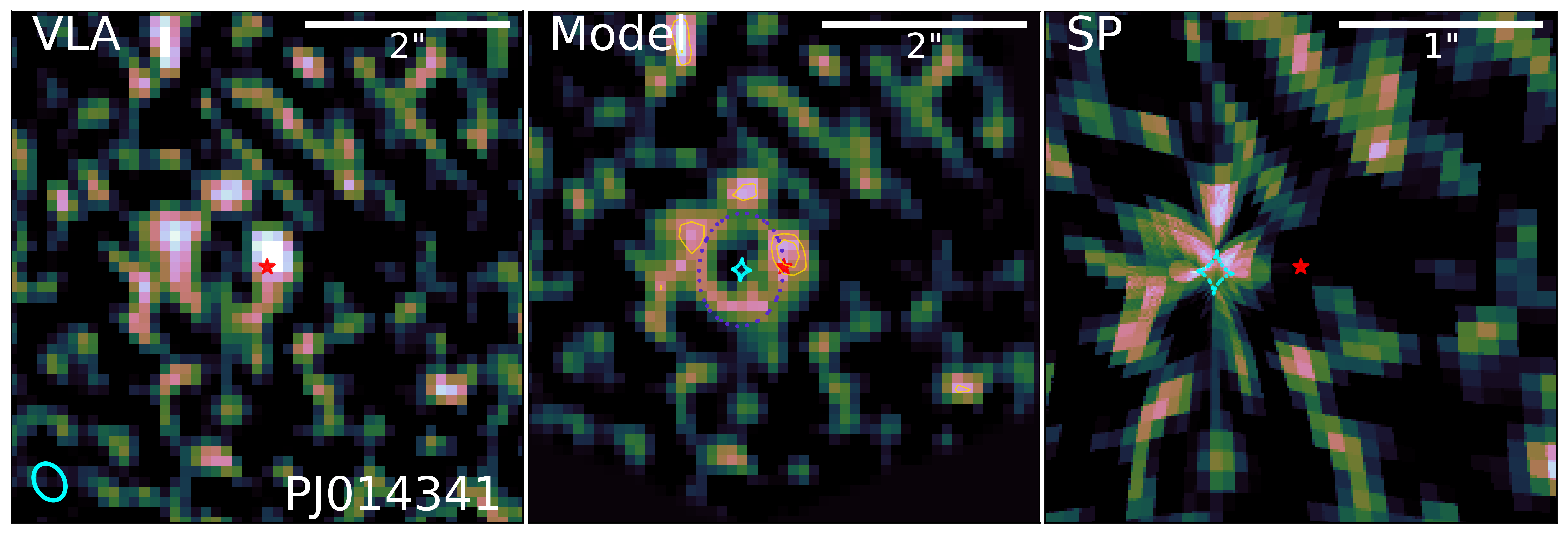}\\
	\includegraphics[width=0.495\textwidth]{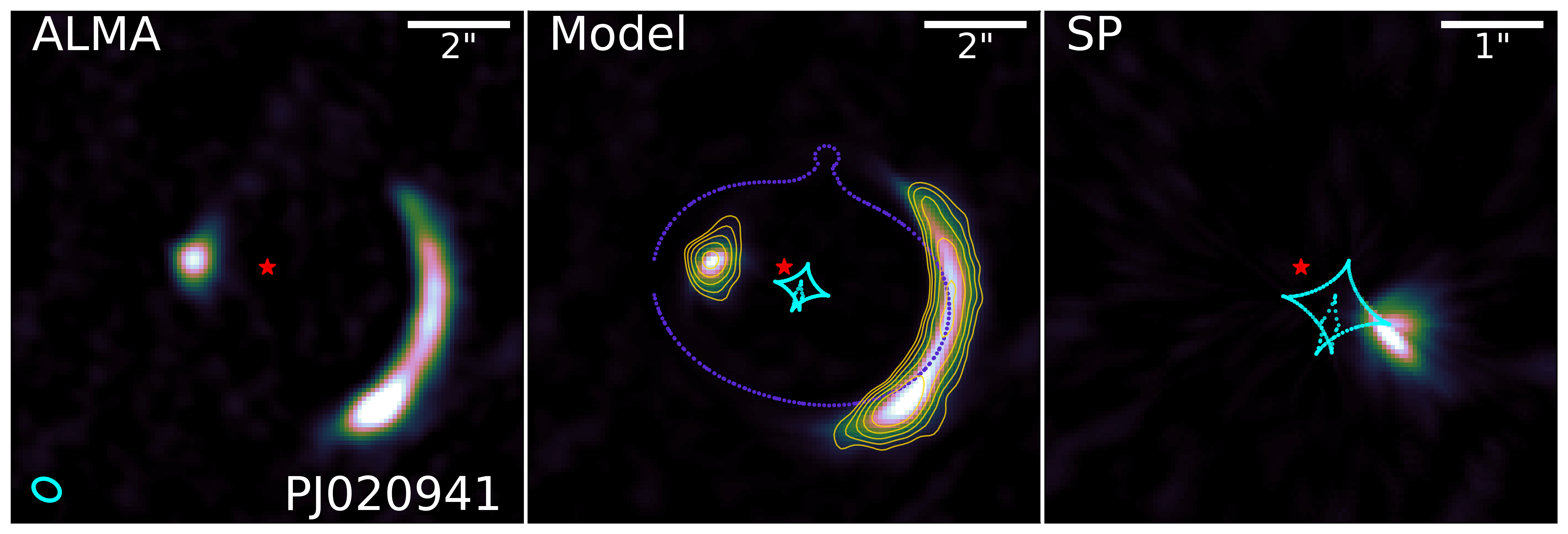}
	&\includegraphics[width=0.495\textwidth]{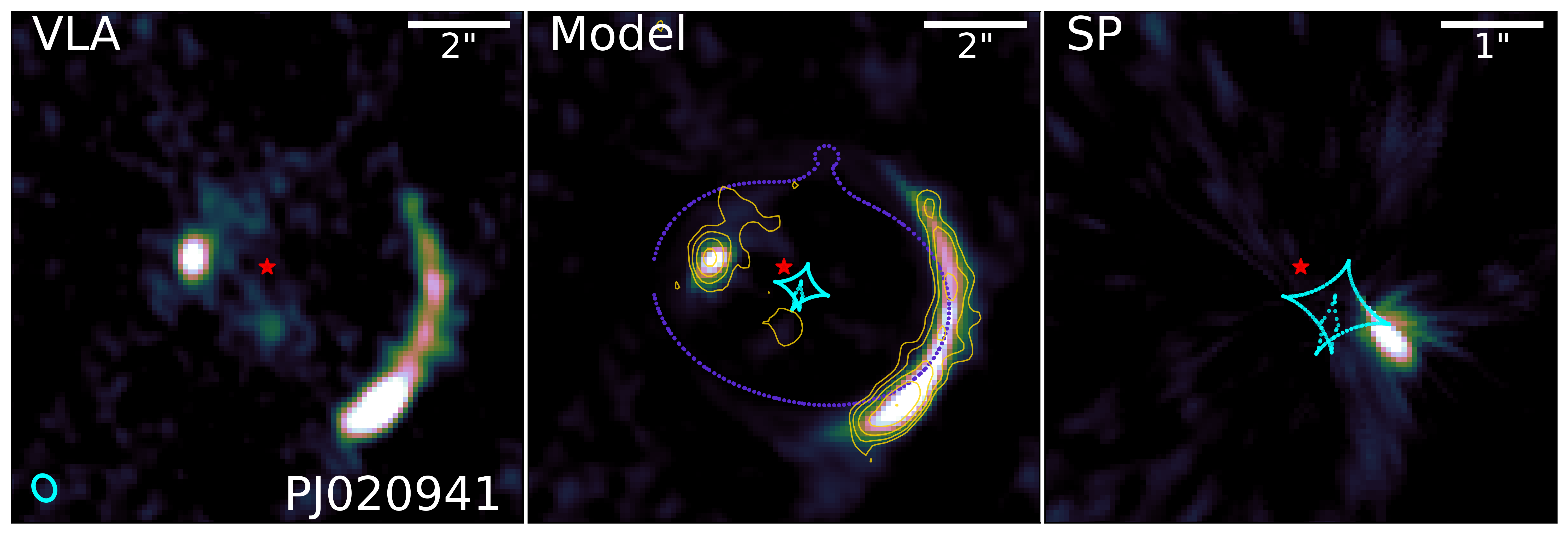}\\
	&\includegraphics[width=0.495\textwidth]{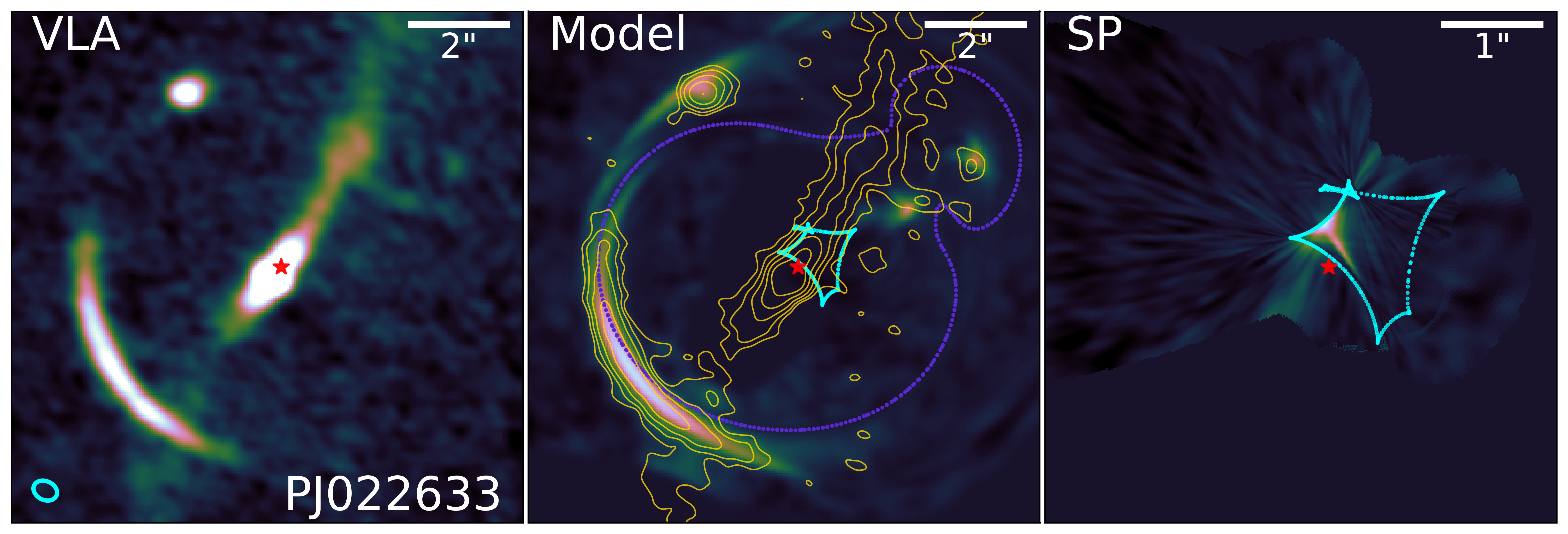}\\
	\includegraphics[width=0.495\textwidth]{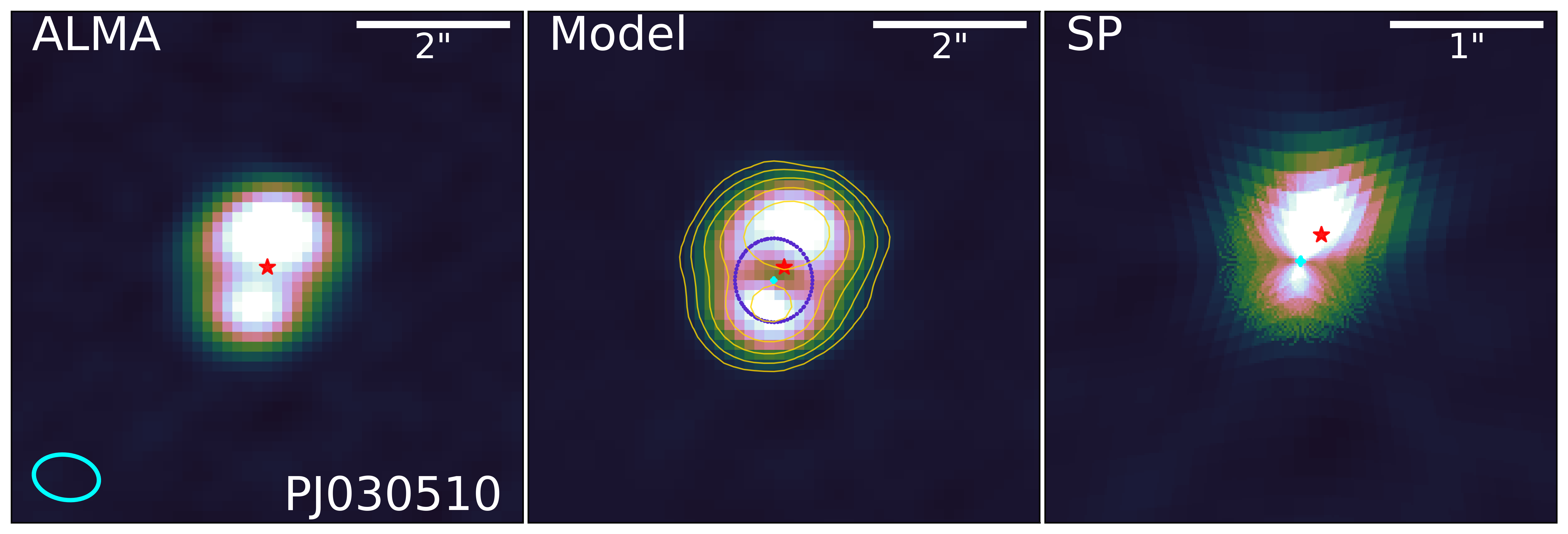}
	&\includegraphics[width=0.495\textwidth]{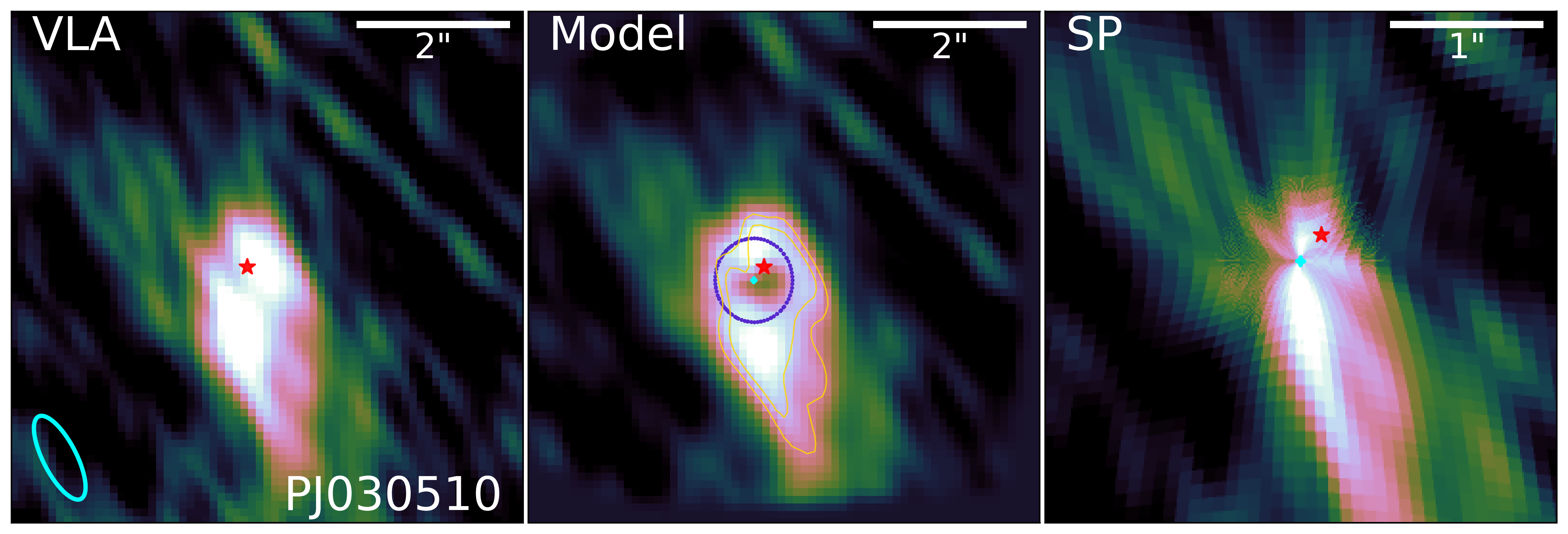}\\
	\includegraphics[width=0.495\textwidth]{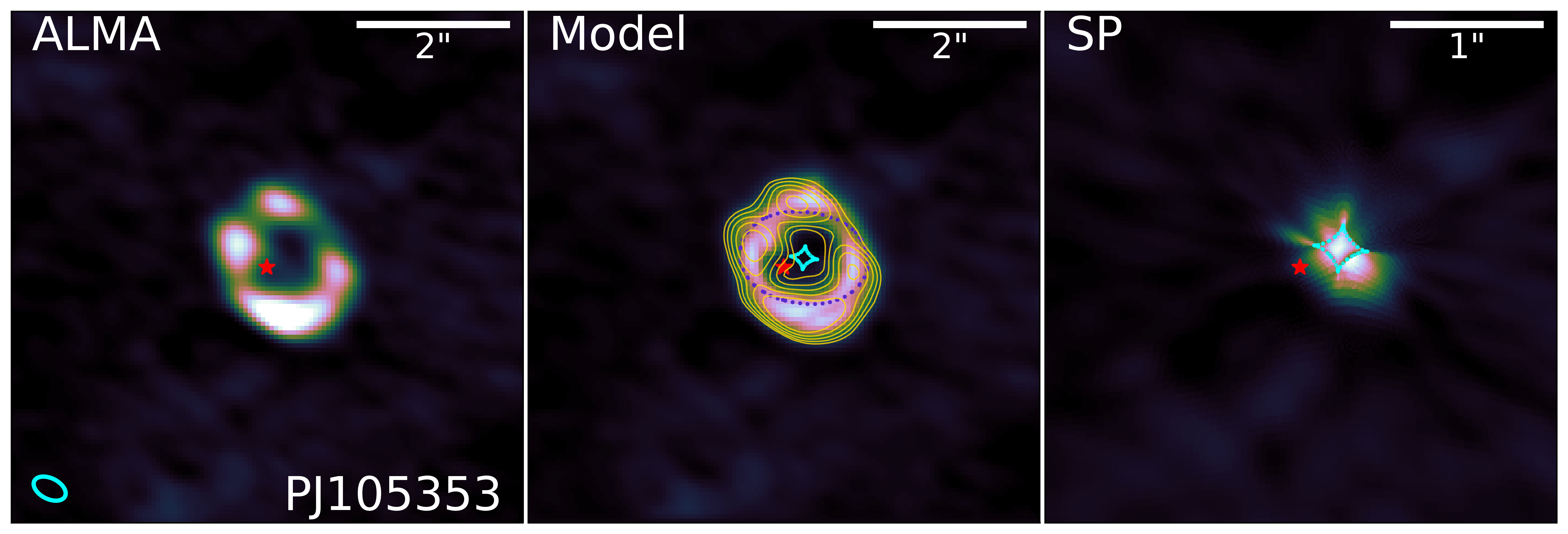}
	&\includegraphics[width=0.495\textwidth]{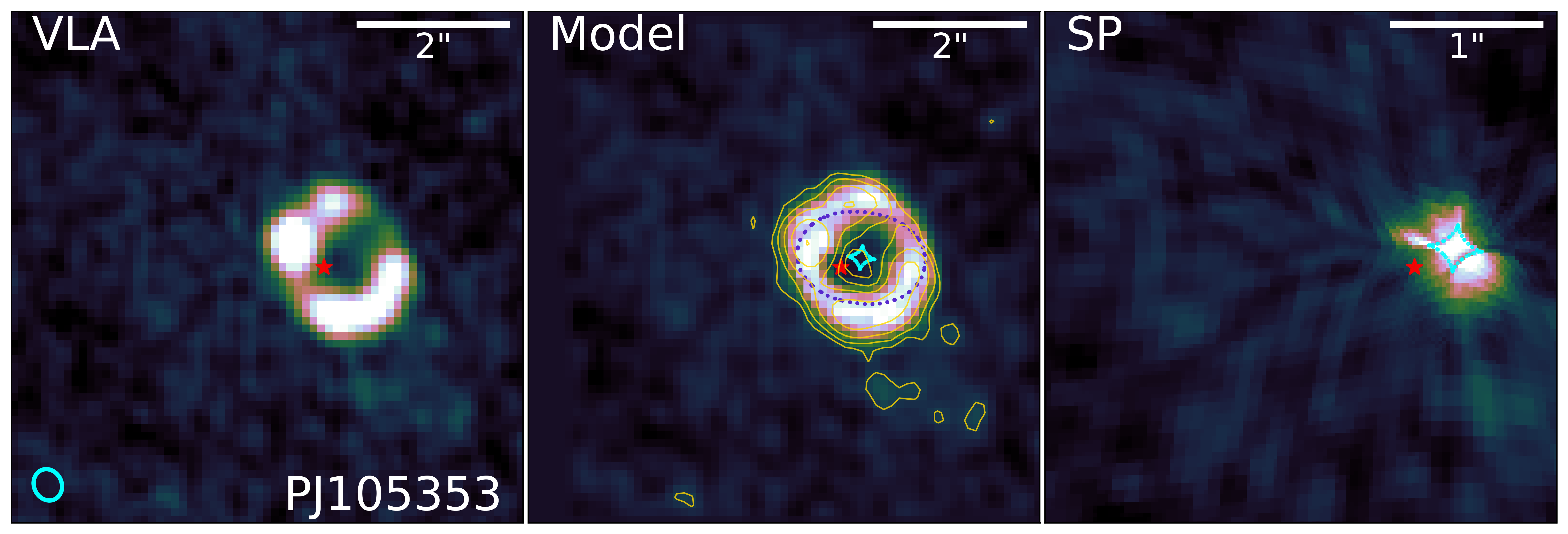}
	\end{tabular}
	}
    \caption{Gravitational lens models for ALMA 1.1 mm (left column) and JVLA 6 GHz (right column) continuum imaging. 
    %(as indicated in upper left). 
    Each row shows the observed image for different PASSAGES members ({\it left}), the model-derived image plane structure as a result of ray-tracing the observed image to the source plane and then back to the image plane, with gold contours showing the observed data for comparison ({\it center}), 
    %and the residual between observed and modeled image-plane structure ({\it right}). 
    and the source-plane reconstruction (for all multiple images combined), zoomed-in to show more detail ({\it right}). 
    All images are shown with the same colorscale limits. 
    Red stars in each panel indicate the location of the WISE centroid, with each set of panels centered on the phase center of the interferometric pointing.
    The synthesized beam is shown in cyan in the lower left of the left panels. The model-derived caustic and critical curves are shown in the center panel in cyan and purple, respectively. 
    The contours in the middle panel are typically chosen to range evenly from $3\sigma$ to the 99th percentile of the image (unless this value is less than $5\sigma$).
    The ALMA image shown for PJ105353 is at Band 7 to ensure a 
    %more consistent resolution.
    resolution more consistent with that of the other fields.
    %{\red To-do: make labels in panels larger.}
    }
    \label{fig:model_SP}
\end{figure*}

\addtocounter{figure}{-1}
\begin{figure*}
	\centering
	{
	\setlength\tabcolsep{2 pt}
	\begin{tabular}{@{}l l}
	&\includegraphics[width=0.495\textwidth]{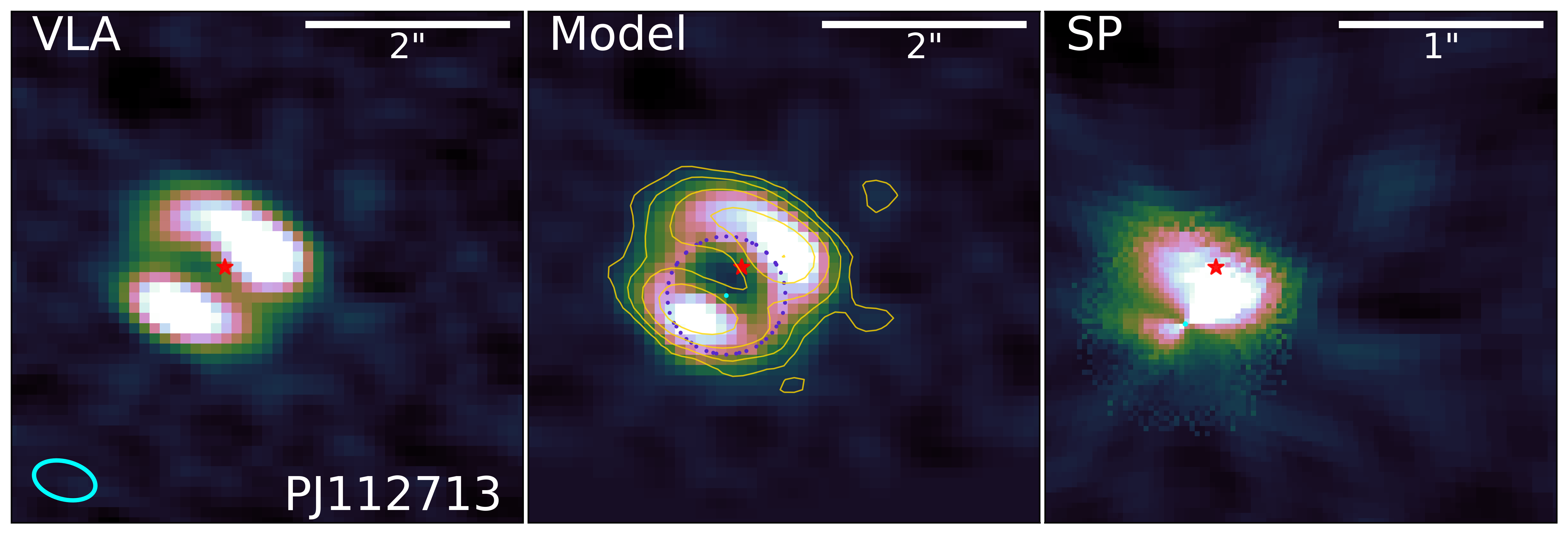}\\
	\includegraphics[width=0.495\textwidth]{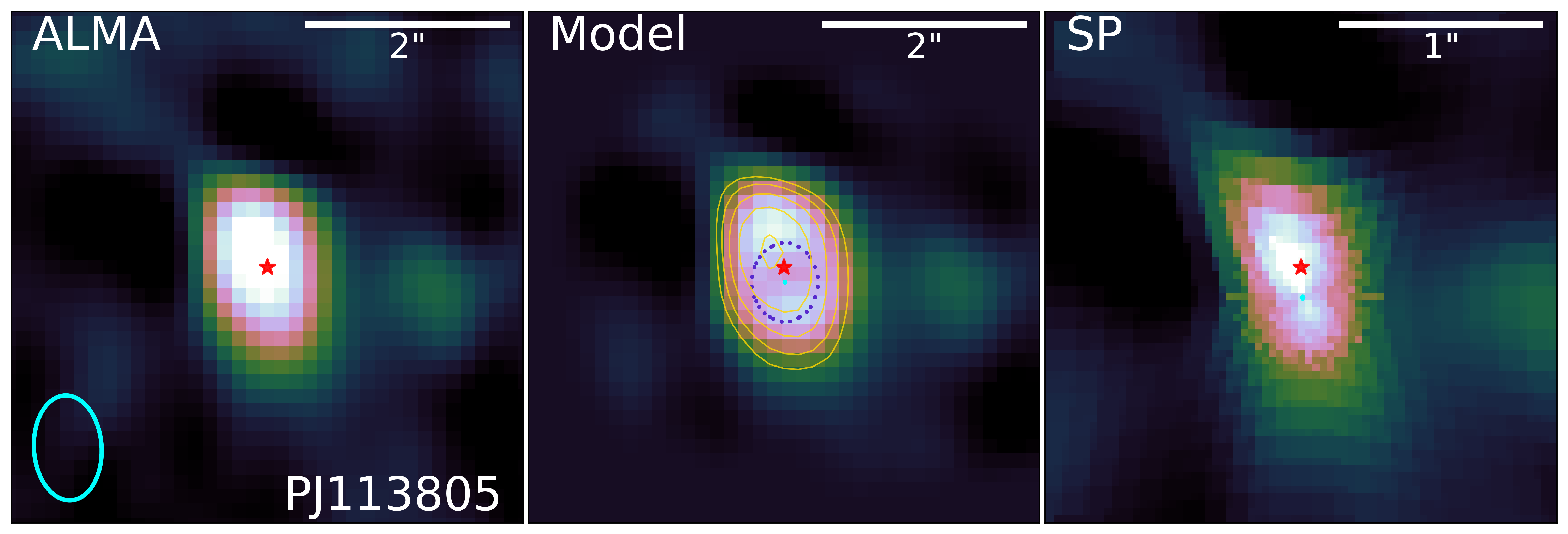}
	&\includegraphics[width=0.495\textwidth]{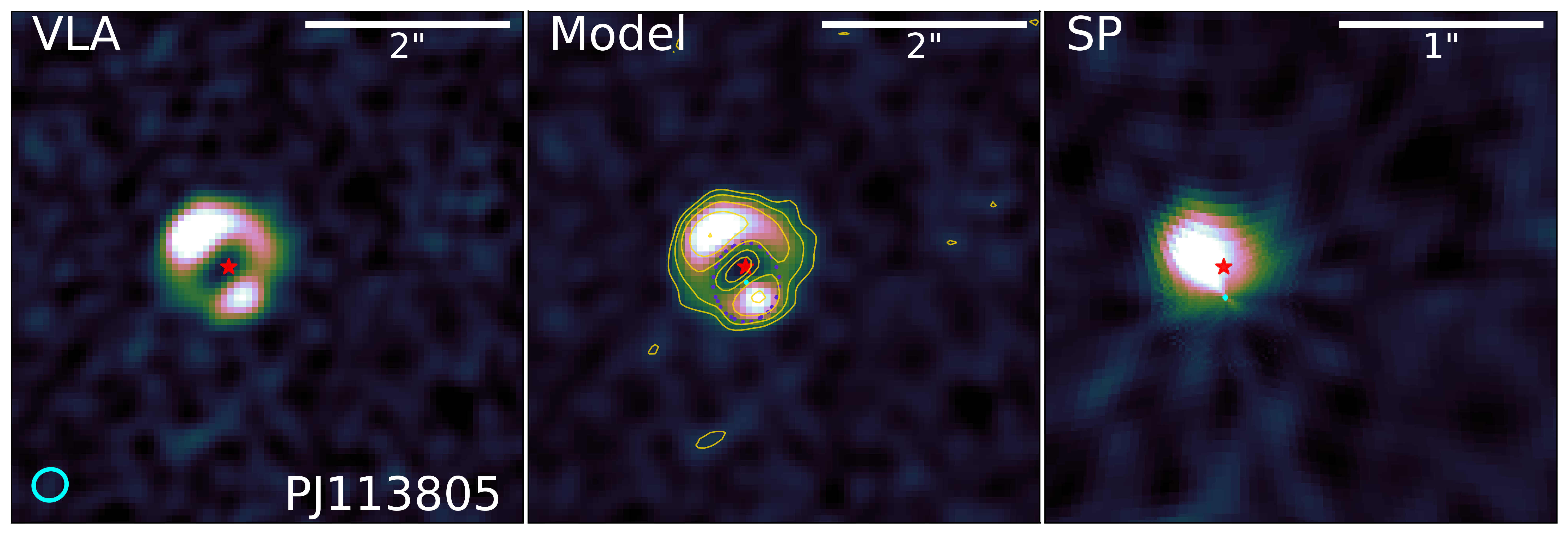}\\
	\includegraphics[width=0.495\textwidth]{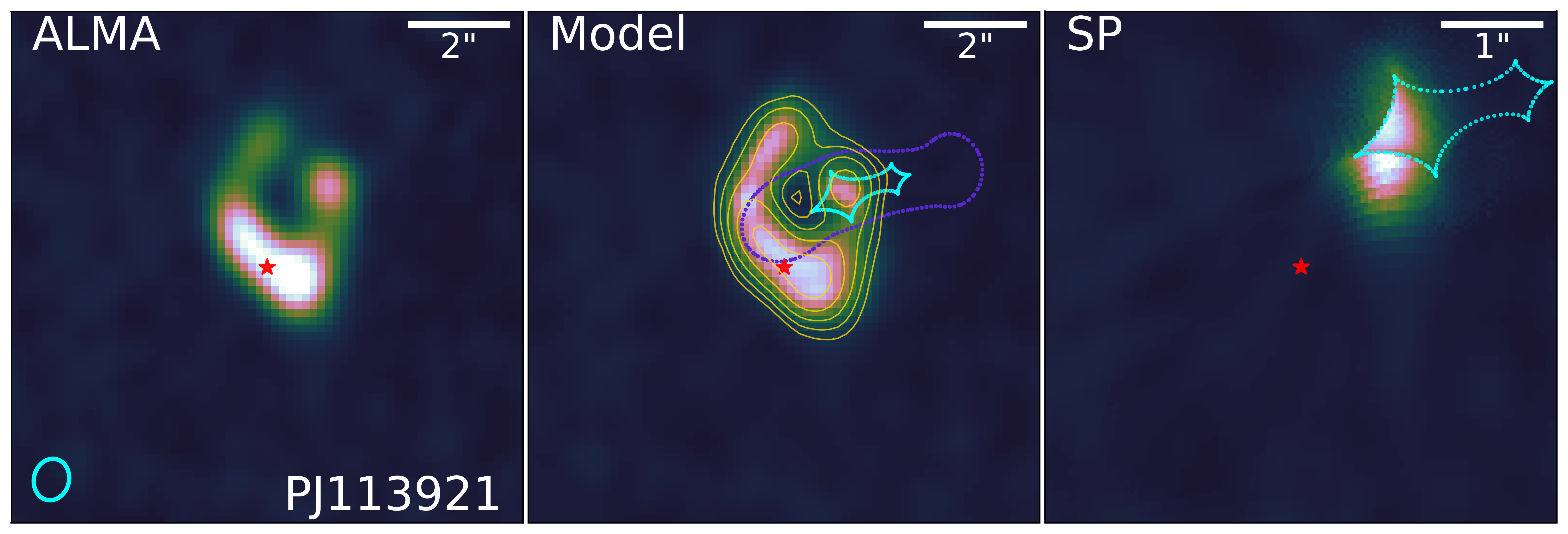}
	&\includegraphics[width=0.495\textwidth]{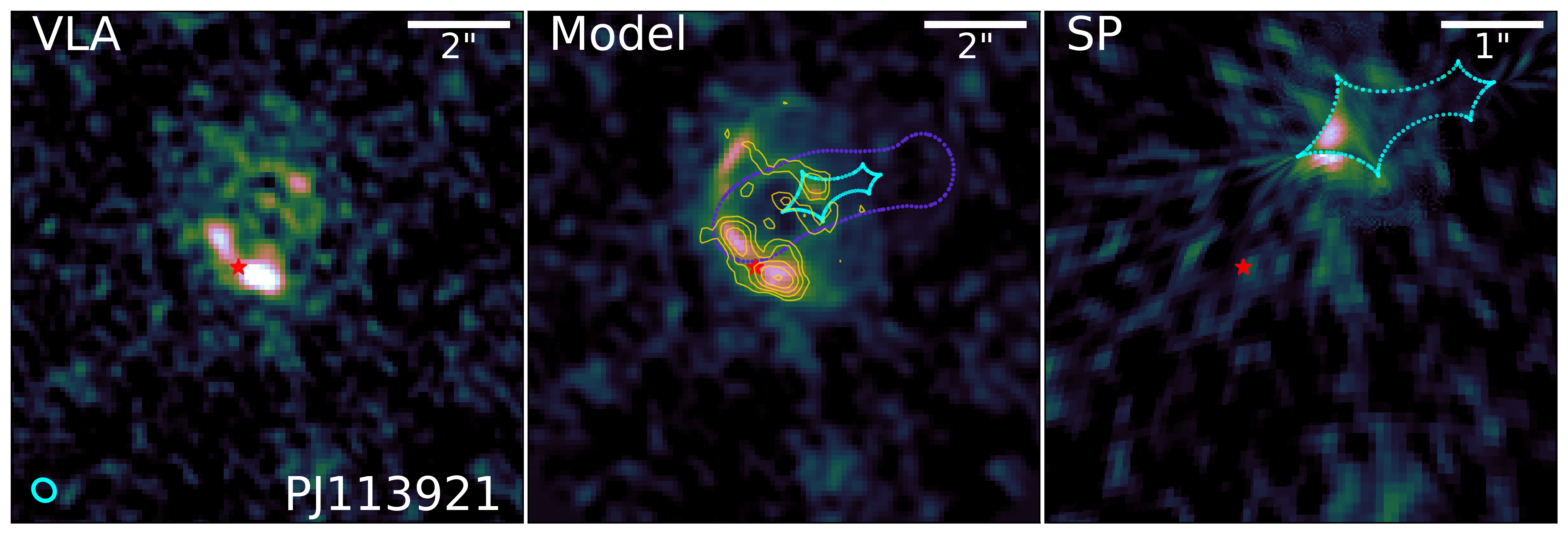}\\
	\includegraphics[width=0.495\textwidth]{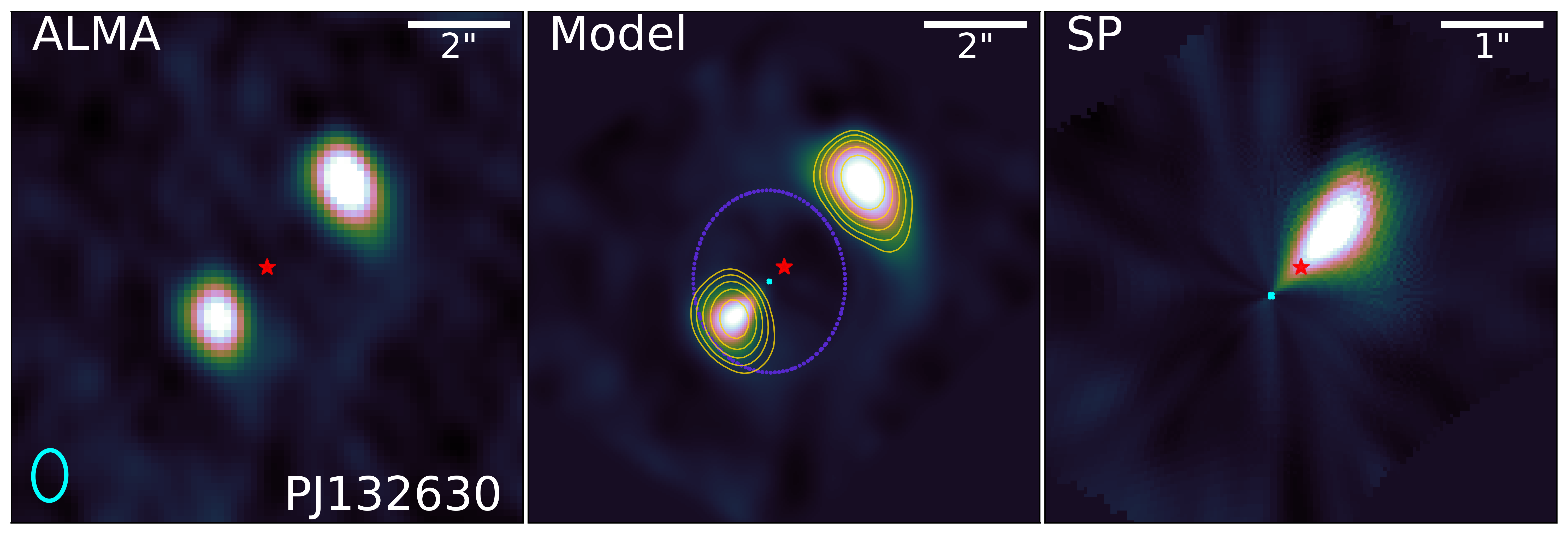}
	&\includegraphics[width=0.495\textwidth]{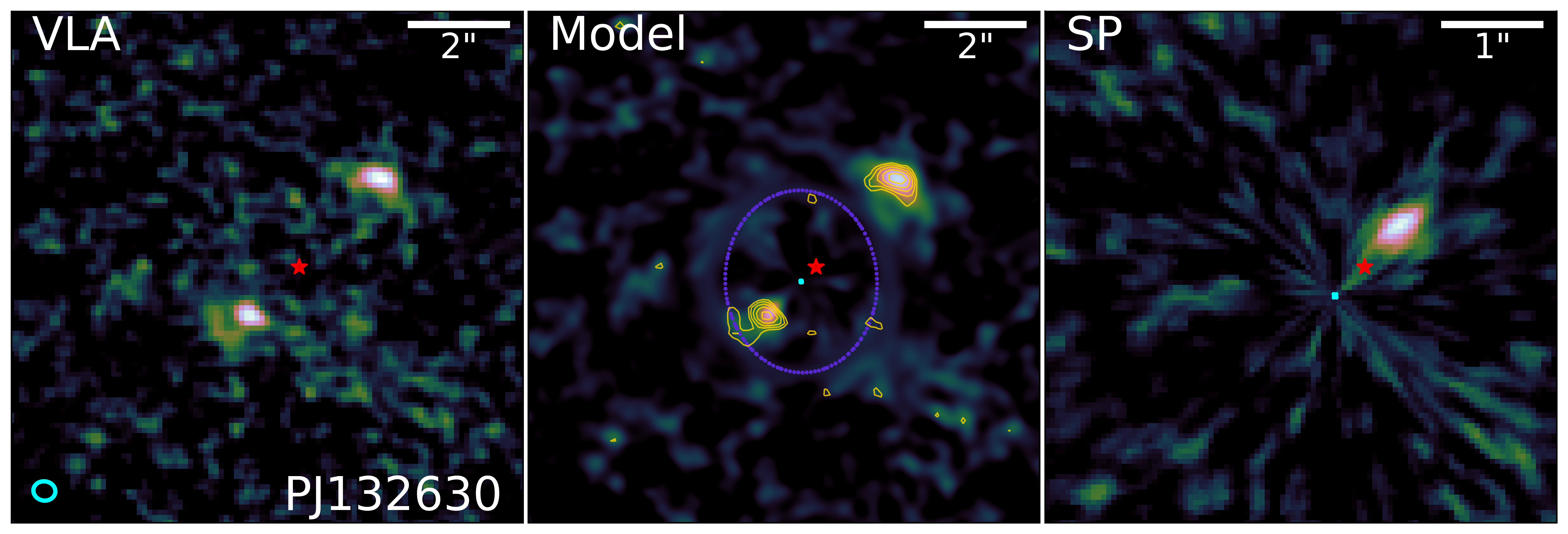}\\
	&\includegraphics[width=0.495\textwidth]{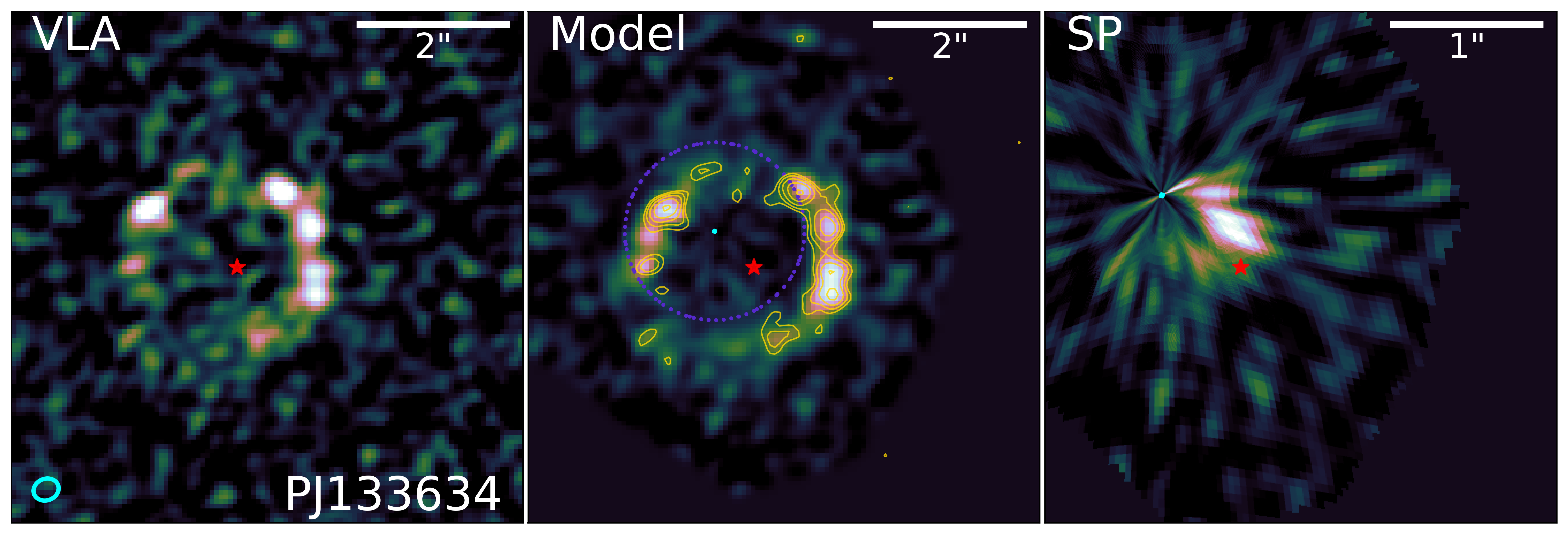}\\
	\includegraphics[width=0.495\textwidth]{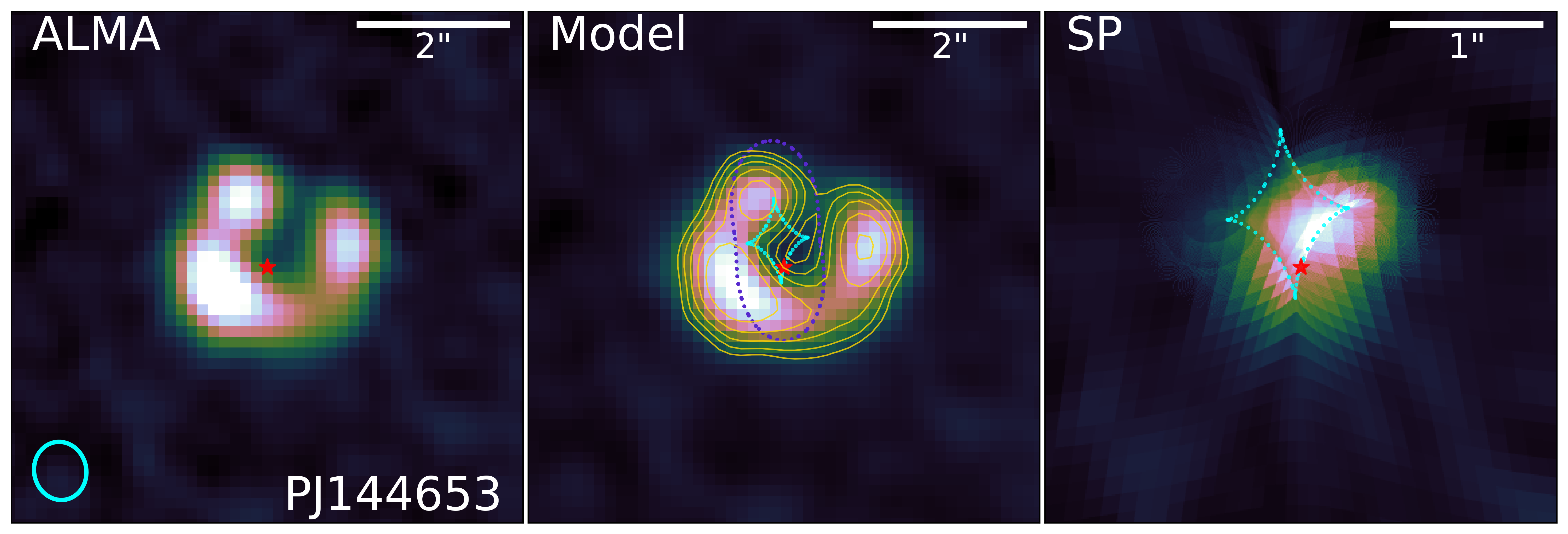}
	&\includegraphics[width=0.495\textwidth]{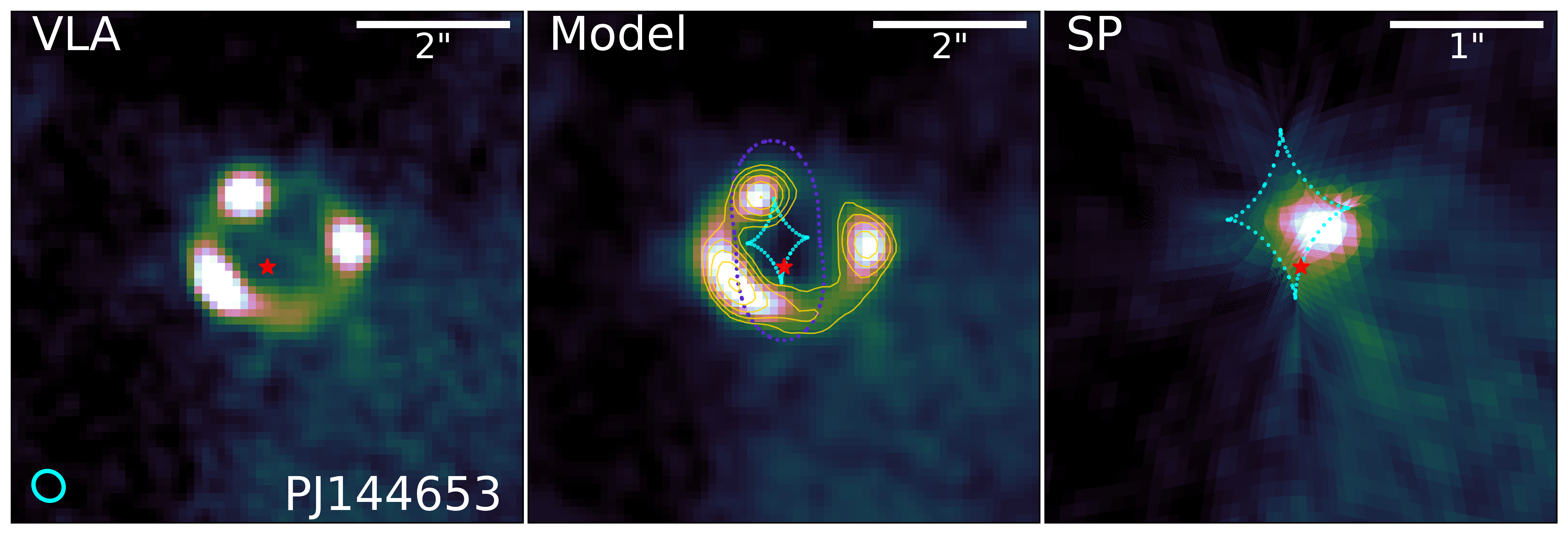}\\
	\includegraphics[width=0.495\textwidth]{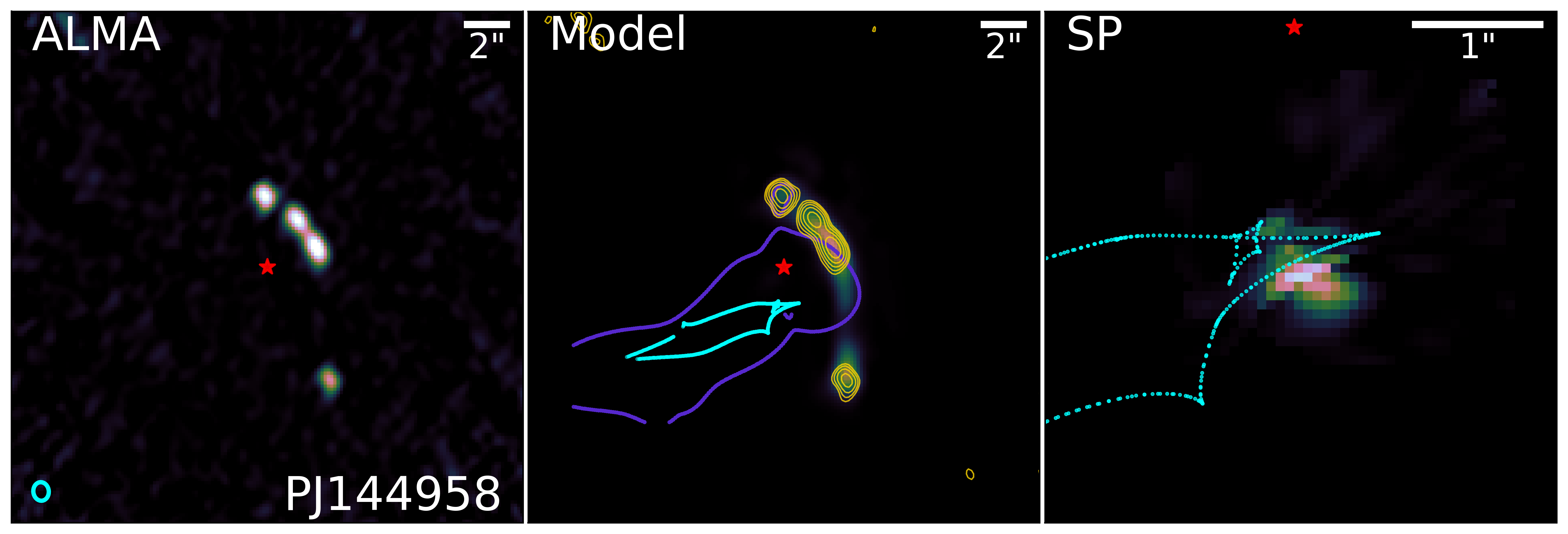}
	&\includegraphics[width=0.495\textwidth]{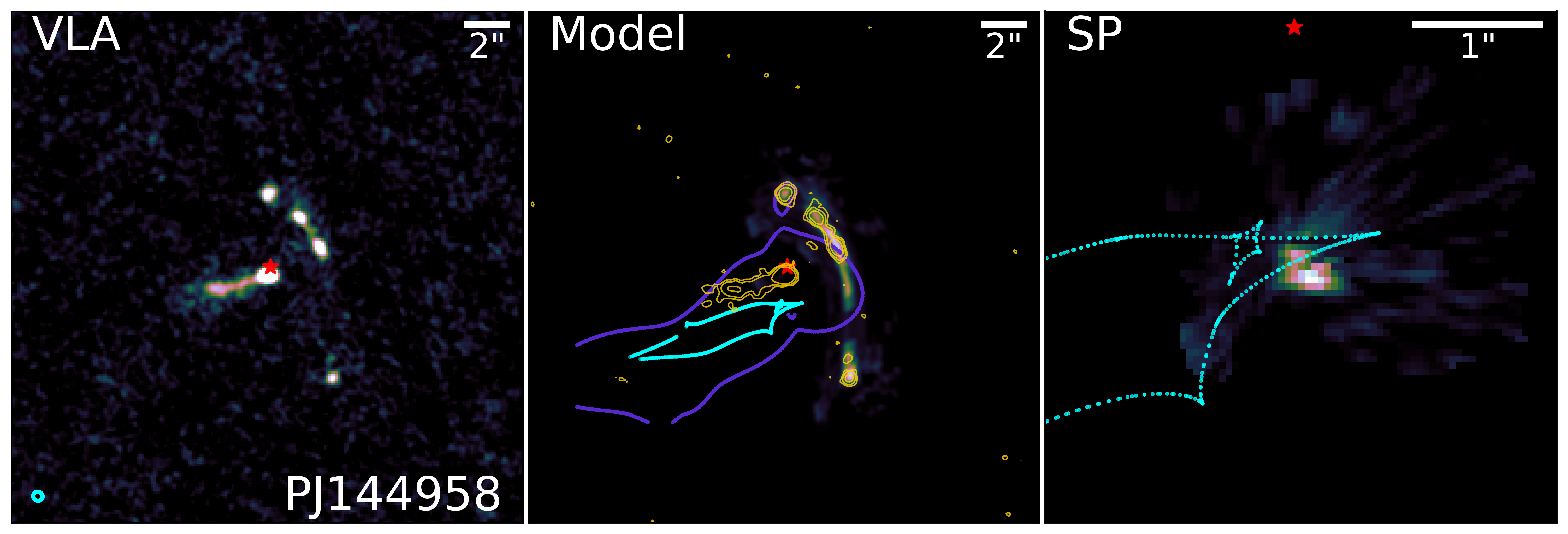}
	\end{tabular}
	}
    \caption{(Continued.) 
    }
\end{figure*}

\addtocounter{figure}{-1}
\begin{figure*}
	\centering
	{
	\setlength\tabcolsep{2 pt}
	\begin{tabular}{@{}l l}
	&\includegraphics[width=0.495\textwidth]{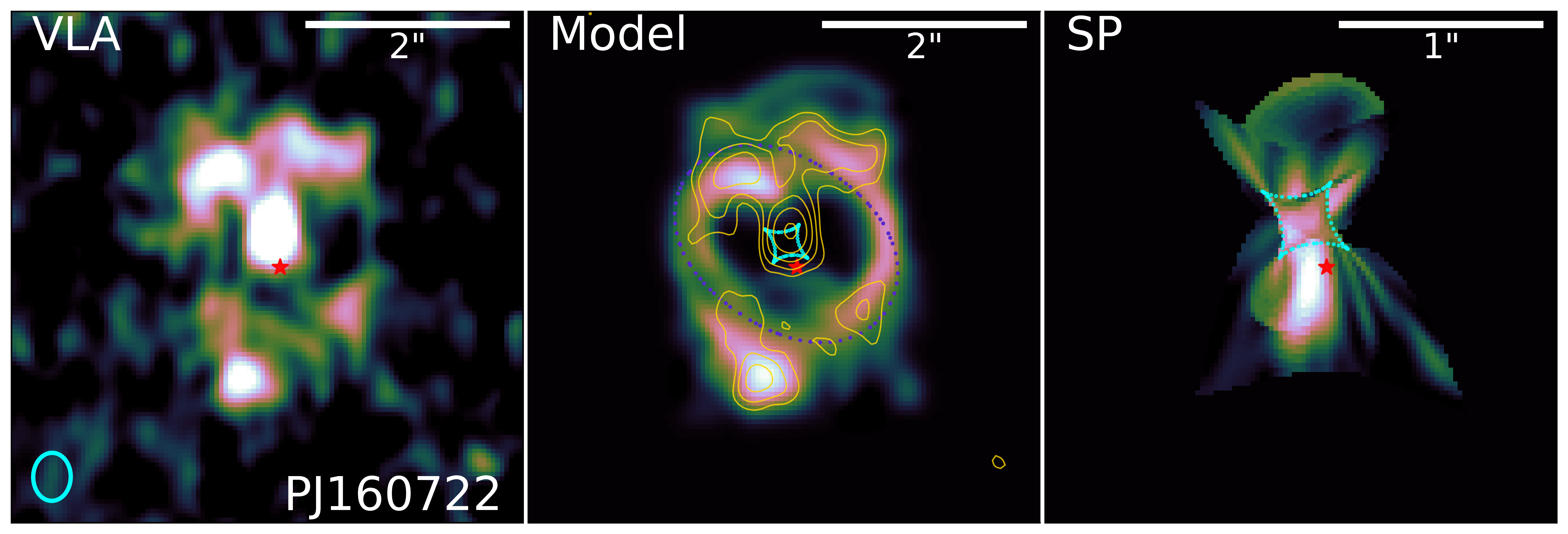}\\
	\includegraphics[width=0.495\textwidth]{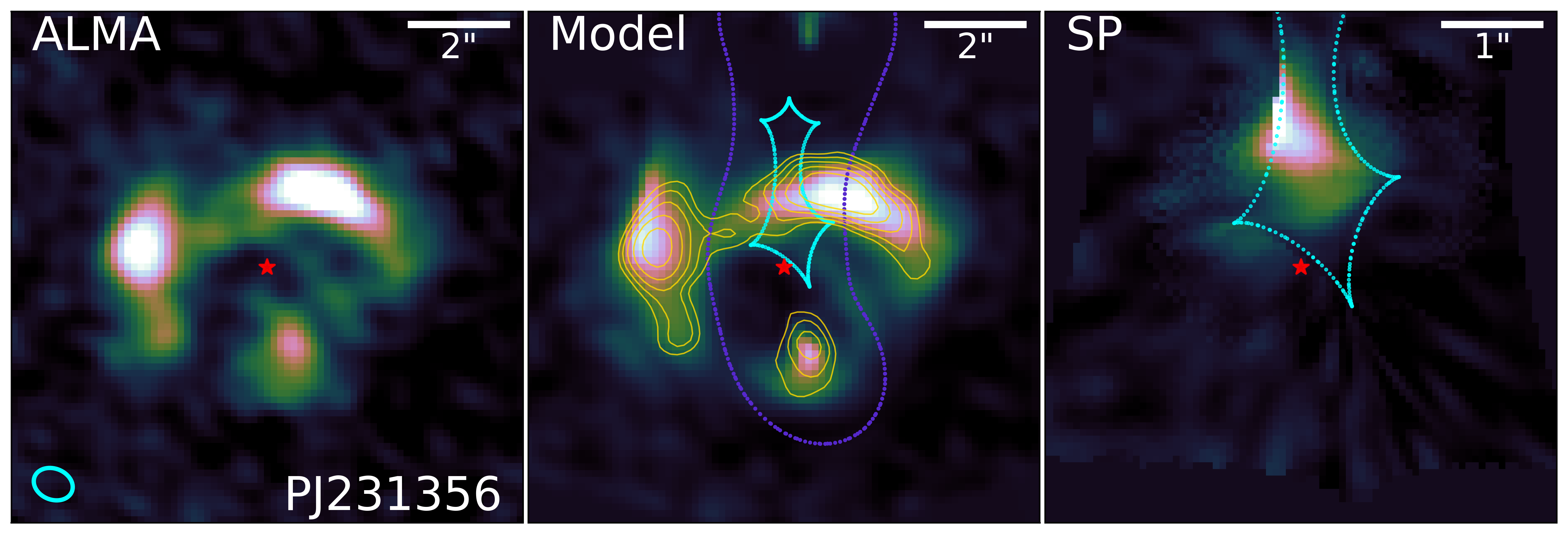}
	&\multicolumn{1}{l}{\includegraphics[width=0.168\textwidth]{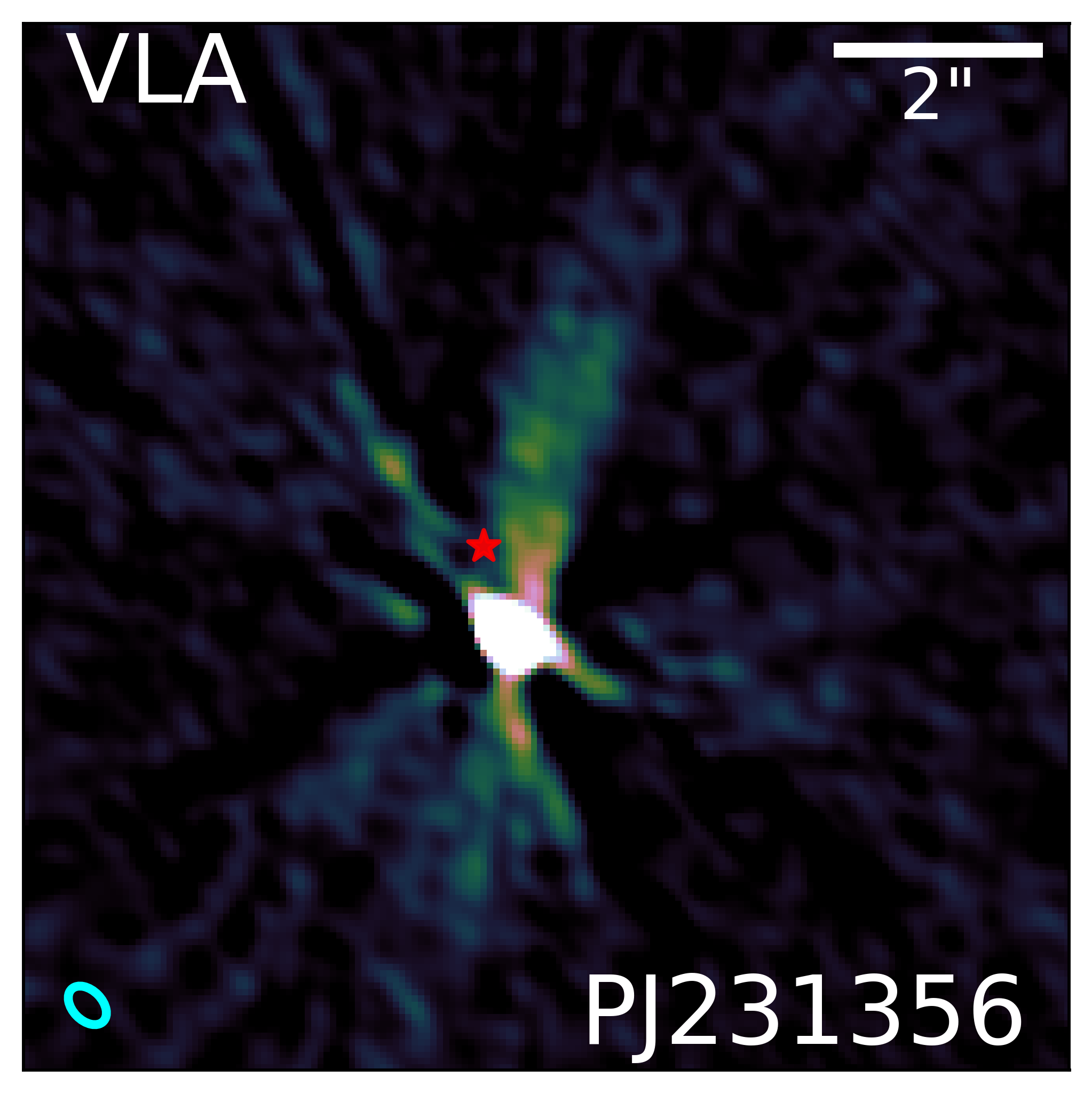}}
	\end{tabular}
	}
    \caption{(Continued.) The 6 GHz image for PJ231356 is not modeled, as the emission is dominated by an apparent AGN and jet in the foreground lensing elliptical, and any background lensing structure is not easily recovered.}
\end{figure*}

\subsection{Visual identification of multiple image families}
\label{sec:family_id}

Identifying multiple image families with long-wavelength, high-resolution radio/millimeter imaging can be easier than at shorter wavelengths, as the ubiquity of dust in the background DSFGs leads to strong obscuration in the optical regime and a much brighter rest-frame far-IR arising from the peak of the dust SED. Moreover, foreground galaxies are largely filtered out at longer wavelengths. There is a notable exception of radio AGN and jets propagating from the massive elliptical galaxies in the lens plane (e.g., \citealt{Fanaroff:1974aa}, \citealt{Smith:1986aa}), which can contaminate image classification. For this reason, our method of identifying image families relies on the combined information of optical/near-IR imaging from \HST, millimeter imaging 
%and spectroscopy 
with ALMA, and radio imaging from the JVLA.
This coverage spanning 0.6 $\mu$m to 6 cm gives us a large number of unique sightlines through foreground mass distributions, which we use to constrain our models.

The interferometric observations by ALMA and the JVLA do not directly probe the sky brightness pattern, but rather the complex visibility, which is a 2-dimensional Fourier transform of the brightness pattern. Translating the visibilities into an image introduces correlated noise, which is a function of the sampling pattern of the interferometer. For this reason, a number of works have performed gravitational lens modeling in the visibility $uv$-plane instead of the surface brightness plane (including \citealt{Bussmann:2012aa, Bussmann:2015aa, Hezaveh:2013aa, Spilker:2016aa}).
In this work, as we are merging information from traditional optical telescopes and radio interferometers, we apply lens modeling to the transformed image plane. This has the additional benefit of improved computational efficiency. As a proof of concept, \citet{Dye:2018aa} used both image-plane and visibility-plane lens modeling towards ALMA imaging of six {\it Herschel}-detected lenses for direct comparison, and found minimal difference in the derived parameters and source reconstructions, and derived total magnifications were in agreement within 1$\sigma$ uncertainties. \citeauthor{Dye:2018aa} note that there might be greater disparities in cases where the visibility sampling is sparse, but we do not expect this to be a major concern in our case. 
Both ALMA and JVLA observations make use of a large number of antennas and long integration times to maximize $uv$-coverage (and as mentioned previously, JVLA observations of fields were carried out in pairs during 3-hour tracks to ensure more distributed $uv$-sampling).

Each lensed image family of multiplicity $n$ provides $2(n-1)$ constraints to be used in the optimization\footnote{The right ascension and declination of each image provide 2 constraints for each image, but one member does not add constraints because the actual source position of all images is unknown.}. If relative fluxes (or equivalently, solid angle) of each family member are included, this adds an extra $n-1$ constraints \citep{Blandford:1992aa}. For most models herein, we use only image positions, which helps to minimize the systematic uncertainty introduced in the model. However, some lensed systems on galaxy- or group-scales with few multiple image constraints are poorly suited for strong lens modeling (see e.g. \citealt{Limousin:2009aa, Verdugo:2014aa}). In such cases, it may be necessary to include additional information in the model, which we discuss in the following section where applicable.

\subsection{Lens model mass profile optimization}
\label{sec:optimization}

\lenstool\ employs Bayesian Markov chain Monte Carlo (MCMC) methods to thoroughly explore the posterior distribution for each supplied parameter in order to find a best-fit solution. While this method is more impervious to local $\chi^2$ minima, it is more computationally expensive than other minimization techniques when the parameter space is complex. Because of the possible parameter degeneracies and non-Gaussian distributions that can arise in the underlying posterior, it is usually highly beneficial to employ MCMC. Performing this optimization in the source plane (i.e. minimizing separation between image family members once ray-traced to the source plane, rather than making the comparison in the image plane) can reduce the requisite computational time. However, this also results in lower precision and poorer estimation of model uncertainties. As the PASSAGES objects in this work are primarily galaxy-scale lenses, fewer free parameters are involved than cluster-scale systems, so we perform all optimization in the image plane unless otherwise noted. Another quantification of goodness-of-fit can be the root-mean-square of the image-plane observed vs. model-reconstructed image locations, as this is independent of the positional uncertainty, therefore facilitating an easier comparison of different lens models (e.g. \citealt{Caminha:2016aa}).

One of the priors provided to \lenstool\ is the functional form of each mass potential, including the relevant free parameters (each of which has their own prior distribution, specified by the user). 
For these models, we use a singular isothermal ellipsoid (SIE) potential as a basis for all mass profiles \citep{Kormann:1994aa}. With this model, the normalized surface mass density, or convergence $\kappa$, takes the form
\begin{equation}
\kappa(x_1, x_2) = 
\frac{
\sqrt{f}
}
{
2\sqrt{x_1^2 + f^2 x_2^2},
}
\end{equation}
where $f \equiv b/a$ is the axis ratio (for semi-major and semi-minor axes $a$ and $b$; $0 < f \leq 1$) and $x_1$ and $x_2$ are normalized Cartesian coordinates. 
As noted by \citet{Treu:2010aa}, the SIE\textemdash a generalization of the singular isothermal sphere (SIS) where 3-dimensional density follows $\rho \sim r^{-2}$\textemdash appears to be the simplest profile that effectively describes galaxy-scale strong lensing configurations (see also work by \citealt{Treu:2004aa, Koopmans:2006aa, Koopmans:2009aa, Barnabe:2009aa}).
A number of studies of lensed DSFGs have had success using this profile, including
\citet{Fu:2012aa},
\citet{Bussmann:2012aa, Bussmann:2015aa},
\citet{Hezaveh:2013aa},
\citet{Calanog:2014aa}, 
and
\citet{Spilker:2016aa}.
The profile can be fully parameterized by position ($\alpha$ and $\delta$, or $x$ and $y$), projected 2-dimensional ellipticity $e = (a^2 - b^2) / (a^2 + b^2)$, position angle (PA or $\theta$; measured counterclockwise from east),
and velocity dispersion, $\sigma_{\rm SIE}$. This velocity dispersion is generally in good agreement with observable stellar velocity dispersions for elliptical galaxy lenses (e.g., \citealt{Bolton:2008ab} for $\sigma\approx 175 - 400$ km s$^{-1}$).
For the majority of the galaxy-scale lenses in the PASSAGES sample, the arcs and rings are visible in the intermediate range between core and cut radii (e.g., Table \ref{tab:lensproperties}), and so these parameters of pseudo-isothermal elliptical mass distributions (PIEMD; \citealt{Kassiola:1993aa}) and dual pseudo-isothermal elliptical distributions (dPIE; \citealt{Eliasdottir:2007aa}) would be poorly constrained.

As galaxy-scale systems have comparatively fewer constraints than clusters that are rich with multiply-imaged arcs, the number of available free parameters is usually quite limited. In some cases, 
%we are restricted to a 
this necessitates a simple, single SIE potential, with free parameters including the location of the profile center, the velocity dispersion, and the ellipticity (and associated position angle). Where possible (and appropriate), a secondary singular isothermal ellipsoidal/spheroidal potential can be added to account for the effect of any secondary deflectors. This 2nd-order correction is similar to including an external shear field induced by underlying large scale structure (see review of weak lensing by \citealt{Bartelmann:2001aa}), which is a common approach to modeling galaxy-scale lenses, including for a number of PASSAGES objects (e.g., \citealt{Geach:2015aa,Geach:2018aa,Rivera:2019aa}). In some galaxy-scale cases, the location of the profile center is fixed to the centroid of the visible baryonic component revealed by \HST, which reduces the number of free parameters\footnote{For cluster-scale lenses, the lens position is kept as a free parameter, although often the coordinates of the BCG may be assumed to represent the cluster center.}. An upper limit is generally imposed on the ellipticity of $e < 0.75$, following \citet{Acebron:2017aa}. Theoretical predictions by \citet{Despali:2017ac} indicate that $\approx 95\%$ of galaxy-scale, $10^{12}~\Msun$ halos at $z\sim0.5$ will have a projected ellipticity less than $e = 0.5$.
%\footnote{A discussion of typical ellipticity values for other submillimeter lens samples is given in Section \ref{sec:magnifications}.}.
In our sample, there are several cases where the optimized lens models have ellipticity values larger than expected (up to $e \sim 0.7$). 
This suggests that these models may benefit from additional complexity and/or substructure (e.g. \citealt{Johnson:2014aa}), but this is often limited by the small number of constraints presently available.

The results of our MCMC optimization are summarized in Table~\ref{tab:lensmodels}, including the positional center of each mass profile (relative to reference coordinates), and the ellipticity $e$, position angle $\theta$, and velocity dispersion $\sigma$ in km s$^{-1}$. For each lensing field, we provide at least two optimized solutions, labeled {\it median} and {\it best}. The {\it median} solution comprises the median values of the MCMC-sampled posterior for each parameter, with associated asymmetric uncertainties from the inner 68\% confidence interval of the distribution, whereas the {\it best} solution is only those MCMC realizations with the highest likelihood (or lowest $\chi^2$). In some cases, the {\it mode} solution is also provided, comprising the peak of the probability density for each parameter. This latter solution may be preferable in situations where the posterior distribution is multimodal, for which the median may lie in troughs between multiple peaks in parameter space. Additionally, the mode solution can be useful when a parameter is only weakly constrained in the model, such that the choice of parameter bounds can have a strong impact on the distribution. Here, the median might be sensitive to these bounds, while the mode might be more resistant. 

In general, we select the median of the posterior distribution, which is less prone to local likelihood maxima, where small perturbations to any parameter can drastically reduce the likelihood. However, in the ideal case, the median, mode, and best solutions would be in agreement within error, but this requires a parameter space that is sufficiently simple.
Altogether, the median $\chi^2$ value for all median solutions\footnote{For fields where the mode solution is also provide, we use the higher $\chi^2$ value of the two.} is $\tilde{\chi}_{\rm med}^2 = {1.9}$, and the median for the best solutions is $\tilde{\chi}_{\rm best}^2 = { 0.3}$. This suggests that the goodness-of-fit of our lens models are reasonable, in aggregate, although the individual $\chi^2$ values are influenced by the complexity of the foreground environment and the amount of lensing evidence available from current observations.
The lensing models we select for analysis are shown in Fig.~\ref{fig:model_SP}, which depicts the observed image-plane structure relative to the model-reconstructed image-plane and source-plane structure, the latter of which we discuss in the next section. Qualitatively, the image-plane reconstructions are in agreement with the observations, mainly in that they reproduce the correct number of multiple images.

\subsection{Source-plane reconstruction}
\label{sec:SP_reconstruction}

With an optimized potential model in hand, we can invert the observed arcs to approximate the unlensed brightness distribution in the source plane using the \cleanlens\ function of \lenstool. Fundamentally, this inversion can be quite trivial, in that it consists only of ray-tracing a (usually sub-sampled) image plane distribution to the lens plane, computing the deflection vectors based on the lens mass potential, and continuing until the source-plane redshift is reached. Typically, sub-pixel sampling is also used for the source plane. The direction and magnitude of the deflection are determined by a surface integral of the surface mass density, weighted by the distance from where each light ray intersects the lens potential (see e.g., \citealt{Schneider:1992aa}, \citealt{Refsdal:1994aa}).

While the optimization and goodness-of-fit measurement with \lenstool\ is done in the image plane, 
one of the tests that we use to evaluate the goodness-of-fit of a lens model is evaluating the spatial coincidence of the source plane reconstruction of individual multiple images. Every observation has an intrinsic point spread function (PSF) with which the image-plane structure is convolved, which is carried through as a distorted, non-uniform PSF in the source plane. This will be slightly different for each individual multiple image (or even dramatically different, in the case of a source near the lensing caustics; see e.g. \citealt{Sharma:2018aa,Sharma:2021aa}). 
These individual reconstructions are compared visually
% with each other 
to ensure that the total reconstruction of all image-plane light is appropriate and self-consistent.

%\startlongtable
\begin{deluxetable}{cccccc}
	\tablecolumns{6}
	%\centering
	\tabletypesize{\scriptsize}
	\tablecaption{Summary of strong lensing properties derived from the models described in Table \ref{tab:lensmodels}, including weighted total magnification $\mu_{\rm tot}$ for different wavelengths and the equivalent Einstein radius $\theta_{\rm Ein, eq}$. Under the {\it Literature} heading, these values are listed for other PASSAGES objects not included in this study that have published lens models. Here, magnifications are derived for 870 $\mu$m, and Einstein radii are defined as the radius of the circle with equivalent area to that inside the critical curve (see discussion in Section \ref{sec:Einstein_radius}).
	\label{tab:lensproperties}
}
	%\begin{threeparttable}
	%\begin{tabular}{lccccc} % four columns, alignment for each
		%\hline
		%\hline
	\tablehead{
		\colhead{Field} & 
		\colhead{$\mu_{\rm 1mm}$} & 
		\colhead{$\mu_{\rm 6 GHz}$} & 
		\colhead{$\theta_{\rm Ein}$} & 
		\colhead{$M_{<\theta_{\rm E}}$}   \\[-1ex] 
		& & & \colhead{[$\arcsec$]} &   
		\colhead{[$10^{11} \Msun$]}  \\[-3ex]
	}	
	\startdata
		%\hline\\[-1.5ex]
		%\hline\\[-2ex]
		PJ011646 	& $7.0^{+0.3}_{-0.3}$ & $7.0^{+0.4}_{-0.3}$ & $2.37^{+0.04}_{-0.03}$ &$15.4 \pm 0.3$\\	
		% old mass: $11.1 \pm 0.3$
		%& & & &$[5.7-16.7]$\\[1ex]	                           
		PJ014341 	& $7.1^{+2.2}_{-2.2}$ & $5.9^{+2.0}_{-1.4}$  & $0.53^{+0.07}_{-0.09}$ & $1.3 \pm 0.3$ \\[1ex]
		PJ020941 	& $10.5^{+4.2}_{-3.2}$ & $12.0^{+3.7}_{-3.3}$ & $2.55^{+0.14}_{-0.20}$ & $6.6 \pm 0.6$ \\[1ex]
		PJ022633 	& \textemdash  & $28.2^{+10.7}_{-11.9}$ & $3.61^{+0.36}_{-0.30}$ & $24.7 \pm 3.2$\\[1ex]
		PJ030510 	& $2.2^{+0.6}_{-0.5}$ & $2.6^{+1.0}_{-1.0}$ & $0.58^{+0.12}_{-0.15}$ & $0.8 \pm 0.3^*$ \\
		%$0.5 \pm 0.2^*$ originally z=0.3
		& & & &$[0.5-1.1]$\\[1ex]	
		PJ105353 	& ${7.6^{+0.5}_{-0.8}}^{**}$ & $7.9^{+0.6}_{-0.8}$  & $0.71^{+0.03}_{-0.05}$ & $3.6 \pm 0.3$  \\[1ex]
		PJ112713 	& \textemdash & $4.5^{+0.5}_{-0.6}$  & $0.58^{+0.07}_{-0.04}$ & $0.8 \pm 0.1$ \\[1ex]
		PJ113805 	& $2.8^{+0.4}_{-0.3}$ & $4.4^{+1.2}_{-1.0}$  & $0.40^{+0.06}_{-0.04}$ & $0.4 \pm 0.1$ \\[1ex]
		PJ113921 	& $4.8^{+1.3}_{-0.9}$ & $8.6^{+2.8}_{-2.6}$ & $0.71^{+0.47}_{-0.39}$  &  $1.3 \pm 1.1$ \\
		%& & & &$[0.9-1.7]$\\[1ex]	
		PJ132630 	& $4.3^{+0.7}_{-0.5}$ & $11.4^{+4.5}_{-2.2}$ & $1.78^{+0.21}_{-0.14}$ & $11.0 \pm 1.5$\\[1ex]
		PJ133634 	& \textemdash & $8.3^{+4.2}_{-1.9}$  & $1.17^{+0.11}_{-0.08}$ & $1.7 \pm 0.2$   \\[1ex]
		PJ144653 	& $3.9^{+0.6}_{-0.7}$ & $7.7^{+1.5}_{-2.0}$ & $0.82^{+0.05}_{-0.04}$ & $1.6 \pm 0.1$ \\[1ex]
		PJ144958 	& $10.1^{+5.7}_{-4.7}$ & $13.6^{+15.7}_{-6.7}$ & $6.52^{+1.31}_{-1.16}$ & $84.0 \pm 22.5^*$ \\
		& & & &$[43.3-126.0]$\\[1ex]	
		PJ160722 	& \textemdash & $6.8^{+2.1}_{-1.3}$ & $1.00^{+0.11}_{-0.08}$ & $4.0 \pm 0.5$ \\[1ex]
		PJ231356 	& $6.1^{+1.2}_{-1.2}$ & \textemdash$^\dagger$ & $2.05^{+0.26}_{-0.23}$ & $11.5 \pm 1.9$ \\[1ex]
		\hline
		\hline\\[-3ex]
		{\it Literature} & $\mu_{870\mu {\rm m}}$ & & $\theta_{\rm Ein}$ & $M_{<\theta_{\rm E}}$   \\
		& & & [$\arcsec$] &   [$10^{11} \Msun$]  \\[0.5ex]
		\hline\\[-1.5ex]
		PJ105322$^a$ 	& $7.6 \pm 0.5$ &  & $5.9$ & 120.3 \\[1ex]
		PJ112714$^b$ 	& $29.4 \pm 5.9$ &  & $13$ & 283.7 \\[1ex]
		PJ132302$^c$ 	& $11.2  \pm 0.7$ &  & $\lesssim 13$ & $\lesssim 380.1$ \\[1ex]
		PJ132934$^d$ 	& $11  \pm 2$ &  & $11.0 \pm0.4$ & $267.9 \pm 13.8$  \\[1ex]
		PJ142823$^e$ 	& $3.0  \pm 1.5$ &  & $0.10 \pm 0.03$ & $0.13 \pm 0.06 $ \\[1ex]
		PJ154432$^f$ 	& $14.7  \pm 0.8$ &  & $7$ & 152.6 \\[1ex]
		PJ160918$^g$ 	& $15.4  \pm 1.0$ &  & \textemdash & \textemdash \\[1ex]
		%\hline\\[-1.5ex]
	\enddata
	%\multicolumn{5}{p{1\columnwidth}}{
	%{\bf Notes.}
	%\begin{tablenotes}\footnotesize
	%\item[$^a$] 
	%\end{tabular}
	\tablecomments{
	$^*$ Only a preliminary photometric redshift measurement (see Table \ref{tab:sample}), so approximate bounds of enclosed masses are also given for a set of redshifts representative of the range of the full sample.
	$^{**}$ 870 $\mu$m (Band 7) measurement (see \S \ref{sec:alma_cont}). The 1 mm magnification is $\mu \approx 11.8 \pm 2.1$ for the $0.07\arcsec$ image, but the 870 $\mu$m value is adopted for $\mu_{\rm dust}$ to more closely match the imaging limitations for the other targets.
	$^\dagger$ 6 GHz image of PJ231356 is not resolved from a possible radio jet, 
	so it is not currently possible to estimate magnification of the background DSFG.
	$^a$ Also known as PLCK\_G145.2+50.9; \citet{Canameras:2018ab,Frye:2019aa}.
	$^b$ Also known as PLCK\_G165.7+67.0; \citet{Canameras:2018ab,Frye:2019aa}.
	$^c$ Also known as PLCK\_G113.7+61.0; \citet{Canameras:2018ab}. Einstein radius upper limit estimated from VLA FIRST survey image \citep{Becker:1995aa}.
	$^d$ Also known as the Cosmic Eyebrow; \citet{Diaz-Sanchez:2017aa, Dannerbauer:2019aa}.
	$^e$ Also known as HBo\"otes3; \citet{Bussmann:2013aa, Borys:2006aa}.
	$^f$ Also known as PLCK\_G080.2+49.8; \citet{Canameras:2018ab,Frye:2019aa}.
	$^g$ Also known as PLCK\_G092.5+42.9; \citet{Canameras:2018ab,Frye:2019aa}.
	}
	%\end{tablenotes}
	%\end{threeparttable}
\end{deluxetable}

\subsection{Total magnification factors}
\label{sec:derive_mags}

Total magnification of each lens is determined as the ratio of the total solid angle subtended in the image plane to that of the source plane (as reconstructed by our lens model). These measurements are wavelength-dependent, as the source-plane structure varies with wavelength.
%The wavelength-dependent magnification is then calculated as the ratio of the total image-plane area to source-plane area. 
%
To estimate model-dependent magnification uncertainties, we follow an approach similar to \citealt{Sharon:2012aa}.
300 samples are drawn from the full MCMC-sampled posterior distribution for each lens mass profile parameter optimized with \lenstool, and the observed image-plane arcs are ray-traced into the source plane for each realization, each resulting in a slightly different magnification, measured as the area above a $3\sigma$ signal-to-noise threshold (for both the source plane and image plane). The median of the distribution (and uncertainty in the form of 16th and 84th percentiles) are reported in Table~\ref{tab:lensproperties} for each object.
However, we caution that these statistical values can underestimate larger systematic uncertainties\textemdash such as mis-identified multiple images and inappropriate choices of parameterized mass distributions\textemdash which are difficult to properly take into account. 
For example, \citet{Limousin:2016aa} considered the effect of using cored vs. non-cored cluster mass models\textemdash i.e., where the isothermal mass profile of cluster members is modified within an inner core radius.
The authors found that systematic uncertainties in the magnification could be nearly an order of magnitude greater than statistical model-specific uncertainties.
Here, this is unlikely to have a strong impact on the predominantly galaxy-scale lenses, which are influenced more by mass at intermediate radii than the typical core radius ($\sim 100$ pc) or cut radius ($\sim 30 - 100$ kpc), as noted by \citet{Canameras:2017aa}.
Regardless, this provides some motivation for model-independent characterizations, such as \citet{Wagner:2016aa} and \citet{Wagner:2018aa}, as magnifications can vary across different works.
We reserve the discussion of the distribution in magnification factors in the PASSAGES sample for Section \ref{sec:discussion_magnifications}.

\subsection{Measuring Einstein radii}
\label{sec:Einstein_radius}

The Einstein radius $\theta_E$ is a characteristic scale relating directly to the total enclosed mass of the foreground lens inside a projected circle in the sky $M(<\theta_E)$, with some additional dependence on the redshift geometry of the lens and source planes:
\begin{equation}
\label{eqn:Eradius}
\theta_E = \sqrt{
\frac{4G M(<\theta_E) }{c^2} 
\frac{
D_{LS}
}{
D_L D_S
}
},
\end{equation}
where $D_L$, $D_S$, and $D_{LS}$ are the angular diameter distances to the lens plane, to the source plane, and from the lens to source plane, respectively \citep{Narayan:1996aa}.
%; $G$ and $c$ are the gravitational constant and speed of light. 
$\theta_E$ is usually quite consistent across different lens models (e.g., \citealt{Kochanek:1991aa,Wambsganss:1994aa}), making it a valuable measurement. Observationally, the Einstein radius is comparable to half the typical separation between multiple images. 

There are a number of 
%popular 
methods for deriving $\theta_E$ from lens models. Here, we opt for the {\it equivalent Einstein radius}, $\theta_{Ein, eq}$, defined as the radius enclosing an average surface mass density equal to the critical surface density (or equivalently, where the average convergence $\kappa$ is equal to 1; see \citealt{Richard:2010aa, Zitrin:2011aa}). Other definitions include the {\it effective Einstein radius} \citep{Redlich:2012aa}, or the radius enclosing an area equal to that interior to the tangential critical curve, and the {\it median Einstein radius}, defined as the median distance from points on the tangential critical to the lens centroid \citep{Meneghetti:2011aa, Meneghetti:2013aa}. Since all definitions are more or less connected to the area interior to the critical curves, they are also related to the lensing cross section, which is the source-plane area inside the caustics (or sometimes defined as the area of the region where the magnification is greater than a certain threshold). 

We report the Einstein radii $\theta_{\rm Ein}$ in Table~\ref{tab:lensproperties}, as the median value and asymmetric $1\sigma$ uncertainties. The Einstein radii are computed analogously to the magnification, where 100 samples are drawn from the modeling-derived posterior distribution, with a new convergence map created for each iteration. Centered on the peak pixel of each convergence map (the point of greatest surface mass density), the radius is increased linearly until the average convergence crosses below 1. 
The distribution of Einstein radii for this sample is discussed in \S \ref{sec:lens_mass}.

\subsection{Source-plane size}
\label{sec:effective_radius}

As our lens-modeling algorithm does not assume a parametric form of the source-plane structure, the intrinsic size of the far-IR continuum-emitting region is measured from the source-plane reconstruction using \casa\ \imfit. 
As a caveat, this measurement is perhaps best made in the visibility-plane in order to deconvolve the effect of the synthesized beam. Since our source-plane reconstruction retains a distorted version of the beam that varies across the source, removing this is not trivial without a $uv$-based forward-modeling approach (e.g. \textsc{visilens} by \citealt{Hezaveh:2013aa,Spilker:2016aa} or \textsc{uvmcmcfit} by \citealt{Bussmann:2012aa, Bussmann:2013aa, Bussmann:2015aa}).
Moreover, since this lens modeling approach does not impose a restriction on the nature of the source plane emission, small but non-negligible offsets might be expected between the source-plane positions of reconstructions from different lensed images (\S \ref{sec:optimization}). We measure the effective radius from the reconstruction of all lensed images together, so these offsets can artificially change the apparent size\footnote{It is not advisable here to measure sizes based on the reconstruction of individual lensed images, as some images may not include the full source-plane structure. As an example, for galaxies where only a small region crosses into a quadruply-imaged region, two of the four images will contain only this region alone (while the remaining two images will show the full galaxy).}. 
To capture some of this uncertainty from lens modeling, we randomly select 10 of the MCMC lens modeling iterations to create different source-plane reconstructions, measure effective radii independently, and take the mean and standard deviation of this distribution. 
Lastly, we convert these angular measurements to physical sizes in kpc given the redshift of the source plane.
Our observing setup with ALMA is not markedly different from that of \citet{Spilker:2016aa} and \citet{Bussmann:2015aa}, for example, and the relative uncertainty in effective radius for this work is likewise similar. For this reason, we consider a direct comparison to be feasible.

However, as an added complication in comparing to the broader literature, there are different approaches to quantifying source size.
%, so comparing between works can be difficult. 
At optical wavelengths, it is standard to use the effective radius or half-light radius within which half the object's light is contained (under an assumption of circular symmetry). This in turn is determined through fitting a S\'ersic model to the radial profile (\citealt{Sersic:1963aa}; e.g., with GALFIT, \citealt{Peng:2002aa}), for which the degree of central concentration is parameterized by a S\'ersic index $n$. For radio imaging, it's more common to assume a two-dimensional Gaussian functional form (equivalent to a S\'ersic model with $n=0.5$). The full-width at half-maximum of the Gaussian is related to the effective radius as ${\rm FWHM} = 2 \times R_{\rm e}$. 
%(e.g. \citealt{Voigt:2010aa})
In our method, we therefore use the geometric mean of the semi-major and semi-minor axes as the effective radius
\footnote{
\citet{Murphy:2017aa} additionally demonstrate that the effective radius of marginally resolved sources can be connected to the deconvolved major axis of an inclined disk as $\theta_{\rm maj} \approx 2.430 \times R_{\rm e}$.
}. 
While the lensing-reconstructed PSF will vary in dimensions for the source plane, we do not account for this here for these global size measurements, as we expect this effect to be smaller than the uncertainties introduced from modeling.

\begin{figure}
	\includegraphics[width=\columnwidth]{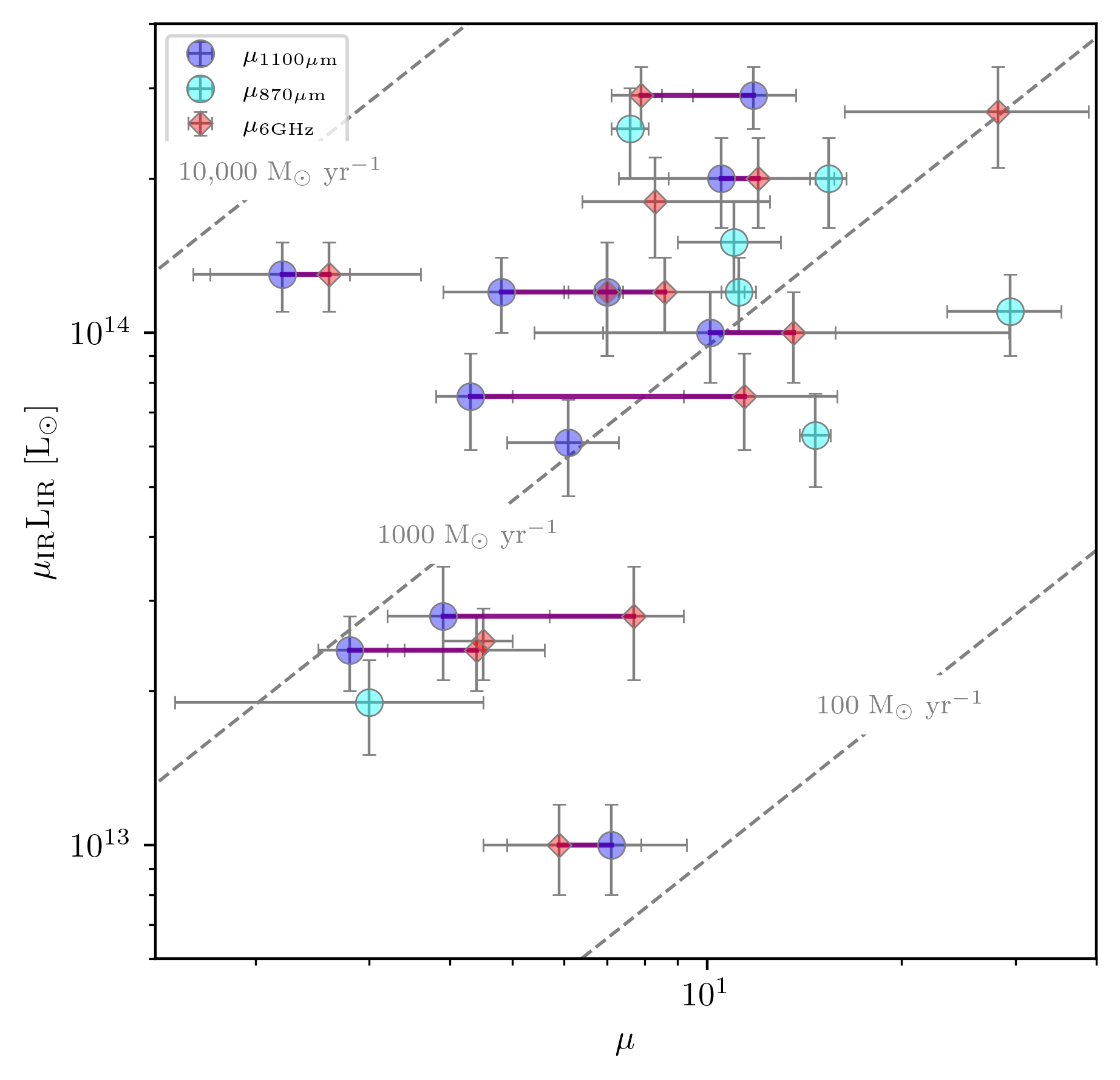}
    \caption{Apparent IR luminosity ($\mu_{\rm IR} L_{\rm IR}$) vs. magnification $\mu$ at 1100 $\micron$ (or 870$\micron$ if not available) and at 6 GHz (see Table~\ref{tab:lensproperties}). Dashed lines indicate intrinsic, magnification-corrected star formation rates from the \citealt{Kennicutt:1998aa} calibration under assumption of a Kroupa initial mass function (IMF), ${\rm SFR} = L_{\rm IR} / (9.4 \times 10^9) \Msun {\rm yr}^{-1}$. 
    Purple connecting lines are drawn between sub-mm and radio magnifications where both are measured.
    While the far-IR magnifications are perhaps more representative of the factor modifying the total IR luminosity than those for the radio continuum, the two are largely consistent, and so the radio magnifications can still offer a representative correction to estimate intrinsic star formation rates.
    }
    \label{fig:corrected_luminosities}
\end{figure}

\section{Discussion}
\label{sec:discussion}

\subsection{PASSAGES sources have modest magnifications}
% and extreme intrinsic infrared luminosities}

\label{sec:discussion_magnifications}

Eleven out of the 15
objects in this work for which we present new lensing models were previously undiscovered\footnote{We note that some were since identified independently by \citet{Trombetti:2021ab}.} before PASSAGES \citep{Harrington:2016aa, Berman:2022aa}.
These new lens models therefore provide many of the first measurements of 
magnification at any wavelength for PASSAGES. 
In our analysis, we utilized \HST\ (1.6\micron\, $0.15\arcsec$ resolution) and JVLA (6 GHz, $0.3-0.7\arcsec$ resolution) observations for all members. For 11 out of 15, we also used ALMA imaging (1 mm, $0.4-0.8\arcsec$ resolution), and for 8 out of 15, we used Gemini $r'$ and $z'$ observations. 
In future work, we intend to refine and apply these lens models towards multi-J CO and [CI] high angular resolution datasets from ongoing campaigns with ALMA, the Submillimeter Array (SMA), and the Northern Extended Millimeter Array (NOEMA).
% in the same sample, including an interpretation of their dynamical structure.
These will be used to interpret the kinematic structure of different gas components and build a systematic set of differential magnification corrections. 
After accounting for this work, only 4 high-$z$ members of the PASSAGES sample lack published lens models: PJ074851.7 (Garcia Diaz et al., in prep.), PJ084650.1 (Foo et al., in prep.), and PJ132217.5 (clusters with background DSFGs at $z=2.76$, $z=2.66$, and $z=2.07$), and the presumed galaxy-galaxy lens PJ141230.5 ($z=3.31$), which currently lacks high-resolution mm/radio follow-up and does not have discernible lensing features in the near-IR.

The PASSAGES objects are detected in near-IR imaging with {\it Hubble} (Lowenthal et al. in prep.), 
in contrast to the often optical faintness of DSFGs, especially at $z>3$ \citep{Franco:2018aa,Wang:2019ab, Barrufet:2022aa, Nelson:2022aa, Zavala:2022ab}.
%, Rodighiero:2023aa}),
%
This latter point is due in part to the fraction of obscured star formation increasing with SFR \citep{Whitaker:2017aa} 
and to the more dust-attenuated shorter wavelengths probed at higher-$z$.
Given the availability of lensing evidence in the rest-frame optical, in complement to the rest-frame far-IR, our approach in using \lenstool\ is less vulnerable to the pitfalls of modeling using solely interferometric imaging, which would typically necessitate direct modeling of $uv$-plane visibilities.

\begin{figure}
	\includegraphics[width=\columnwidth]{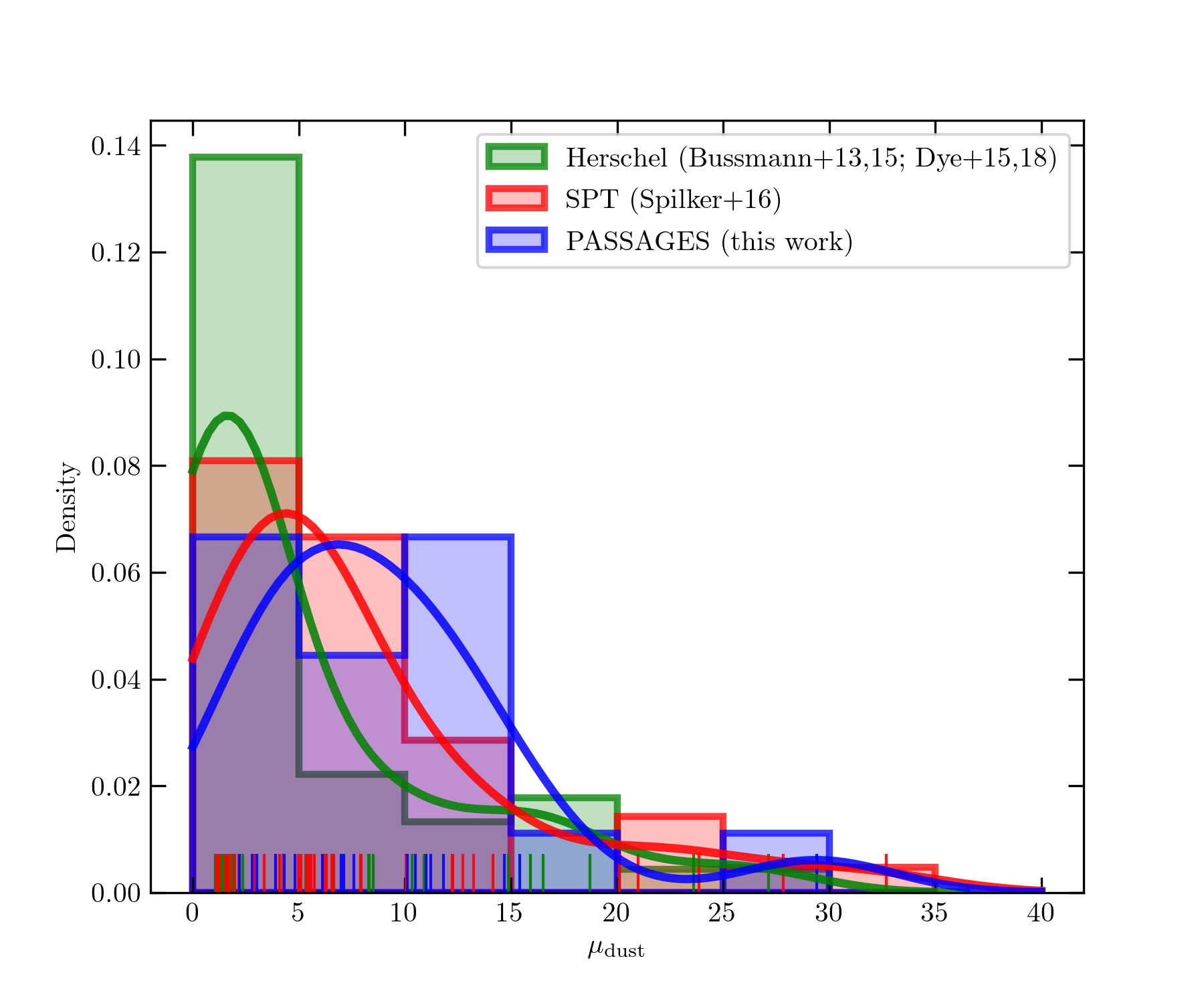}
    \caption{Histogram of magnifications for the PASSAGES sample (derived from lens models presented in this work and in the references included in Table \ref{tab:lensproperties}), compared with \Herschel\ lenses from \citet{Bussmann:2013aa, Bussmann:2015aa}, \citet{Dye:2015aa,Dye:2018aa}, and \citet{Massardi:2018aa}; and with SPT lenses from \citet{Spilker:2016aa}. 
    ``Barcode" lines at the bottom indicate the actual distribution values, for which kernel density estimations are shown as solid curves. 
    %{\red What kernel?}
    The smoothing kernel is chosen according to Scott's Rule \citep{Scott:1992aa}.
    We caution that, crucially, lens models have not been developed for all PASSAGES members, and that the magnification distribution may be skewed towards lower values due to several more complex, cluster-scale halos being excluded from this analysis at this juncture. Nonetheless, from the information currently available, the PASSAGES selection does not appear to be subject to drastically different lensing systematics from similar works, as discussed in \S \ref{sec:discussion_magnifications}.
     }
    \label{fig:magnifications}
\end{figure}

The distribution of magnifications for (sub-)mm continuum in this work is largely consistent with those from similar lensed surveys defined from {\it Herschel} and SPT, as shown in Figure \ref{fig:magnifications}. 
This may be surprising given the larger apparent luminosities of PASSAGES.
However, we caution that there are still several objects in the PASSAGES sample that lack detailed lens modeling, so this comparison is not robust at this point. 
In addition, there are external factors at play in this distribution: in this work, we have largely reserved the time-intensive modeling of cluster lenses for subsequent analyses. 
Likewise, the lenses identified with \Herschel\ preferentially exclude cluster-scale deflectors, as noted by \citet{Bussmann:2015aa}, due to their deconvolution of the \Herschel\ Spectral and Photometric Imaging REceiver (SPIRE; \citealt{Griffin:2010aa}) PSF (beam FWHM $18 - 36\arcsec$), which removes extended objects close to the size of the beam.
The SPT sample, however, contained four cluster-lensed objects, but these were not analyzed by \citet{Spilker:2016aa} as not all images were captured within the ALMA field-of-view.
Both this work and others are limited to objects with sufficient high-resolution, high-sensitivity data that are suitable for modeling.
%

%We acknowledge that the 
The resolution of data can have some effect on the derived magnifications (e.g., \citealt{Hezaveh:2012aa}). For example, the current $\sim 0.4\arcsec$ resolution comfortably resolves the lensed images in most cases, but does not offer much detail of substructure inside each of the images. Higher resolution ($\sim 0.1\arcsec$) might reveal chance alignments of compact, bright star-forming clumps with high-magnification areas near the caustic curve. Lower resolutions that are blind to these details could bias the magnification factors. As an added example, two well-separated clumps lying on opposite sides of the critical curve might result in an over-estimated magnification factor if they are blended in lower-resolution imaging, thus appearing to cross the critical curve contiguously. Such effects can not be mitigated completely, but future higher-resolution observations will be key to ruling out these biases.

\citet{Bussmann:2013aa} remarked on lower-than-expected magnifications for \Herschel-identified lenses, given the objects' apparent luminosities. Theoretical predictions from \citet{Wardlow:2013aa} based on submillimeter galaxy number counts to estimate the average magnification factor as a function of 500$\micron$ flux density were consistently higher than model-derived magnifications in all but three cases. 
In their case, an assumption of a single S\'ersic profile to represent the source plane might systematically underestimate magnifications.
\citet{Bussmann:2015aa} offered an explanation that the resolution of their SMA and ALMA observations ($\theta \approx 0.5\arcsec$) was insufficient to spatially resolve emission within individual lensed images, leading to half-light radius estimates skewed to larger values, in turn giving smaller magnification factors. Indeed, later works showed that higher-resolution imaging ($\approx 100$ milliarcsecond) of the same objects \citep{Dye:2015aa,Rybak:2015aa,Tamura:2015aa} could lead to a factor of $1.5-2$ increase in magnification \citep{Bussmann:2015aa}.

\subsection{Intrinsic IR luminosities of highly-magnified DSFGs and their relation to the main sequence}
\label{sec:HyLIRGs}

Apart from a different modeling approach, it is also possible that the
all-sky {\it Planck} selection used in PASSAGES (contrasted with the survey area-limited {\it Herschel} and SPT counterparts) is simply more advantageous for selecting 
the rarest objects: intrinsically extreme starbursts that are also serendipitously subject to strong lensing.
While it is clear that identifying lensing candidates by setting simple submillimeter flux thresholds is very efficient, there is not an obvious relation between flux and magnification for the resultant sample.
With a larger ensemble, it will be logical to compare both the distribution functions of magnifications and the luminosity functions of the de-lensed objects with unlensed populations, but that is beyond the scope of this current work.
In the next section, we discuss the apparent balance for a source size that is large enough to sustain intense star formation but small enough to avoid the dilution of magnification due to size bias.

Using the lens model-derived magnifications at 1.1 mm, we can derive intrinsic, de-lensed IR luminosities (given in Table \ref{tab:effective_radii} and shown in Figure \ref{fig:corrected_luminosities}), which range from $L_{\rm IR} = 0.2 - 5.9 \times 10^{13}~\Lsun$ (median $1.4 \times 10^{13}~\Lsun$).
Since we expect the ALMA 1.1 mm dust continuum to capture the structure of the IR-emitting regions of the source plane, we use this magnification factor as a proxy in place of $\mu_{\rm IR}$. Deriving the intrinsic IR SED would ideally involve de-lensing each of the IR photometric measurements, but the resolution of \Planck, \WISE, \Herschel, and AzTEC in the mid-IR and far-IR is dramatically coarser than the high-resolution ALMA 1.1 mm imaging. Using the magnification from a representative high-resolution rest-frame far-IR image remains the best approach. 
However, for the objects modeled in this work and in others summarized in Table~\ref{tab:lensproperties}, four are lacking comparable ALMA imaging (due primarily to being inaccessible northern sources). In Figure \ref{fig:corrected_luminosities}, we include the magnifications measured from 6 GHz radio continuum imaging at similar angular resolution, which are consistent within 40\% relative difference for the majority of cases. These $\mu_{\rm 6 GHz}$ factors can thus be used to roughly estimate intrinsic $L_{\rm IR}$, as indicated in Table \ref{tab:effective_radii}.
However, because of these possible differences in magnification, we do not include the objects lacking far-IR magnifications in the subsequent discussion on source sizes and star formation rate surface densities.

Despite their high inferred SFRs of $170 - 6300~\Msun~{\rm yr}^{-1}$ (median $1500~\Msun~{\rm yr}^{-1}$; Table~\ref{tab:effective_radii}), it is not clear where exactly the PASSAGES members fall in relation to the star-formation main sequence, 
%\citep{Harrington:2021aa}, 
a tight correlation between SFR and $M_\star$, the normalization of which evolves with redshift (e.g. \citealt{Brinchmann:2004aa, Noeske:2007aa, Elbaz:2007aa, Daddi:2007aa, Whitaker:2012aa, Speagle:2014aa}). 
The peak and Rayleigh-Jeans tail of their dust SEDs are well sampled by \Planck, \Herschel, AzTEC, and ALMA photometry, leading to secure SFR estimates; extensive multi-line/continuum modeling by \citet{Harrington:2021aa} results in robust molecular gas estimates. Yet, the dearth of stellar mass estimates makes it impossible to reliably place these other properties in context. 
%
%It remains uncertain where 
This is not a novel problem for DSFGs:
% lie in relation to this sequence: 
it has not been settled whether they largely represent abnormal outliers forming stars at elevated rates compared to other objects of similar mass, or if their stellar masses are proportionally higher to match their extreme SFRs \citep{Casey:2014aa}. Typically poor constraints on rest-frame optical luminosities due to heavy dust attenuation make the assessment of $M_\star$ difficult for this population. This is additionally complicated by both the higher sensitivity to the initial mass function (IMF), as shorter-wavelength optical and UV fluxes predominantly trace the contribution of the most massive stars, and the less-constrained contribution of AGN to the near-IR continuum \citep{Hainline:2011aa}. Recent work with other submillimeter-selected samples\textemdash primarily with $\log_{10}[{\rm SFR} / \Msun~{\rm yr}^{-1}] \sim 2.5 - 3.5$,
%300 - 3000,
in line with PASSAGES\textemdash by \citealt{Miettinen:2017aa} (for $z \sim 1 - 7$ DSFGs in the COSMOS field; \citealt{Scoville:2007aa}) and \citealt{Barrufet:2020aa} (for $z \sim 1 - 5$ DSFGs in the North Ecliptic Pole field) suggested that around 60\% of DSFGs might be consistent with the main sequence at their respective redshifts, 
%but this is in direct contrast with other work by 
whereas \citet{Ikarashi:2017aa}
%, which 
suggested 72\% were safely above the main sequence ($z \approx 1.4 - 2.5$).
%, in contrast.

%
\citet{Harrington:2021aa} and \citet{Berman:2022aa} highlighted the PASSAGES sample's short gas depletion times ($\tau_{\rm dep}=100-400$ Myr), but these values are not particularly meaningful without an estimate of the average depletion time of a main sequence galaxy at the same redshift and of similar stellar mass (e.g. \citealt{Scoville:2017aa,Tacconi:2018aa}).
%, in contrast to the $\sim 1$ Gyr timescales seen for local star-forming galaxies (e.g. \citealt{Leroy:2013aa}), as evidence that they can safely be classified as starbursts. 
%
%
Following \citet{Liu:2019ad}, 
\citet{Wang:2022ac} recently derived a best-fit relationship for molecular gas depletion time $\tau_{\rm mol} \equiv M_{\rm mol} / {\rm SFR}$ as a function of stellar mass, redshift/cosmic age, and distance to the main sequence, $\Delta {\rm MS} \equiv {\rm log}_{10} ({\rm SFR}/{\rm SFR}_{\rm MS})$ (see their equation 9).
For fiducial stellar masses\footnote{These values are chosen in line with the characteristic mass (and $1\sigma$ confidence interval) of \citealt{Leslie:2020aa} that contributes most to the cosmic SFR density at $z=2.5$. Moreover, abundance matching by \citet{Behroozi:2013aa} suggests an average upper stellar mass limit of $5\times10^{11}~M_\odot$ within halos of mass $10^{12}-10^{13}~M_\odot$.} of $10^{10.5} M_\odot$, $10^{10.75} M_\odot$, and $10^{11} M_\odot$ at a representative redshift of $z=2.5$, this yields main-sequence depletion times of 340, 360, and 380 Myr, respectively.
Conversely, for our observed range of depletion times\textemdash with an inner 68\% confidence interval of $\tau \sim 150 - 350$ Myr\textemdash we obtain $\Delta {\rm MS} = -0.02 - 0.69$ for $M_\star = 10^{10.5}~M_\odot$, $\Delta {\rm MS} = 0.02 - 0.78$ for $M_\star = 10^{10.75}~M_\odot$, and $\Delta {\rm MS} = 0.08 - 0.88$ for $M_\star = 10^{11}~M_\odot$.
This suggests that a clear majority of the PASSAGES objects should have specific star formation rates (sSFR $\equiv {\rm SFR}/M_\star$) that lie at least above the center of the main sequence, but raises doubts that they could all be classified uniformly as traditional starbursts. Specifically, a median depletion time and stellar mass (250 Myr and $M_\star = 10^{10.75}~M_\odot$) yields $\Delta {\rm MS} = 0.32$, which falls below the $\Delta {\rm MS} > 0.6$ dex threshold for starbursts proposed by \citet{Rodighiero:2011aa}.

Another independent approach is to use a relation between star-formation rate surface density $\Sigma_{\rm SFR}$ and $\Delta {\rm log}_{10}({\rm sSFR})_{\rm MS} \equiv {\rm log}[{\rm sSFR}/{\rm sSFR}_{\rm MS}(M_\star, z)] (= \Delta {\rm MS})$, as proposed by \citet{Jimenez-Andrade:2019aa}. For the highest-redshift bin\footnote{We note that the next highest redshift bin, $1.3 < z \le 1.75$, follows effectively the same relation, but where the vertical offset is 1.1 instead of 1.2.}, $1.75 < z \le 2.25$, \citeauthor{Jimenez-Andrade:2019aa} find a best-fit power-law relation of ${\rm log}(\Sigma_{\rm SFR}) = [1.5 \pm 0.2] \times \Delta {\rm log}({\rm sSFR})_{\rm MS}+ [1.2 \pm 0.1]$ for galaxies with ${\rm log}_{10} (M_\star / M_\odot) \gtrsim 10.5$.
For this work, the inner 68\% confidence interval in star-formation surface density is $\Sigma_{\rm SFR} \approx 20 - 160$ (see \S \ref{sec:Eddington_SFR}), suggesting distances to the main sequence of $\Delta {\rm log}_{10}({\rm sSFR})_{\rm MS} = 0.07 - 0.67$, in rather close agreement with the values estimated from depletion times. However, this relation seems to evolve somewhat from $z=0.5 - 2$, where the slope of the relation becomes shallower and the offset increases with redshift, meaning that surface densities are higher in the Universe, with smaller discrepancies in $\Sigma_{\rm SFR}$ between starbursts and main sequence galaxies, so its predictive power may be weakened for higher-$z$ PASSAGES members.

Forthcoming dynamical measurements of the gravitational potentials for PASSAGES will offer helpful context to this discussion, but
at present,
their actual specific star formation rates 
%currently 
remain an open question.
Sensitive
%, next-generation 
telescopes like \JWST\ will also offer the opportunity to better constrain the rest-frame UV and optical SED of such DSFGs, rendering stellar mass measurements slightly less uncertain.

To reiterate, based on the gravitational lens modeling in this work, we have found that the PASSAGES sample overwhelmingly consists of objects that are intrinsically luminous and not extraordinarily amplified by lensing. 
This discovery prompts a more thorough exploration of their intrinsic properties, which we discuss in the next section.

\subsection{Source-plane reconstructed galaxy sizes}

In \S \ref{sec:effective_radius}, we discuss our computation of galaxy size or effective radius as the geometric mean of the semi-major and semi-minor axes of a best-fit Gaussian. 
At 1 mm, this results in dust continuum sizes of $R_{\rm e} = 1.7 - 4.3$ kpc (median 3.0 kpc; Table~\ref{tab:effective_radii}). As we discuss in this section, the PASSAGES galaxies are larger at this wavelength than most DSFGs studied thus far, but to a degree that is consistent with their larger luminosities. 
However, as these are lensed objects selected by their image-plane fluxes, selection effects may bias the intrinsic sizes to which we are sensitive.

\subsubsection{Possible source size bias?}
\label{sec:size_bias}

\begin{figure}
	\includegraphics[width=\columnwidth]{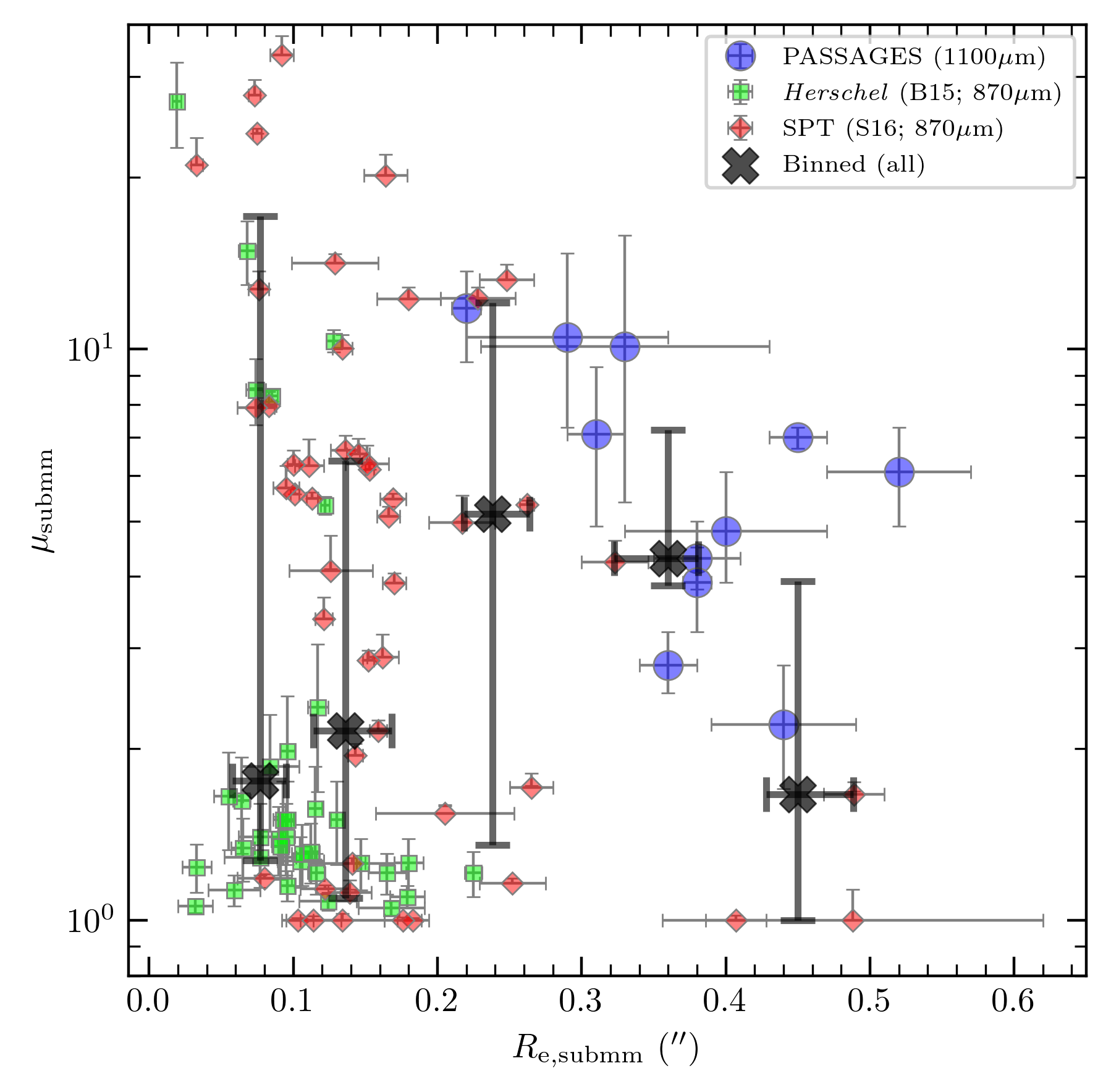}
    \caption{
    Our model-derived 1.1 mm magnification factors vs. (angular) effective radii for the PASSAGES members of this work (closely following Fig.~5 of \citealt{Spilker:2016aa}), in the context of matching measurements at 870 $\mu$m for the {\it Herschel} \citep{Bussmann:2015aa} and SPT \citep{Spilker:2016aa} samples. 
    All measurements are made consistently for $\approx0.5\arcsec$ ALMA imaging, but 
    %acknowledge that 
    \citet{Bussmann:2013aa,Bussmann:2015aa} and \citet{Spilker:2016aa} use a parameterized source-plane in their lens modeling approaches. 
    Black markers indicate binned values of $R_e$ and $\mu$ for the three samples together (with error bars showing the 16th and 84th percentiles for each bin).
    An apparent trend where larger magnifications are correlated with smaller intrinsic source sizes would indicate the presence of a possible size bias. While there is not a clear such correlation at all scales, it does appear to be the case that the highest-magnification objects are preferentially more compact, as noted by \citet{Spilker:2016aa}. 
    With the addition of our sample, it is also evident that the most extended objects tend to have lower magnifications; all objects with $R_e > 0.3\arcsec$ have magnifications $\mu \leq 10$.
    }
    \label{fig:size_bias}
\end{figure}

The intrinsic sizes of the DSFGs that make up the PASSAGES sample have implications for both the maximum lensing magnifications and the maximum star formation surface densities that can be observed. 
For the former, there is a small region of the source plane capable of producing high magnifications (e.g. $\mu > 10$), especially in the case of galaxy-scale lenses where the caustic network is comparable in size to that of typical background objects. 
For EAGLE-simulated lenses at $z_{\rm lens} \approx 0.37$, \citet{Robertson:2020aa} found source-plane cross-section solid areas at $z_s = 2$ satisfying $|\mu| > 10$ to be $\sigma^{\rm S}_{\rm lens} \approx 0.008$ and 0.08 arcsec$^2$ for halo masses of $10^{12}$ and $10^{13}~M_\odot$, respectively. For large magnification thresholds ($\mu_0 \gg 1$), this cross-section drops off as $\sigma^{\rm S}_{\rm lens} \propto \mu_0^{-2}$ (see also \citealt{de-Freitas:2018aa}).

In the vicinity of a caustic curve (where magnification diverges), increasing source size has the effect of diluting the overall magnification, as the outer extent of the source reaches into lower magnification regions (such as the local minimum inside the caustic curves, or the asymptotically-decreasing magnifications exterior to the caustics). 
However, in simulations to study the predicted size bias of flux-selected lens samples, \citet{Hezaveh:2012aa} found the intriguing result that intermediate total magnifications ($\mu\sim10$) could preferentially magnify {\it diffuse} (or extended) components in the source plane, contrary to the common intuition (e.g. \citealt{Serjeant:2012aa}) that more compact components are amplified more strongly (which \citeauthor{Hezaveh:2012aa} did indeed find to be the case for smaller and larger magnifications).
This latter effect may be due to the nuanced consideration that a larger source still has a greater chance of being modestly magnified ($\mu < 10$). In other words, as described by \citet{Robertson:2020aa}, the maximum possible magnification (optimized as a function of source position) decreases for larger source sizes (as a decreasing fraction of the source will be some arbitrary distance from the caustic), but the probability of some part of the source being near to the caustic network increases for larger sources. 
In the case of a singular isothermal sphere acting as a lens, \citet{de-Freitas:2018aa} determined that the cross section for $\mu > 10$ was optimally maximized when the source-plane object had an angular size of $\approx 30\%$ the Einstein radius (in contrast to $\approx 15\%$ for the threshold $\mu > 20$).
We make further comparisons to these predictions in \S \ref{sec:lens_mass}.

\begin{figure}
	\includegraphics[width=\columnwidth]{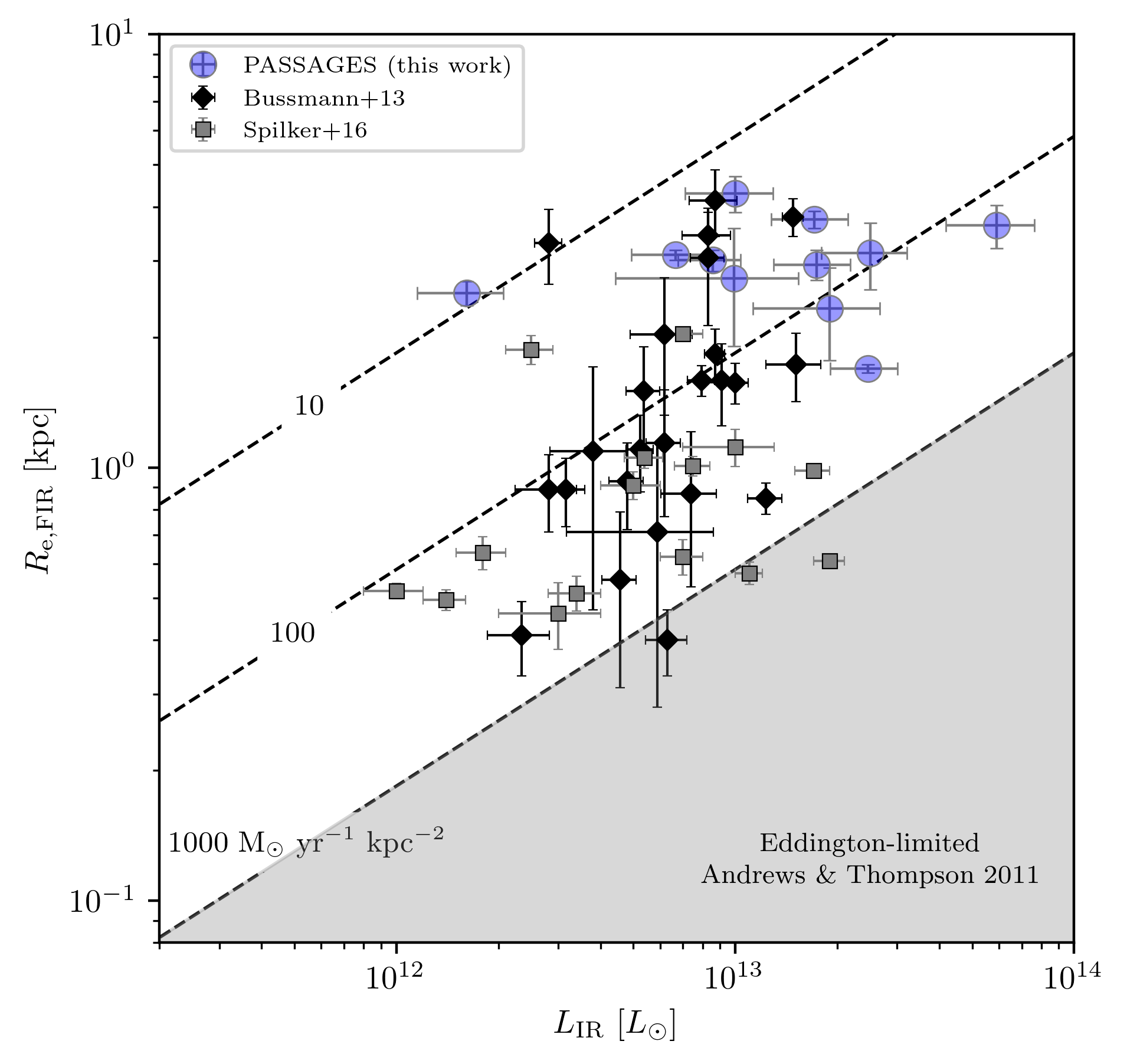}
    \caption{
   % {\red Note: one version for comparison with lensed samples, one for unlensed.}
    Relation between far-IR effective radius $R_e$ (measured at 1 mm) and the magnification-corrected, intrinsic IR luminosity (see Table~\ref{tab:effective_radii}) for the 15 PASSAGES objects studied in this work. 
    The gray shaded area shows the Eddington-limited region (1000 $\Msun$ yr$^{-1}$ kpc$^{-2}$) from \citet{Andrews:2011ab}. 
    For comparison, we also plot data points from similar samples of lensed DSFGs studied by \citet{Bussmann:2013aa} and \citet{Spilker:2016aa}, with corrected $L_{\rm IR}$ values from \citealt{Reuter:2020aa} for the latter.
    For additional comparison, see Figure 5 of \citealt{Enia:2018aa} (not shown).
    }
    \label{fig:LIR_Reff_lensed}
\end{figure}

\begin{figure}
	\includegraphics[width=\columnwidth]{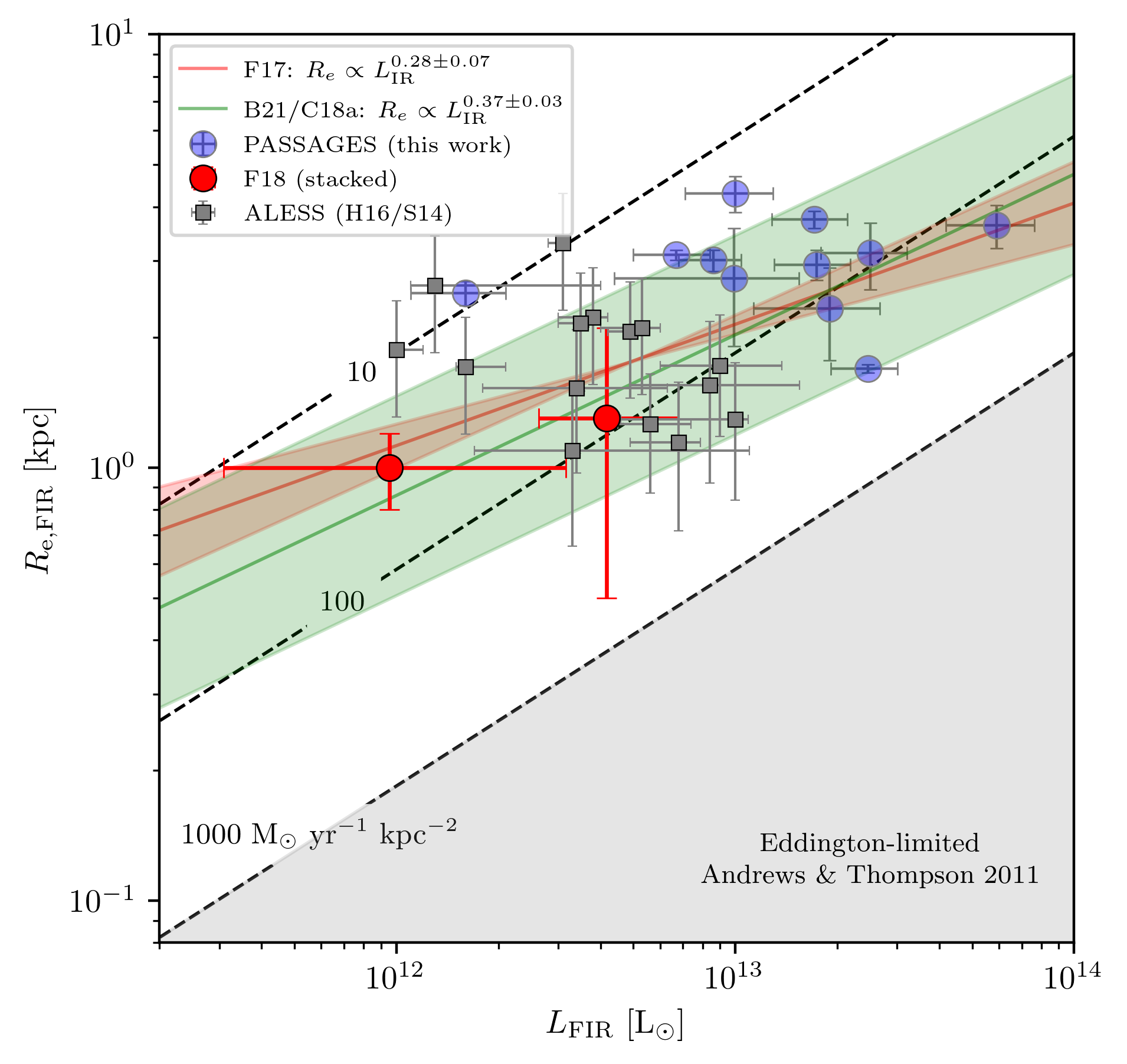}
    \caption{
    As with Fig.~\ref{fig:LIR_Reff_lensed}, but with unlensed samples shown for comparison. Objects are drawn from the ALESS sample \citep{Hodge:2016aa,Swinbank:2014aa} and from \citet{Fujimoto:2018aa}, comprising 33 stacked ASAGAO objects (ALMA twenty-Six Arcmin$^2$ survey of GOODS-S at One-millimeter) at $L_{\rm IR} \approx 10^{12}~ L_\odot$ and the median of 12 other bright ALMA sources ($L_{\rm IR} \approx 10^{12.6}~ L_\odot$). 
    The shaded red region illustrates the best-fitting power-law relation found by \citet{Fujimoto:2017aa} in a large sample of $\approx 500$ archival 1 mm maps from ALMA, $R_e \propto L_{\rm IR}^{0.28\pm0.07}$. Actual dispersion in the sample is much larger than the shaded region, and is consistent with that of the PASSAGES objects. For a secondary comparison, the green shaded region found by \citet{Burnham:2021aa} (using the $L_{\rm IR}-\lambda_{\rm peak}$ trend from \citealt{Casey:2018ab}) is consistent but slightly steeper.
    The hypothesized Eddington limit from \citet{Andrews:2011ab} is again shown as a gray forbidden region, and star formation surface densities of 10, 100, and 1000 $\Msun$ yr$^{-1}$ kpc$^{-2}$ are shown as dashed lines.
    The sample from this work is located at higher luminosities ($L_{\rm IR} \gtrsim 10^{13} L_\odot$), but effective radii are in general correspondingly larger, and thus consistent with the previous trends. 
    }
    \label{fig:LIR_Reff}
\end{figure}

This lensing size bias effect has been tested empirically by many works, including \citet{Serjeant:2012aa}, 
%Wardlow:2013aa}, 
and \citet{Spilker:2016aa}.
In the case of the latter, \citet{Spilker:2016aa} showed that 870$\micron$ magnifications for SPT lenses, along with \Herschel\ lenses from \citealt{Bussmann:2013aa} and \citealt{Bussmann:2015aa}, did not appear to correlate strongly with source size across the entire range of observed sizes or magnifications. In fact, the largest sources covered a wide range of magnifications consistent with the remaining objects. However, the authors did note that the highest-magnification objects ($\mu_{870\mu m} > 10$) were preferentially more compact. 
Notwithstanding this, \citeauthor{Spilker:2016aa} found that the distribution of source sizes in the SPT and \Herschel\ lens samples was statistically consistent with that of unlensed DSFGs,
such as the 850 $\micron$ SCUBA-2 Cosmology Legacy Survey (\citealt{Simpson:2015ab,Simpson:2015aa}; median FWHM $0.30 \pm 0.04\arcsec$) and 1.1-mm AzTEC sources (\citealt{Ikarashi:2015aa, Ikarashi:2017aa}; median FWHM $0.20 \pm 0.04\arcsec$ and $0.31 \pm 0.03$).
Together, these tests suggested that the hypothesized size bias was not observable with the then-available sample size.

In Fig.~\ref{fig:size_bias}, we similarly test for the presence of any size bias for the PASSAGES sample (extending Fig. 5 of \citealt{Spilker:2016aa}). 
%
%While there is not a strong anti-correlation 
%
In our case, there is some evidence of anti-correlation between magnification and effective radius, 
especially for $R_e \gtrsim 0.2\arcsec$,
a region of the parameter space for which our sample contributes significantly.
Objects with sizes larger than $0.3\arcsec$ all have lower magnifications, $\mu \leq 10$, whereas smaller objects exhibit a large range of magnifications.
These findings raise the question of the nature of selection biases in the definition of PASSAGES \citep{Harrington:2016aa, Berman:2022aa}, as objects capable of producing observed luminosities $\mu L_{\rm IR} \sim 10^{14} L_\odot$ must have some combination of high total lensing magnification and high intrinsic luminosity. 
If galaxy-scale lenses were to preferentially magnify more compact objects, then there would be a negative power-law slope between $\mu$ and $R_e$. In tandem with the positive slope of $L_{\rm IR}$ vs. $R_e$, it might be the case that the selection function of this sample identifies objects of an optimal size that maximizes the product of $\mu$ and $L_{\rm IR}$ (e.g. \citealt{Lutz:2014aa}). 
Of course, the distribution of intrinsic luminosities is also subject to the luminosity function at high-redshift, as there is an effective cutoff on the high-luminosity end.

\begin{deluxetable*}{ccccccccc}
	%\centering
	\tablecolumns{9}
	\tabletypesize{\small}
	\tablecaption{
	Lensing-corrected source-plane effective radii at 1.1 mm and intrinsic total IR luminosities $L_{\rm IR}$, using $\mu_{\rm 1.1mm}$ magnification factors as representative of dust emission magnification. 
	The continuum synthesized beam parameters for each field are shown for context.
	The star formation rate surface density, $\Sigma_{\rm SFR}$ is given as both a global average value (total SFR divided by the area of the disk with given effective radius), and as a peak value given in brackets (see \S \ref{sec:Eddington_SFR}).
	\label{tab:effective_radii}
	}
	%\begin{threeparttable}
	%\begin{center}
	%\begin{tabular}{ccccccccc} % four columns, alignment for each
		%\hline
		%\hline
	\tablehead{
		\colhead{Field} & 
		\colhead{$R_{{\rm e},{1100\mu{\rm m}}}$} & 
		\colhead{$R_{{\rm e},{1100\mu{\rm m}}}$} & 
		\colhead{$\theta_{\rm maj}$} & 
		\colhead{$\theta_{\rm min}$} & 
		\colhead{PA} & 
		\colhead{${L_{\rm IR}}^a$} & 
		\colhead{${{\rm SFR}_{\rm IR}}^b$} & 
		\colhead{$\Sigma_{\rm SFR}$ [$\Sigma_{\rm SFR, peak}$]}  \\[-1ex]
		& 
		\colhead{[$\arcsec$]} & 
		\colhead{[kpc]} & 
		\colhead{[$\arcsec$]} &   
		\colhead{[$\arcsec$]} & 
		\colhead{[\degrees]} & 
		\colhead{[$10^{13} \Lsun$]}  & 
		\colhead{[$\Msun ~{\rm yr}^{-1}$]} & 
		\colhead{[$\Msun~{\rm yr}^{-1}~ {\rm kpc}^{-2}$]}  \\[-2.5ex]
		%\hline\\[-1.5ex]
	}
	\startdata
		%\hline\\[-2ex]
		PJ011646 	& $0.45\pm0.02$ 					&  	$3.7 \pm 0.2$	& 0.43 			& 		0.38 		& -83.2 		& $1.7 \pm 0.4$ 		& $1820 \pm 460$			&	$42 \pm 11$  [41]				\\[1ex]	
		%		 	& $0.41\pm0.01$ 					&  	$3.4 \pm 0.1$	& 0.20 			& 		0.12 		&  73.8 		&  \textemdash 			& \textemdash				&	$50 \pm 13$ 		 	 	\\[1ex]			                                 
		PJ014341 	& $0.31\pm0.02$ 					&  	$2.5 \pm 0.2$	& 0.49 			&		0.40 		&  67.7 		& $0.16 \pm 0.05$  		& $170 \pm 50$				&	$9  \pm 3$ [16]					\\[1ex] 
		PJ020941 	& $0.29\pm0.07$ 					&  	$2.3 \pm 0.6$	& 0.54 			&		0.40 		&  63.8 		& $1.9 \pm 0.8$  		& $2030 \pm 820$			&	$119 \pm 56$ [136]				\\[1ex] 
		%		 	& $0.17\pm 0.03$					&  	$1.4 \pm 0.2$	& 0.19 			& 		0.12 		& -83.3 		&   \textemdash 		& \textemdash				&	$346 \pm 153$ 				\\[1ex]
		PJ022633 	& \textemdash  						&  	\textemdash		& \textemdash	&	\textemdash 	& \textemdash 	& $1.0 \pm 0.4^\dagger$ & $1020 \pm 470^\dagger$	&	\textemdash					\\[1ex] 
		PJ030510 	& $0.44\pm0.05$ 					&  	$3.6 \pm 0.4$	& 	0.85 			&		0.59 			& 80.9 			& $5.9 \pm 1.7$ 		& $6290 \pm 1850$			&	$153 \pm 48$ [187] 				\\[1ex] 
		PJ105353 	& $0.22\pm0.01^{\dagger \dagger}$	&  	$1.7 \pm 0.0$	&0.45 & 		0.27 		&  63.1  		& $2.5 \pm 0.6$			& $2610 \pm 590$			&	$290 \pm 66$ [264]				\\[1ex]
		%			& $0.12\pm0.02$ 					&  	$0.9 \pm 0.2$	&  	0.08 		&	0.07 			& -73.0 		& \textemdash	  		& \textemdash				&	$975 \pm 273$ 				\\[1ex] 				
		PJ112713 	& \textemdash 						&  	\textemdash		& \textemdash 	&	\textemdash 	& \textemdash 	& $0.6 \pm 0.1^\dagger$   & $590 \pm 120^\dagger$		&	\textemdash					\\[1ex] 
		PJ113805 	& $0.36\pm0.02$ 					&  	$3.0 \pm 0.2$	& 	1.03 		&		0.66 		&    3.5 		& $0.9 \pm 0.2$  		& $910 \pm 190$				&	$32 \pm 7$ [37]					\\[1ex] 
		PJ113921 	& $0.40\pm0.07$ 					&  	$3.1 \pm 0.5$	& 	0.82 		&		0.69 		& -14.0 		& $2.5 \pm 0.7$ 		& $2660 \pm 750$			&	$87 \pm 29$ [79]				\\[1ex] 
		%			& $xx\pm xx$ 						&  	$x.x \pm 0.$	& 		xxx 	& 		xxx 		&  xxx  		&  \textemdash 			& \textemdash				&	$xx \pm xx$ 				\\[1ex]	
		PJ132630 	& $0.38\pm0.03$ 					&  	$2.9 \pm 0.2$	& 	0.99 		&		0.64 		&  -2.2  		& $1.7 \pm 0.4$ 		& $1860 \pm 470$			&	$68 \pm 18$ [61]				\\[1ex] 
		PJ133634 	& \textemdash 						&  	\textemdash	& \textemdash 	&	\textemdash 	& \textemdash 	& $2.2 \pm 0.9^\dagger$ 	& $2310 \pm 990^\dagger$		&	\textemdash 				\\[1ex] 
		PJ144653 	& $0.38\pm0.01$ 					&  	$3.1 \pm 0.1$	& 	0.76 		&		0.68 		&  11.7 		& $0.7 \pm 0.2$ 		& $710 \pm 180$				&	$24 \pm 6$ [24]					\\[1ex] 
		PJ144958 	& $0.33\pm0.10$ 					&  	$2.7 \pm 0.8$	& 	0.81 		&		0.67 		&  6.3  		& $1.0 \pm 0.6$ 		& $1050 \pm 580$			&	$45 \pm 28$ [78]				\\[1ex] 
		PJ160722 	& \textemdash 						&  	\textemdash		& 	\textemdash &	\textemdash 	& \textemdash 	& $0.2 \pm 0.1^\dagger$ 	& $220 \pm 110 $							&	\textemdash 				\\[1ex] 
		PJ231356 	& $0.52\pm0.05$ 					&  	$4.3 \pm 0.4$	& 	0.78 		&		0.61 		&  68.0 		& $1.0 \pm 0.3$ 		& $1060 \pm 310$			&	$18 \pm 6$ [29]					\\[1ex]
		%			& $xx\pm xx$ 						&  	$x.x \pm 0.$	& 	xxx 		& 		xxx 		&  xxx  		&  \textemdash 			& \textemdash				&	$xx \pm xx$ 				\\[1ex]	
		\hline
		\hline\\[-1.5ex]
		{\it Literature} & $R_{{\rm e},{870\mu{\rm m}}}$ & $R_{{\rm e},{870\mu{\rm m}}}$  &  $\theta_{\rm maj}$ & $\theta_{\rm min}$ & PA & ${L_{\rm IR}}$ & ${{\rm SFR}_{\rm IR}}^c$ & $\Sigma_{\rm SFR}$  \\ 
		&  [$\arcsec$] & [kpc] &  [$\arcsec$] &   [$\arcsec$]  & [\degrees] & [$10^{13} \Lsun$]  & [$\Msun ~{\rm yr}^{-1}$] & [$\Msun~{\rm yr}^{-1}~ {\rm kpc}^{-2}$] \\[0.5ex]
		\hline\\[-1.5ex]
		PJ105322$^d$ 	&  \textemdash				& \textemdash 		& \textemdash 	& \textemdash	& \textemdash  	& $2.9 \pm 0.2$ 		& $3090 \pm 210$			&	\textemdash 				\\[1ex]
		PJ112714$^e$ 	& $0.17 \pm 0.04$ 			&  $1.4\pm0.3$		& 0.90 		& 0.75		& 39 			& $0.4 \pm 0.1$ 		& $430   \pm 110$			&	$70\pm20$  				\\[1ex]
		PJ132302$^d$ 	&  \textemdash				& \textemdash 		& \textemdash 	& \textemdash	& \textemdash	& $0.9 \pm 0.1$ 		& $960   \pm 110$			&	\textemdash 				\\[1ex]
		PJ132934$^f$ 	&  \textemdash				& \textemdash 		& \textemdash 	& \textemdash	& \textemdash 	& $1.3 \pm 0.1$ 		& $1380 \pm 110$			&	\textemdash 				\\[1ex]
		PJ142823$^g$ 	&  $0.08\pm0.05$ & $0.71\pm0.43$ 	& $\approx 0.7$&$\approx 0.7$	& \textemdash  	& $0.6 \pm 0.3$& $640 \pm 320$			&	$400\pm320$ 				\\[1ex]
		PJ154432$^d$ 	&  \textemdash				& \textemdash 		& \textemdash 	& \textemdash	& \textemdash 	& $0.31 \pm 0.02$ 		& $330 \pm 20$			&	\textemdash 				\\[1ex]
		PJ160918$^d$ 	&  \textemdash				& \textemdash 		& \textemdash 	& \textemdash	& \textemdash 	& $1.6 \pm 0.1$ 		& $1710 \pm 130$			&	\textemdash 				\\[1ex]
		%\hline\\[-0.5ex]
	%\end{tabular}
	\enddata
	\tablecomments{
	%\multicolumn{9}{p{0.9\textwidth}}{
	%{\bf Notes.}
	%\begin{tablenotes}\footnotesize
	%\item[$^a$] 
	$^a$ Lensing-corrected intrinsic luminosities derived by dividing apparent far-infrared luminosities ($8 - 1000\micron$) from Table 6 of \citet{Berman:2022aa} (or Table 4 of \citealt{Harrington:2016aa}) by magnification factors in Table \ref{tab:lensproperties}.
	$^b$ Lensing-corrected IR-inferred star formation rates derived by dividing Starburst SED star formation rates from Table 6 of \citealt{Berman:2022aa} (or Table 4 of \citealt{Harrington:2016aa}) by magnification factors in Table \ref{tab:lensproperties}.
	$^c$ Star formation rates derived from \citealt{Kennicutt:1998aa} calibration, ${\rm SFR} = L_{\rm IR} / (9.4 \times 10^9) \Msun {\rm yr}^{-1}$.
	$^d$ De-lensed $L_{\rm IR}$ from \citet{Canameras:2015aa, Canameras:2018ab}.
	$^e$ De-lensed $L_{\rm IR}$ from \citet{Canameras:2015aa, Canameras:2018ab, Canameras:2018aa}, using latest magnification factor from \citet{Pascale:2022aa}.
	$^f$ De-lensed $L_{\rm IR}$ from \citet{Diaz-Sanchez:2017aa}.
	$^g$  De-lensed $L_{\rm IR}$ from \citet{Bussmann:2013aa}.
	$^\dagger$ Approximate intrinsic IR luminosity estimated using 6 GHz magnification factor (as high-resolution 1 mm/870 $\mu$m image is not available, see discussion in \S \ref{sec:HyLIRGs}).
	$^{\dagger \dagger}$ Low-resolution ALMA Band 7 ($870\mu$m) measurement from program 2019.1.01636.S (PI: M. Yun). 
	}
	%\end{center}
\end{deluxetable*}

\subsubsection{Intrinsic far-IR source size}
\label{sec:FIR_size}

With this work, we seek to place the 
PASSAGES objects modeled in this work 
%to this consideration.
in the context of other submm-selected lensed systems, as seen in Figures~\ref{fig:size_bias} and \ref{fig:LIR_Reff_lensed}.
In comparison to the {\it Herschel} and SPT lensed DSFGs, the PASSAGES DSFGs encompass larger intrinsic source sizes in physical units. 
However, the largest PASSAGES object is comparable in size to the largest member of the \citet{Bussmann:2013aa} sample (J091305.0-005343 or SDP.130, $R_e = 4.14 \pm 0.72$).
Yet, despite occupying the upper end of luminosities covered by the other samples, PASSAGES objects show notable evidence for lower star-formation rate surface densities, which are primarily between $10 - 100~\Msun$ yr$^{-1}$ kpc$^{-2}$, consistent with the vast majority of over 1000 dusty starbursts studied by \citet{Fujimoto:2017aa}.

To further elucidate this discrepancy, we extrapolate our analysis to include unlensed DSFGs. Figure~\ref{fig:LIR_Reff} reveals that PASSAGES is in accordance with a scaling relation of $L_{\rm IR}-R_e$ found by \citet{Fujimoto:2017aa}, but both larger and more luminous than the physical quantities derived by \citealt{Bussmann:2013aa} and \citealt{Spilker:2016aa} (in members for which redshifts are measured). 
\citet{Berman:2022aa} found that the PASSAGES DSFGs account for an average number density of about $10^{-2}~{\rm deg}^{-2}$, in comparison with the {\it Herschel} samples' density of $0.02-0.1~{\rm deg}^{-2}$ \citep{Vieira:2013aa, Wardlow:2013aa,Weiss:2013aa}, suggesting that PASSAGES includes some of the very rarest all-sky objects. It is thus not entirely surprising that they would have higher intrinsic luminosities (when considering comparable lensing magnifications) and larger source sizes than other DSFG samples, which probe deeper into the regime of less extreme DSFGs but are restricted to a smaller survey area.

It might also be the case that other factors are at play. 
In the case of \citet{Spilker:2016aa} in particular, the more compact sizes might be due to their higher redshift ($z\sim 3.5-5$ primarily). While the angular size does not evolve significantly for a given physical size between $z\sim4$ and $z\sim2$, \citet{Fujimoto:2017aa} find a 
%rather strong 
detectable
evolution in physical size for a given $L_{\rm IR}$ ($L_{\rm IR} \sim 10^{12} - 10^{13}~\Lsun$), albeit with a small sample size ($<10$ galaxies at $z>4$). On the other hand, \citet{Enia:2018aa} do not observe a clear redshift evolution, and note that some apparent evolution could be due to selection effects biased against the decrease in surface brightness (for fixed luminosity and size) at higher redshifts. 
Additionally, the lens modeling approach for PASSAGES in this work is fundamentally different from \citet{Bussmann:2013aa,Bussmann:2015aa}, and \citet{Spilker:2016aa}, which all optimize model parameters directly over interferometric visibilities. This requires an adoption of a parametric form of the source-plane distribution (in contrast to our method, described in Section \ref{sec:effective_radius}).
\citet{Enia:2018aa} note that estimating source size from parametric modeling consistently resulted in systematically lower values, in comparison with an approach of measuring the area of the source above a certain signal-to-noise threshold. This is especially true for clumpy morphologies, where a Gaussian or S\'ersic profile might underestimate the true spatial extent.

An ample number of recent studies have revealed the presence of significant sub-kpc inhomogeneous, clumpy substructure that can dominate the sub-mm flux of DSFGs \citep{Swinbank:2010aa, Swinbank:2011aa, Danielson:2011aa, Hodge:2012aa, Hodge:2019aa, Alaghband-Zadeh:2012aa, Menendez-Delmestre:2013aa, Hatsukade:2015aa, Iono:2016aa, Canameras:2018aa, Tadaki:2018aa, Dessauges-Zavadsky:2019aa, Ramasawmy:2021aa, Spilker:2022aa}, although some counter-evidence to the ubiquity of clumps in DSFGs does exist (e.g. \citealt{Hodge:2016aa, Ivison:2020aa}).
These clumpy sites of star formation appear to be somewhat akin to larger ($\sim$100pc), more luminous analogs to giant molecular clouds \citep{Swinbank:2010aa,Dessauges-Zavadsky:2019aa}. 
Similar substructure is also apparent in rest-frame UV/optical imaging of star-forming galaxies (e.g. \citealt{Forster-Schreiber:2011aa, Iani:2021aa}, and references therein).
%
%Yet, for 
While the angular resolution of this current ALMA imaging ($\sim 0.5 \arcsec$)
is consistent with that of the {\it Herschel} and SPT lenses, 
it is challenging at present to assess the degree of clumpiness at sub-kpc scales (Fig.~\ref{fig:model_SP}). In the future, we intend to compare these results with higher-resolution ALMA observations, which enables physical resolutions closer to 100 pc with lensing, at which point we can adequately map the spatial variation of $\Sigma_{\rm SFR}$ (e.g. \citealt{Hodge:2019aa}).

The intrinsic far-IR continuum sizes in this study span a range of {$1.7 - 4.3$ kpc}, consistent with the typical predicted size of $\sim3$ kpc from radiative transfer modeling by \citet{Harrington:2021aa}, especially when adopting a revised fiducial magnification of $\mu\sim10-20$.
For optically-thick regions, the Stefan-Boltzmann Law applied to typical ULIRG luminosities and temperatures gives
$R_e = 1.09 \cdot [L_{\rm IR} / 10^{12}~\Lsun]^{1/2} \cdot [T_{\rm dust} / 35~{\rm K}]^{-2}$ kpc \citep{Scoville:2013aa}, so the luminosities for this sample ($L_{\rm IR} \approx (2 - 60) \times 10^{12} \Lsun$) 
translate to expected sizes of $R_e \approx 1.5 - 8.5$ kpc.
The far-IR physical sizes we find are comparatively a bit larger than those found by recent studies of submillimeter-bright galaxies, including \citealt{Simpson:2015ab} (median $R_e = 1.2\pm0.1$ kpc, interquartile range $1.8-3.2$ kpc), \citealt{Tadaki:2015aa} ($R_e = 0.7\pm0.1$ kpc), \citealt{Barro:2016aa} and \citealt{Oteo:2016aa} ($R_e \sim 1$ kpc), \citealt{Hodge:2016aa} (median $R_e = 1.8\pm0.2$ kpc), \citealt{Rujopakarn:2016aa} (median $R_e = 2.1\pm0.9$ kpc),
\citealt{Ikarashi:2017aa} ($R_e = 1.6\pm0.3$ kpc),
\citealt{Tadaki:2017ab} ($R_e \sim 1.5$ kpc),
and \citealt{Gomez-Guijarro:2022ab}
(median $R_e = 0.73\pm0.29$ kpc).
As emphasized by radiative transfer modeling of FIRE-2 \citep{Hopkins:2014aa,Hopkins:2018aa} zoom-in cosmological simulations by \citet{Cochrane:2019aa}, the spatial extent of single-band dust continuum emission is determined by a combination of the spatial extent of dust mass and of recent star formation, given the mechanism of dust heating. \citeauthor{Cochrane:2019aa} confirmed that spatially-resolved dust temperature gradients (measured via spectral indices of multiple sub-mm bands) is an effective method to disentangle these combined effects (e.g. as done with local galaxies by \citealt{Galametz:2012aa}).

A consistent picture has arisen that the far-IR emission of high-$z$ star-forming galaxies originates from a substantially more compact spatial region (by a typical factor of $2-4$) than that traced by rest-frame UV and optical imaging (e.g. \citealt{Calanog:2014aa,Simpson:2015ab,Chen:2015aa,Ikarashi:2015aa,Barro:2016ab,Talia:2018aa,Tadaki:2020aa,Pantoni:2021ab}). 
This is likewise generally the case for radio sizes as well, which are comparable to far-IR sizes (e.g. \citealt{Bondi:2018aa,Jimenez-Andrade:2019aa}).
%Biggs:2008aa, 
%
This compact phase of star formation is consistent with the evolutionary schemes for progenitors of massive elliptical/quiescent galaxies (e.g. \citealt{Barro:2013aa,Toft:2014aa, Lapi:2018aa}).
Intriguingly, \citet{Ikarashi:2017aa} found that DSFGs with evidence for composite star-forming/AGN growth were more compact than star formation-dominated or AGN-dominated objects ($R_e=1.0\pm 0.2$ kpc vs. $R_e=1.6\pm 0.3$ for star-formation dominated or $R_e=1.5\pm 0.6$ for AGN-dominated), which the authors suggest might be due to supermassive black hole growth during a compact star-forming phase of merger coalescence \citep{Springel:2005aa}. In contrast, extended FIR sizes of star-formation-dominated DSFGs may arise from an intermediary merger stage, while extended morphology of AGN-dominated DSFGs may be the result of positive feedback inducing star formation at larger radii \citep{Ishibashi:2012aa}. 
At present, though, there is not overwhelming evidence to support or rule out this theory.
%
%On the other hand, 
As an added caveat, the extended rest-frame UV/optical sizes could be the result of a radial dependence of dust attenuation (e.g. \citealt{Nelson:2016aa}), where a strongly centrally-concentrated distribution could result in artificially-extended half-light radii larger than half-mass radii (e.g., \citealt{Tacchella:2018aa,Bondi:2018aa, Suess:2019aa}). Early studies with \JWST\ have begun to test this hypothesis, and initial results (e.g. \citealt{Suess:2022aa}) indicate the important role that spatial variations in dust attenuation and stellar population ages can play.

The elevated dust-emitting sizes of PASSAGES starbursts may be explained primarily by the larger intrinsic IR luminosities. 
\citet{Fujimoto:2017aa} uncovered a power-law scaling relation between FIR effective radius and luminosity, $R_e \propto L_{\rm FIR}^{0.28 \pm 0.07}$ for $L_{\rm FIR} \gtrsim 10^{12} L_\odot$, in studying a large sample of several hundred 1 mm ALMA detections, with effective radii spanning physical scales of 
%$0.3 - 3.0$ kpc. 
$0.2 - 5.0$ kpc. 
\citet{Burnham:2021aa} offered some confirmation with a comparable power-law slope of $R_e \propto L_{\rm FIR}^{0.37 \pm 0.03}$ (in a much smaller sample of 18 DSFGs, median $R_e = 0.32\pm0.09$ kpc).
However, these trends have yet to be tested for a large ensemble of objects at $L_{\rm IR} \sim 10^{13} L_\odot$, i.e. HyLIRGs.
As we demonstrate in Fig.~\ref{fig:LIR_Reff}, the PASSAGES objects in this work are consistent with the previously-discovered trends.
%, 
%
Intriguingly, in comparison with the {\it Herschel} and SPT lensed DSFGs of \citet{Bussmann:2013aa} and \citealt{Spilker:2016aa}, as shown in Fig.~\ref{fig:LIR_Reff_lensed}, the PASSAGES objects appear to be larger on average for a given IR luminosity. This may be a result of the lower median redshift of PASSAGES, as discussed in the next section.

Some works have applied the Stefan-Boltzmann law at galaxy scales, where the IR luminosity can be related to effective galaxy size and dust temperature:
%Applying the Stefan-Boltzmann law to galaxy-wide scales suggests 
$L_{\rm IR} \approx 4\pi R_e^2 \sigma T_d^4$, for constant $\sigma = 5.670 \times 10^{-5}$ erg s$^{-1}$ cm$^{-2}$ K$^{-4}$. 
\citet{Yan:2016aa} suggested the use of a generalized, temperature-dependent $\tilde{\sigma}(T) = \sigma \times 10^{-3} (-3.03T^{1.5} + 45.55 T -127.53)$
in the case of a modified blackbody spectrum,
%,
whereas \citet{Ma:2015ab} used a simpler power-law relation to arrive at
$L_{\rm IR} = 4\pi R_e^2 \sigma T_d^{4.32}$.
% for a modified blackbody.
%
Here we adopt $\tilde{\sigma}(T) \propto T^{0.8}$, which closely approximates the \citeauthor{Yan:2016aa} relation as a single power-law for the primary range of interest, $T_d = 20 - 50$ K, such that 
$L_{\rm IR} \propto R_e^2 T_d^{4.8}$.
Empirically, \citet{Fujimoto:2017aa} and \citet{Burnham:2021aa} found a relation of $L_{\rm IR} \propto R_e^{2.7 - 3.6}$, with which the PASSAGES objects concur (Fig.~\ref{fig:LIR_Reff}).
In keeping with the Stefan-Boltzmann relation, this would necessitate 
that $T_d \sim R_e^{0.2 - 0.3}$, 
indicating a weak dependence, but this is contrary to some expectations that dust temperature decreases with size. For example, for 16 objects studied by \citealt{Hodge:2016aa}, there is an approximate correlation of $\log[T_d] / \log[R_e] \approx -2$ to $-1$ (although there is also an anti-correlation of temperature with redshift, which may play a larger role).
Regardless, it may be the case that this simple framework does not adequately describe the infrared emission of dusty star-forming objects, as it assumes that they radiate as optically-thick, spherical blackbodies. For this work in particular, sizes are measured at observed-frame 1 mm, in the Rayleigh-Jeans regime often assumed to be optically thin\footnote{
\citet{Harrington:2021aa} tested this assumption with masses derived from large velocity gradient (LVG; \citealt{Goldreich:1974aa, Scoville:1974aa}) radiative transfer models, finding that the single-band estimation can systematically over-predict mass (see also \citealt{Forster-Schreiber:2018aa, Liu:2019ae}).} 
(which fortuitously allows for the direct estimation of ISM mass from flux; \citealt{Scoville:2016aa, Scoville:2017aa}).

\begin{table*}
	\centering
	\caption{
	%{\red Potentially save high-res. for later paper.}
	Properties of JVLA 6 GHz continuum imaging, including natural-weighting flux ($S_{6 {\rm GHz}}$, uncorrected for lensing), inferred 1.4 GHz luminosity, and the derived radio-FIR correlation parameter, $q_{\rm FIR}$. Also included are the lensing-corrected source-plane effective radii (as with Table~\ref{tab:effective_radii}), and image-plane synthesized beam parameters for the image weightings used in the size calculation (and as presented in Fig.~\ref{fig:model_SP}).
	}
	\label{tab:effective_radii_VLA}
	%\begin{threeparttable}
	%\begin{center}
	\centering
	\begin{tabular}{ccccccccc} 
		\hline
		%\hline
		Field & $S_{6{\rm GHz}}$   & $\log[L_{\rm 1.4 GHz}]$ & $q_{\rm FIR}$  & $R_{{\rm e},{6 GHz}}$ & $R_{{\rm e},{6 GHz}}$ & $\theta_{\rm maj}$ & $\theta_{\rm min}$ & PA   \\ 
		& [$\mu$Jy] & $[{\rm W~Hz}^{-1}]$ & & [$\arcsec$] & [kpc] & [$\arcsec$] &   [$\arcsec$] & [\degrees]   \\[0.5ex]
		\hline\\[-1.5ex]
		%\hline\\[-2ex]
		PJ011646 	& $664	 \pm 37	$			&	24.9				$\pm$ 0.1 	& 		2.3	$\pm$ 0.2			& $0.67\pm0.01$ 					&  	$5.5 \pm 0.1$	& 0.73 		& 		0.36 		&  15.9 		\\[1ex]			                                 
		PJ014341 	& $76	 \pm 7	$			&	23.4				$\pm$ 0.3 	& 		2.8	$\pm$ 0.4			& $0.32\pm0.05$ 					&  	$2.6 \pm 0.4$	& 0.38 		&		0.29 		&  29.9 		\\[1ex] 
		PJ020941 	& $909	 \pm 37	$			&	25.0				$\pm$ 0.3 	& 		2.3	$\pm$ 0.5			& $0.23\pm0.05$ 					&  	$1.9 \pm 0.4$	& 0.51 		&		0.40 		&  25.0 		\\[1ex] 
		PJ022633 	& $1295	 \pm 59	$			&	25.0				$\pm$ 0.4 	& 		2.0	$\pm$ 0.6			& $0.57\pm0.39$  					&  	$4.3 \pm 3.0$	& 0.48 		&		0.37 		&  64.3 		\\[1ex] 
		PJ030510 	& $221	 \pm 22	$			&	24.9				$\pm$ 0.4 	& 		2.9	$\pm$ 0.5			& $0.79\pm0.03$ 					&  	$6.5 \pm 0.2$	& 1.21 		&		0.43 		&  27.2 		\\[1ex] 
		PJ105353 	& $874	 \pm 46	$			&	25.3				$\pm$ 0.1 	& 		2.1	$\pm$ 0.3			& $0.25\pm0.01$						&  	$1.9 \pm 0.1$	& 0.42 		& 		0.36 		&  24.6 		\\[1ex]				
		PJ112713 	& $678	 \pm 35	$			&	24.6				$\pm$ 0.1 	& 		2.2	$\pm$ 0.2			& $0.27\pm0.01$ 					&  	$2.3 \pm 0.1$	& 0.61		& 		0.37 		&  74.6 		\\[1ex]
		PJ113805 	& $525	 \pm 27	$			&	25.0				$\pm$ 0.3 	& 		2.0	$\pm$ 0.3			& $0.21\pm0.03$ 					&  	$1.7 \pm 0.2$	& 0.33 		&		0.30 		& -67.7		\\[1ex] 
		PJ113921 	& $249	 \pm 14	$			&	24.7				$\pm$ 0.3 	& 		2.7	$\pm$ 0.4			& $0.53\pm0.05$ 					&  	$4.1 \pm 0.4$	& 0.44 		&		0.39 		&  47.2 		\\[1ex] 
		PJ132630 	& $92	 \pm 6	$			&	24.2				$\pm$ 0.3 	& 		3.1	$\pm$ 0.4			& $0.34\pm0.11$ 					&  	$2.7 \pm 0.9$	& 0.43 		&		0.38 		&  80.9  		\\[1ex] 
		PJ133634 	& $403	 \pm 24	$			&	25.0				$\pm$ 0.4 	& 		2.3	$\pm$ 0.6			& $0.44\pm0.21$ 					&  	$3.3 \pm 1.5$	& 0.33 		&		0.28 		& -69.2 		\\[1ex] 
		PJ144653 	& $901	 \pm 49	$			&	24.3				$\pm$ 0.2 	& 		2.5	$\pm$ 0.4			& $0.27\pm0.03$ 					&  	$2.2 \pm 0.2$	& 0.41 		&		0.37 		&  49.4 		\\[1ex] 
		PJ144958 	& $400	 \pm 16	$			&	24.4				$\pm$ 0.8 	& 		2.6	$\pm$ 1.0			& $0.30\pm0.18$ 					&  	$2.5 \pm 1.5$	& 0.41 		&		0.38 		&  61.7  		\\[1ex] 
		PJ160722 	& $162	 \pm 9	$			&	23.9				$\pm$ 0.3 	& 		2.4	$\pm$ 0.6			& $0.34\pm0.05$						&  	$2.9 \pm 0.4$	& 0.47 		&		0.37 		&  -1.1  		\\[1ex] 
		PJ231356 	& $3244	 \pm 163^\dagger$	&\textemdash$^\dagger$ 		&\textemdash$^\dagger$			& \textemdash$^\dagger$	 			&  	\textemdash$^\dagger$ & 0.43 	&		0.29 		&  44.2  		\\[1ex] 
		\hline\\[-0.5ex]
	\end{tabular}
	%\multicolumn{6}{p{\columnwidth}}{
	\tablecomments{
	%{\bf Notes.}
	%\begin{tablenotes}\footnotesize
	%\item[$^a$] 
	$^\dagger$ Lensed images of PJ231356 are blended with a foreground double-lobed radio jet at 6 GHz, so the 6 GHz flux and effective radius of the lensed DSFG cannot be determined. The reported flux includes significant contribution from the bright foreground object.
	}
	%\end{center}
	\end{table*}

\begin{figure}
	\includegraphics[width=\columnwidth]{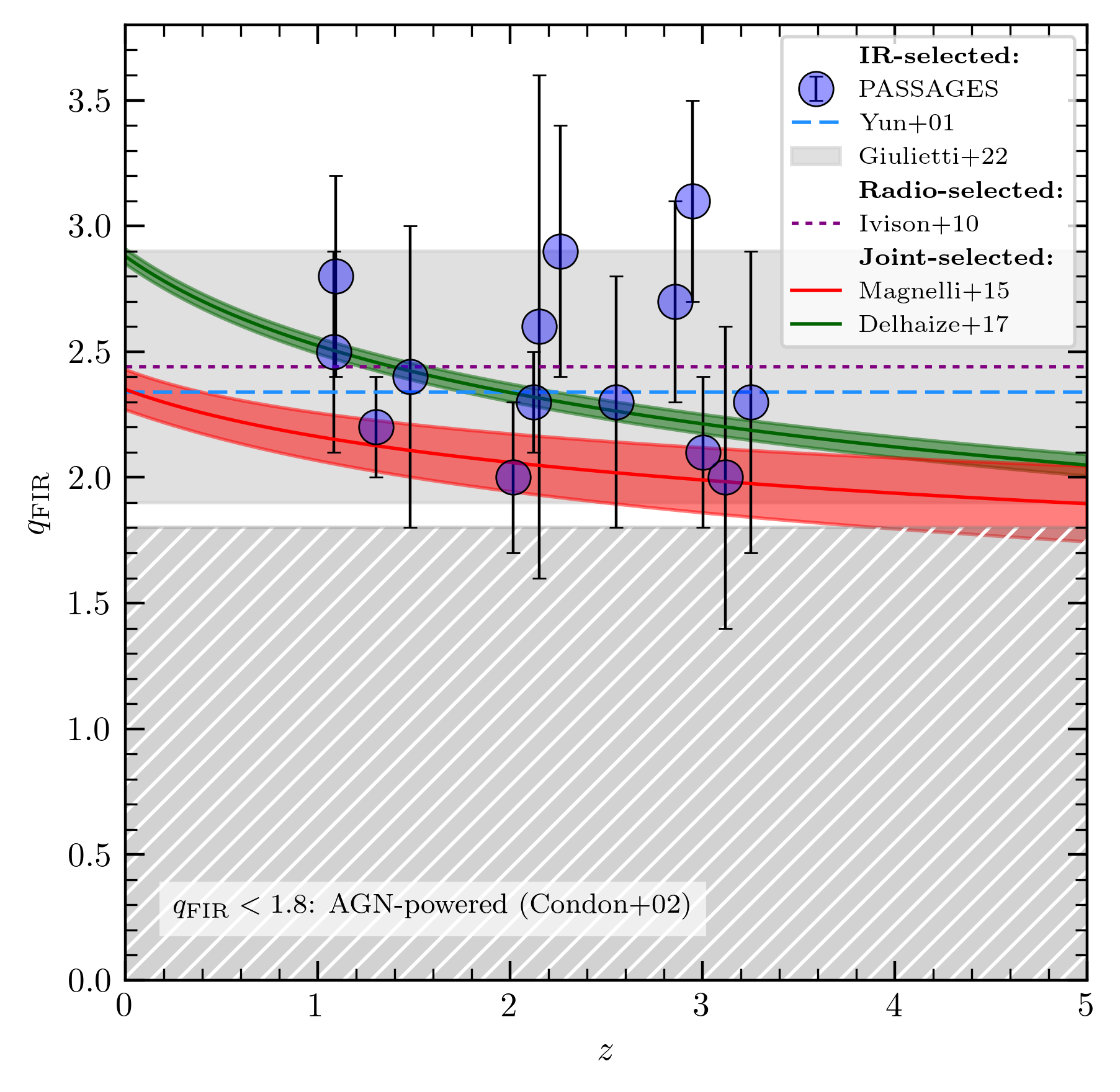}
    \caption{
    		The far-IR-radio correlation parameter $q_{\rm FIR}$ as a function of redshift for the PASSAGES sample. The 2$\sigma$ range of $q_{\rm FIR}$ for {\it Herschel}-selected lensed DSFGs (i.e., also selected in the IR) from \citet{Giulietti:2022aa} are shown as a shaded solid gray region. The median value from the IR-selected local sample from \citet{Yun:2001aa} and from the radio-selected sample of \citet{Ivison:2010ac} are shown as a blue dashed and purple dotted line, respectively. 
		Best-fit redshift evolutions derived from joint radio- and IR-selected samples from \citet{Magnelli:2015aa} and \citet{Delhaize:2017aa} are shown as red and green lines, respectively, with shaded regions showing 1$\sigma$ uncertainties.
		The gray hatch-shaded region at $q_{\rm FIR}$ is the proposed threshold by \citet{Condon:2002aa} for radio-loud galaxies powered by AGN, rather than star formation.
	   }
    \label{fig:radioFIRcorr_z}
\end{figure}

\subsubsection{Radio-FIR correlation and intrinsic radio source size}
\label{sec:radio_size}

The well-studied correlation between radio and FIR flux (e.g., \citealt{Helou:1985aa, Condon:1992aa,Yun:2001aa, Bell:2003aa, Murphy:2006aa, Murphy:2006ab}) is interpreted to be the result of the far-IR probing heated dust surrounding star-forming regions and the radio capturing 
%non-thermal 
synchrotron emission from relativistic electrons originating from supernova remnants. As the latter are the end-product of massive star formation, these two proxies are tightly correlated when averaged over galactic scales. 
The correlation even holds seemingly at sub-galactic scales (e.g. \citealt{Tabatabaei:2007aa, Dumas:2011aa}),
which may be explained by the more efficient propagation of relativistic cosmic rays in the dense ISM, with its higher magnetic field density.
The correlation parameter $q_{\rm FIR}$ is computed as 
\begin{align}
q_{\rm FIR} = & \log[L_{\rm FIR} / (3.75 \times 10^{12}~ {\rm W})] \nonumber \\ 
& - \log[L_{\rm 1.4~GHz} / (1~{\rm W~Hz}^{-1})] 
\end{align}
where we determine $L_{\rm 1.4~GHz}$ from 6 GHz flux as
\begin{align}
\frac{L_{\rm 1.4~GHz}}{{\rm W~Hz}^{-1}} = & \frac{1}{\mu_{\rm radio}} \frac{4\pi}{(1+z)^{1-\alpha}} \bigg( \frac{D_L(z)}{1~ {\rm m}}\bigg)^2 \\
& \times 
\bigg(\frac{S_{\rm 6 GHz}}{10^{32}~\mu{\rm Jy}} \bigg) \bigg(\frac{\rm 1.4~GHz}{\rm 6~GHz}\bigg)^{-\alpha} 
\end{align}
for luminosity distance $D_L$ in meters and spectral index $\alpha$ (assumed here to be 0.8; \citealt{Condon:1992aa}) such that $S_\nu \propto \nu^{-\alpha}$.
These luminosities (corrected for lensing magnification $\mu_{\rm radio}$) are included in Table \ref{tab:effective_radii_VLA}, along with the corresponding values of $q_{\rm FIR}$. 
We find values ranging from $q_{\rm FIR} = 2.0 - 3.1$, with a median of 2.3 and $1\sigma$ dispersion of 0.3, in agreement with the mean found by \citet{Yun:2001aa} for local IR-selected galaxies, $q_{\rm FIR} = 2.34$. This is also in good agreement with more recent results for {\it Herschel}-selected lenses from \citet{Giulietti:2022aa}, who found a $2\sigma$ dispersion of $q_{\rm FIR} \approx 1.9 - 2.9$. 
Unsurprisingly, none of the star-formation-dominated PASSAGES galaxies fall below the $q_{\rm FIR} < 1.8$ threshold for AGN-powered radio sources proposed by \citet{Condon:2002aa}.

As shown in Fig.~\ref{fig:radioFIRcorr_z}, there is not an evident decline in the $q_{\rm FIR}$ parameter with redshift for PASSAGES galaxies.
%,
%
As discussed extensively by \citet{Sargent:2010aa}, any observed trend with redshift is heavily dependent on selection biases from objects selected only in the IR or radio. Accounting for this affect, \citeauthor{Sargent:2010aa} found no evolution in the correlation out to at least $z\sim1.4$ (see also \citealt{Sargent:2010ab}).
However, \citet{Magnelli:2015aa} and \citet{Delhaize:2017aa}, both using a joint radio/FIR selection, observed similar redshift evolution in $q_{\rm FIR}$ to smaller values. 
Moreover, for {\it Herschel}-selected lensed DSFGs, \citet{Giulietti:2022aa} remarked upon a weak but detectable trend in line with previous results. 
\citet{Delhaize:2017aa} consider the possibility that contributions of AGN in the radio regime alone for star-forming galaxies could result in a steepening of the evolution with redshift. 
Alternatively, it is possible the calculation of $q_{\rm FIR}$ for PASSAGES galaxies is weakened by the assumption of a single radio spectral index $\alpha$, rather than direct measurement for each galaxy.
As also noted by \citet{Delhaize:2017aa}, for these higher-frequency measurements at 6 GHz, free-free emission (following instead $S_\nu \propto \nu^{-0.1}$) may contribute non-trivially at rest-frame $\nu \gtrsim 20$ GHz \citep{Condon:1992aa}.

Under a different assumption of a flatter spectral index $\alpha = 0.5$ (similar to that of extreme local starbursts, e.g., \citealt{Condon:1991aa, Clemens:2008aa}), we would find values of $q_{\rm FIR}$ greater than those reported in Table \ref{tab:effective_radii_VLA} by a small amount, $q_{{\rm FIR}, \alpha=0.5} -  q_{{\rm FIR}, \alpha=0.8} = 0.3 - 0.4$. 
This change in $q_{\rm FIR}$ is not insignificant, but it is smaller than the dispersion we find in $q_{\rm FIR}$ for our sample, and not much larger than the uncertainty on $q_{\rm FIR}$ for any individual object, so we conclude that the possible effect on our interpretation is minimal.

\begin{figure}
	\includegraphics[width=\columnwidth]{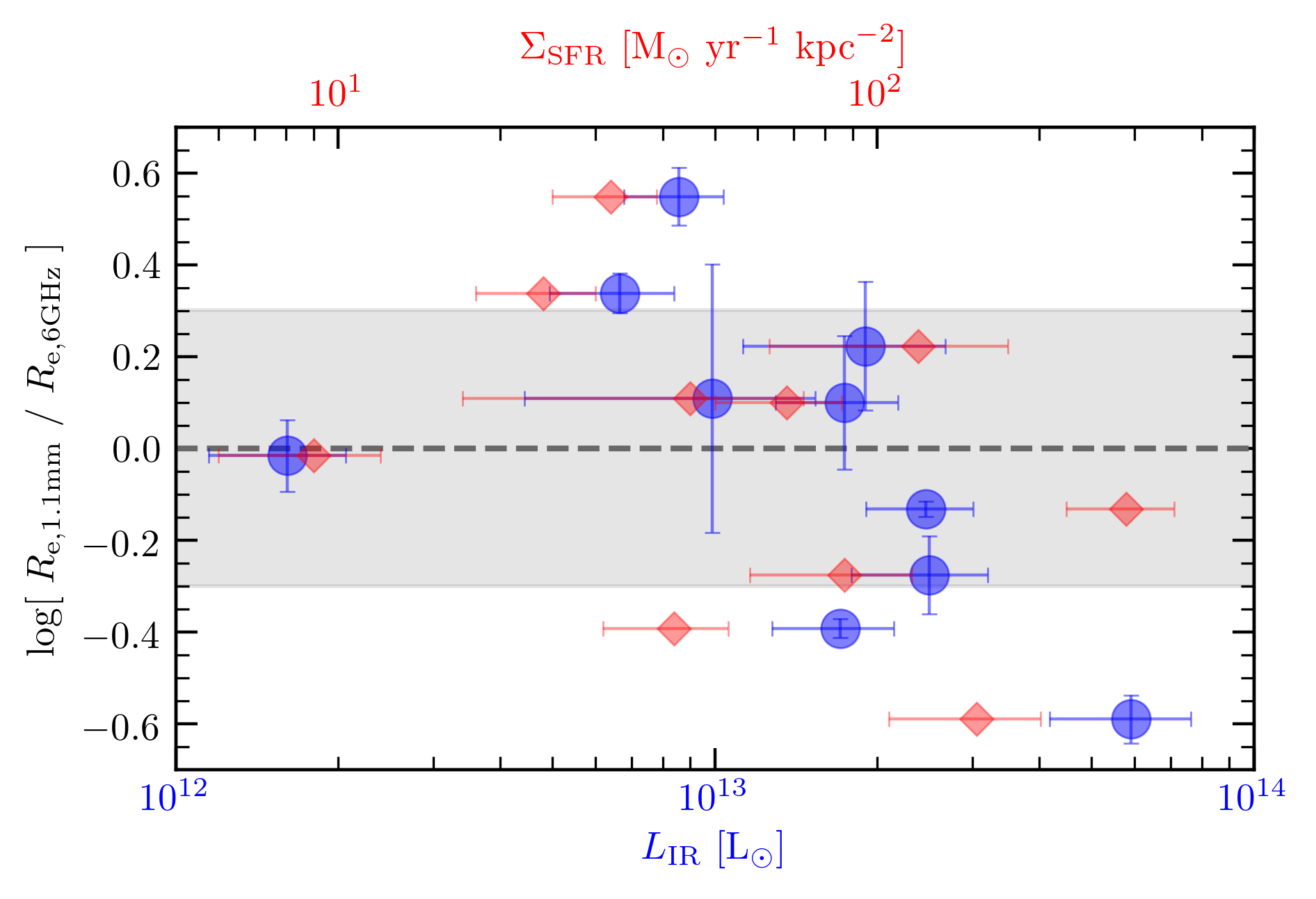}
	\includegraphics[width=\columnwidth]{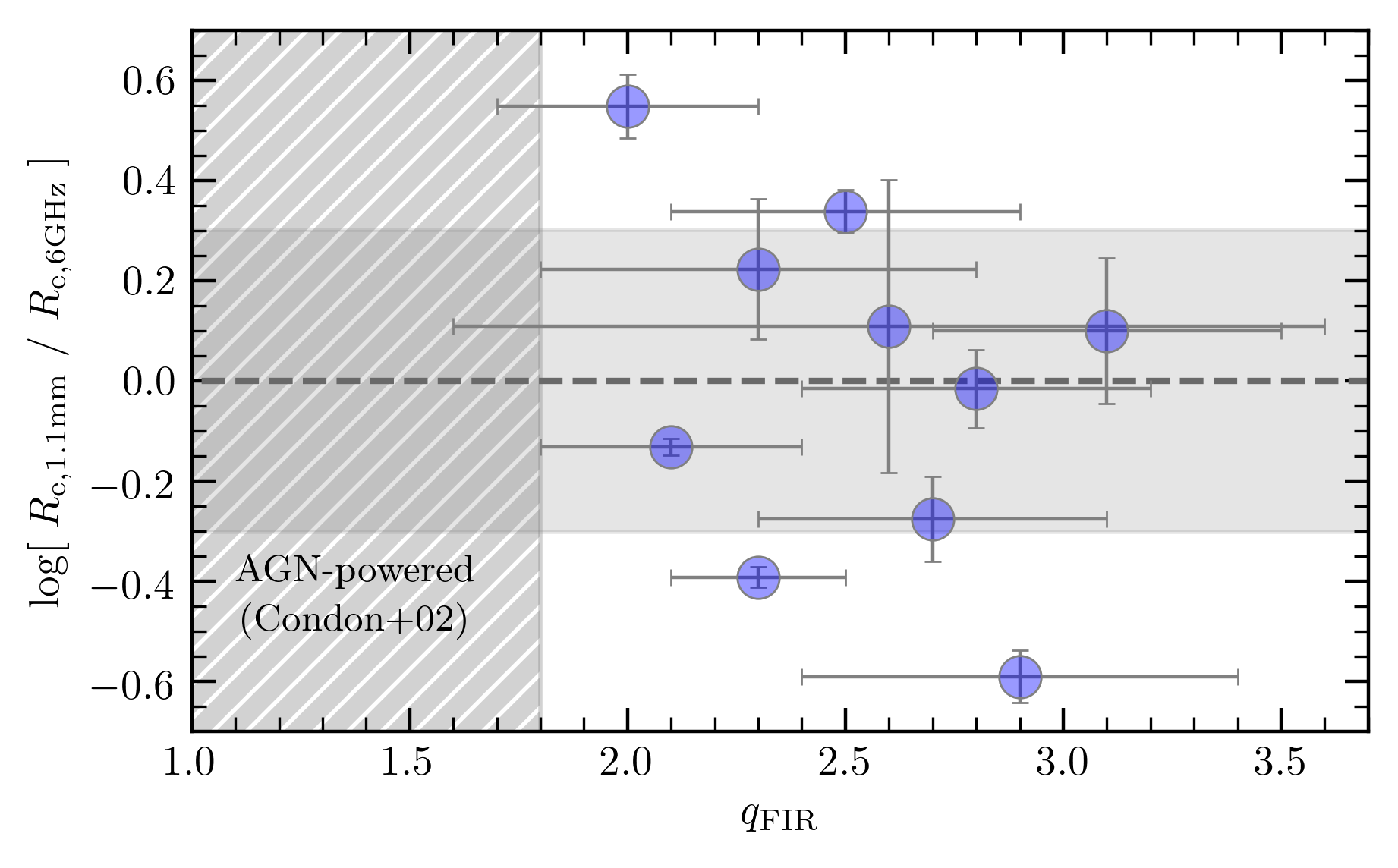}
    \caption{
    	{\it Top:}
	The ratio of 1 mm to 6 GHz continuum size, for the 10 objects in this work with both measurements available, vs. magnification-corrected infrared luminosity (blue circles, bottom axis) and vs. star formation rate surface density $\Sigma_{\rm SFR}$ (red diamonds, top axis). Vertical errorbars are not shown for the latter for legibility, but they are equal to those for the blue points.
	The gray shaded region shows delineates a factor of $>2$ difference in size.
	A possible trend where the more luminous, more densely star-forming objects have a preferentially more compact dust emission region is detectable, but small sample size makes interpretation difficult.
	{\it Bottom:}
	The same ratio of sizes is shown relative to the far-IR-radio correlation parameter $q_{\rm FIR}$. 
	The radio excess (AGN-powered) regime from Fig.~\ref{fig:radioFIRcorr_z} is again shown as a hatch-shaded region.
   }
    \label{fig:Reff_comparison}
\end{figure}

\begin{figure}
	\includegraphics[width=\columnwidth]{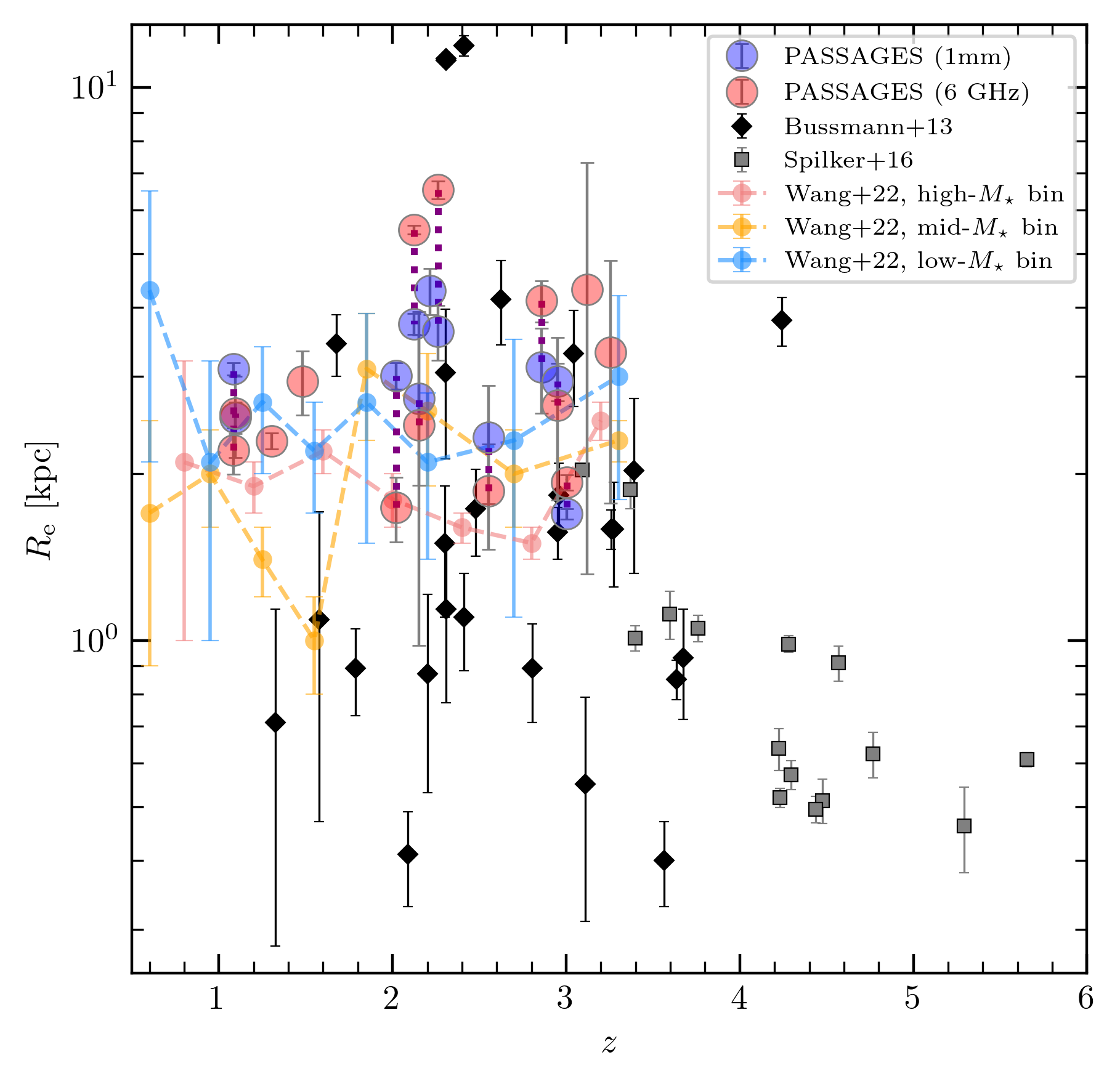}
    \caption{
	Effective radius vs. redshift for the PASSAGES sample (Tables \ref{tab:effective_radii} and \ref{tab:effective_radii_VLA}) and the other lensed DSFGs also shown in Fig.~\ref{fig:LIR_Reff_lensed} ({\it Herschel}, \citealt{Bussmann:2013aa}; SPT, \citealt{Spilker:2016aa}).
    There is some evidence for a downwards trend in galaxy size with increasing redshift, but this is illustrated mostly by the higher-$z$ SPT objects (likely owing to their selection at a longer wavelength of 1.4mm; \citealt{Weiss:2013aa}).
    We compare with the FIR sizes of stacked main sequence galaxies measured by \citet{Wang:2022ac}, in stellar mass bins of high ($11.0 \le {\rm log}_{10} M_\star < 12.0$), mid ($10.5 \le {\rm log}_{10} M_\star < 11.0$), and low ($10.0 \le {\rm log}_{10} M_\star < 10.5$). For $0.4 < z < 3.6$, \citeauthor{Wang:2022ac} find minimal size evolution with redshift, in accordance with the lack of a strong evolution of radio continuum sizes found by \citet{Jimenez-Andrade:2019aa}.
   }
    \label{fig:z_Reff}
\end{figure}

In Figs.~\ref{fig:Reff_comparison} and \ref{fig:z_Reff}, we examine the distribution of effective radii for our sample in both the rest-frame far-IR and radio, in comparison with far-IR sizes from \citet{Bussmann:2013aa} and \citet{Spilker:2016aa}. 
As both are expected to delineate the extent of ongoing massive star formation, the radio and far-IR sizes are reasonably consistent (where both are measurable), with discrepancies possibly explained partly by different observing resolutions. 
While not for the same set of galaxies, \citealt{Murphy:2017aa} found 10 GHz radio sizes comparable to the median dust-emission sizes of DSFGs from \citet{Ikarashi:2015aa} and \citet{Simpson:2015ab}; see also \citet{Jimenez-Andrade:2019aa} at 3 GHz. 
As a caution, however, the 1.4 GHz radio sizes for the same DSFGs from \citealt{Simpson:2015ab} (presented in \citealt{Biggs:2008aa}) are typically twice as large as the 870 $\mu$m sizes. 
This comparison of continuum size at different wavelengths could instead be the result of different optical depths, or even different surface brightness sensitivities between the two sets of observations.
\citet{Simpson:2015ab} suggested that the larger diffusion scale for cosmic rays (of order $\sim1-2$ kpc; \citealt{Bicay:1990aa, Marsh:1998aa, Murphy:2006ab}) might be responsible, relative to the smaller $\sim100$ pc diffusion length for the far-IR photons (which also explains why the radio-FIR correlation begins to decouple at resolved sub-kpc scales).
Fig.~\ref{fig:Reff_comparison} shows the ratio of 1 mm to 6 GHz effective radii vs. $L_{\rm IR}$ and $\Sigma_{\rm SFR}$ for the 10 objects with both radii measured, and there is a discernible trend where the more luminous, more densely star-forming galaxies have preferentially more compact dust emission regions (or conversely, perhaps more extended radio emission). When excluding the lower-luminosity object PJ014341, this negative correlation with luminosity is significant at $p\lesssim0.01$. 
\citet{Murphy:2006ab} suggested that galaxies with larger IR surface densities had undergone a more recent episode of star formation, such that young cosmic rays would only have had time to travel $\sim 100$ pc, leading to smaller radio scale lengths.
Curiously, this appears not to be the case for our sample, 
which instead shows larger radio scales with larger $\Sigma_{\rm SFR}$, but it's not currently practical to draw conclusions based on this sample.
However, in local edge-on galaxies, \citet{Wiegert:2015aa} also found preliminary evidence for a correlation of radio halo size with $\Sigma_{\rm SFR}$, such that a compact star formation distribution was advantageous to creating radio halos (also \citealt{Dahlem:2006aa}).  
Yet, \citet{Heesen:2018aa} did not find a strong correlation between radio scale height and SFR, $\Sigma_{\rm SFR}$, or galaxy rotation speed, for both diffusion- vs. advection-dominated (or wind-driven) modes of cosmic ray transport.
Lastly, in Fig.~\ref{fig:Reff_comparison}, we examine the same size ratio as a function of the radio-FIR correlation parameter, $q_{\rm FIR}$. There is again some weak indication of a negative correlation, but not with sufficient statistical significance given uncertainties. 
Such a correlation might reveal the influence of an AGN in driving towards more compact radio half-light sizes.

\subsubsection{Redshift evolution of galaxy size}
\label{sec:redshift_evolution}

\citet{Fujimoto:2017aa} found modest evolution in the $R_e-L_{\rm FIR}$ relationship between $z=1-2$ and $z=2-4$, and a more significant evolution (by a factor of $\sim 2$) to $z=4-6$. This decrease in effective radius with redshift for a given luminosity is consistent with rest-frame UV/optical samples.
For example,
\citet{Shibuya:2015aa} found that the UV/optical effective radius of UV-luminous star-forming galaxies evolves with redshift as $R_{e, {\rm UV}}/{\rm kpc} \propto (1+z)^{-0.84 \pm 0.11}$. While the slope between $R_e$ and UV luminosity is consistent with redshift, the normalization decreases for the range $z=0-8$.
At longer wavelengths, \citet{Jimenez-Andrade:2021aa} found at 3 GHz that $R_{e, {\rm radio}}/{\rm kpc} \propto (1+z)^{-0.3 \pm 0.3}$ from $z\approx 0.3-3$, while \citealt{Lindroos:2018aa} found steeper evolution at 1.4 GHz of $R_{e, {\rm radio}}/{\rm kpc} \propto (1+z)^{-1.7}$ out to $z\approx3$.
These results suggest that the correlation between luminosity and effective radius holds across cosmic time, but the size of an object with fixed luminosity decreases with lookback time. In turn, the surface density of star formation appears to increase with lookback time.

In Fig.~\ref{fig:z_Reff}, we 
%observe very minimal detectable 
do not observe a clear indication of
redshift evolution for the PASSAGES sample alone, covering only $z=1-3.5$, as is also the case for the {\it Herschel}-selected objects at a similar redshift range \citep{Bussmann:2013aa}.
While the average far-IR size is elevated for PASSAGES relative to the \citeauthor{Bussmann:2013aa} sample, there is significant overlap in the observed size range.
This minimal evolution in far-IR size matches the results found by \citet{Wang:2022ac} for stacked main-sequence galaxies from $z = 0.4 - 3.6$, in three stellar mass bins: high ($11.0 \le {\rm log}_{10} M_\star < 12.0$), mid ($10.5 \le {\rm log}_{10} M_\star < 11.0$), and low ($10.0 \le {\rm log}_{10} M_\star < 10.5$). 
We find also that these main-sequence sizes are typically smaller than those found for the DSFGs in this work, but not much further than the level of $1\sigma$ uncertainties of each bin.
On the other hand, for the higher-redshift SPT sample ($z=3-6$), there is hardly any overlap in effective radius with PASSAGES galaxies, the latter of which are systematically larger in size. 
%
%It is tempting to explain this 
This systematic discrepancy in sizes above vs. below $z\sim 3$
%well 
be a signpost of redshift evolution in DSFGs, but 
this direct comparison is 
%difficult, owing to 
complicated slightly by
the different selection effects at play for each sample.

\subsection{Eddington-limited star-formation surface densities?}
\label{sec:Eddington_SFR}

As the DSFGs in this sample are among the most strongly star-forming systems presently known \citep{Berman:2022aa}, they offer key insight into effective maximum thresholds of star formation (e.g. \citealt{Elmegreen:1999aa, Tacconi:2006aa}).
In particular, stellar mass growth is understood to be a self-regulating process: as gas collapses under self-gravity to form stars, the short-lived, massive stars inject a sizable radiation pressure in opposition to dust grains, which are coupled to the remaining gas \citep{Thompson:2005aa}. 
The concept that radiation pressure can have significant influence on star formation is not novel (e.g. \citealt{Elmegreen:1983aa, Scoville:2001aa, Scoville:2003aa}); it can even induce star formation in other regions of a cloud (e.g. \citealt{Elmegreen:1977aa}).
\citet{Scoville:2001aa} 
%and \citet{Scoville:2003aa} 
derived that a luminosity-to-mass ratio of a star cluster of $\sim 500 ~ L_\odot / M_\odot$ would provide a radiation pressure in excess of its self-gravity and halt accretion to the cloud core\textemdash analogous to the so-called Eddington limit\textemdash and by extension, on galaxy-wide scales as well. This is a conservative theoretical limit, assuming only free-fall collapse and no other turbulence- or magnetic-driven obstacles to molecular core accretion. It also does not consider the likely possibility of asymmetric cloud geometries reducing the importance of radiation pressure.
Additionally, this limit does not account for mechanical feedback contributed by stellar winds, which may well be an important factor (e.g. \citealt{Tan:2001aa, Harper-Clark:2009aa, Rogers:2013aa}).

The vast majority of DSFGs and luminous IR galaxies appear to form stars at sub-Eddington rates \citep{Tacconi:2006aa, Hodge:2015aa, Hodge:2019aa} with the exception of some local LIRGs and ULIRGs (e.g. \citealt{Barcos-Munoz:2017aa}). One example is the prototypical Arp 220 \citep{Scoville:1997aa, Downes:1998aa}, which has star formation surface densities\textemdash defined as $\Sigma_{\rm SFR} = {\rm SFR}/(2\pi {R_{\rm eff, dust}}^2)$\textemdash on the order of $10^4~{M}_\odot~{\rm yr}^{-1}~{\rm kpc}^{-2}$, likely the highest-known value \citep{Barcos-Munoz:2015aa}.
Here, $\Sigma_{\rm SFR} \approx 1000~{M}_\odot~{\rm yr}^{-1}~{\rm kpc}^{-2}$, or $\Sigma_{\rm IR} \approx 10^{13}~L_\odot~{\rm kpc}^{-2}$ \citep{Thompson:2005aa,Andrews:2011ab}, is the typical Eddington limit
%, above which subsequent star formation should be quenched.
for a star-forming system in equilibrium.
There is some evidence for near-Eddington star formation for a small number of high-$z$ objects, including the submillimeter-bright galaxies 
GN20 and AzTEC-1 \citep{Younger:2008aa,Daddi:2009aa}; 
AzTEC2-A \citep{Jimenez-Andrade:2020aa};
AzTEC-3 \citep{Riechers:2014aa}; 
HFLS3 \citep{Riechers:2013aa}; 
the Cosmic Eyelash \citep{Thomson:2015aa};
SGP38326 \citep{Oteo:2016aa};
SPT0346-52 \citep{Ma:2016aa};
ALMACAL-1 and ALMACAL-2 \citep{Oteo:2017ac};
ADFS-27 \citep{Riechers:2017aa};
and, more recently, the Sunrise Arc \citep{Vanzella:2019aa,Welch:2022ab}.
Some quasar hosts have also shown evidence for such compact, maximal starbursts, 
at $z\sim2$ \citep{Stacey:2021ab}
and $z>6$ \citep{Walter:2009aa}.
However, angular resolution plays a critical role in this measurement, as it may be possible for objects to be sub-Eddington on global, galaxy-integrated scales but super-Eddington in local regions of dense star formation, as suggested by \citet{Simpson:2015ab} and \citet{Barcos-Munoz:2017aa}. 
For example, for local ULIRGs, \citet{Song:2022aa} found radio continuum clumps smaller than 100 pc with large surface densities up to $\Sigma_{\rm SFR} \approx 1600 ~{M}_\odot~{\rm yr}^{-1}~{\rm kpc}^{-2}$.
%
%

%SDP.81 - 190 Msun... \citealt{Rybak:2015aa}
Yet, a vexing result has been the discovery that high-redshift DSFGs almost universally do not exceed the Eddington limit even at $\sim$500pc scales, as accessed by very high-resolution ALMA imaging and/or strong lensing (e.g. \citealt{Bussmann:2012aa, Bussmann:2013aa, Rybak:2015aa, Enia:2018aa, Hodge:2019aa, Dudzeviciute:2020aa}).
In particular, in the case of \citet{Rybak:2015aa}, the star formation surface density of SDP.81 is mapped at sub-50pc scales (unprecedented at high-$z$), giving a maximum of $190\pm20~{M}_\odot~{\rm yr}^{-1}~{\rm kpc}^{-2}$, significantly under the theoretical limit.
This also appears to be the case for the PASSAGES sample studied so far, as shown in Fig.~\ref{fig:LIR_Reff_lensed} and \ref{fig:LIR_Reff} and Table~\ref{tab:effective_radii}.
In contrast, turbulence-based ISM modeling of the full sample by \citet{Harrington:2021aa} revealed several objects approaching $L_{\rm IR} / M_{\rm ISM} = 500 \Lsun/\Msun$, but most of these are coincidentally excluded from our measurement, owing to their lack of either ALMA observations or fully-developed lens models.

In line with \citet{Hodge:2015aa, Hodge:2019aa}, we also estimate the peak values of $\Sigma_{\rm SFR}$ in Table~\ref{tab:effective_radii} by normalizing the 1 mm flux density to equal the SFR, effectively assuming that the spatial distribution at sub-mm wavelengths directly traces the distribution of star formation (also \citealt{Hatsukade:2015aa, Tadaki:2018aa, Sharon:2019ab}).
Unlike \citeauthor{Hodge:2019aa}, however, this calculation applied towards lensed galaxies is complicated by the varying source-plane PSF. To mitigate this, we compute the 99.7th percentile (i.e. the 3$\sigma$ confidence interval of pixel values) as a lower bounds on $\Sigma_{\rm SFR, peak}$. 
This helps to mitigate any spurious non-physical artifacts introduced by the lensing reconstruction, and also accounts partially for the correlation of adjacent pixels within a beam size.
In future work, with higher angular resolution observations, it will be worthwhile to construct smoothed (i.e. uniform PSF) maps of $\Sigma_{\rm SFR}$, but the current ALMA data are only marginally able to resolve this distribution. For this reason, the peak values given in Table~\ref{tab:effective_radii} are usually consistent with globally-averaged values, which is what one would expect for source sizes comparable in extent to the resolving beam.

While it is clear that 100pc-scale imaging is essential to properly assess these maximum surface densities, 
there may be other factors responsible. 
One such factor, as pointed out by \citet{Hodge:2019aa}, might be the assumption of a single dust temperature, leading to a possible underestimation of the overall SFR \citep{Berman:2022aa}. For example, \citet{Calistro-Rivera:2018aa} find evidence for linear temperature gradients in four DSFGs, decreasing from the center to outer parts of the disk, which would directly affect the peak values of $\Sigma_{\rm SFR}$.
Still, it is not likely that this effect can fully account for the order-of-magnitude discrepancy in $\Sigma_{\rm SFR}$ below the Eddington limit. $\Sigma_{\rm SFR}$ scales linearly with star-formation rate, but scales with inverse-square dependence on galaxy size, so the latter is a more facile explanation.
Additionally, \citet{Andrews:2011ab} and \citet{Murray:2010aa} point out that intermittency and non-uniformity of star formation over the extent of a galaxy can lead to a global star-formation surface density that appears significantly more sub-Eddington than if higher spatial and temporal resolution were achieved.
Ultimately, a larger sample of sub-100pc dust continuum observations of DSFGs is required to make more substantive claims,
and we intend to pursue this in the future with a more thorough, higher-resolution analysis of the PASSAGES sample.

\begin{figure}
	\includegraphics[width=\columnwidth]{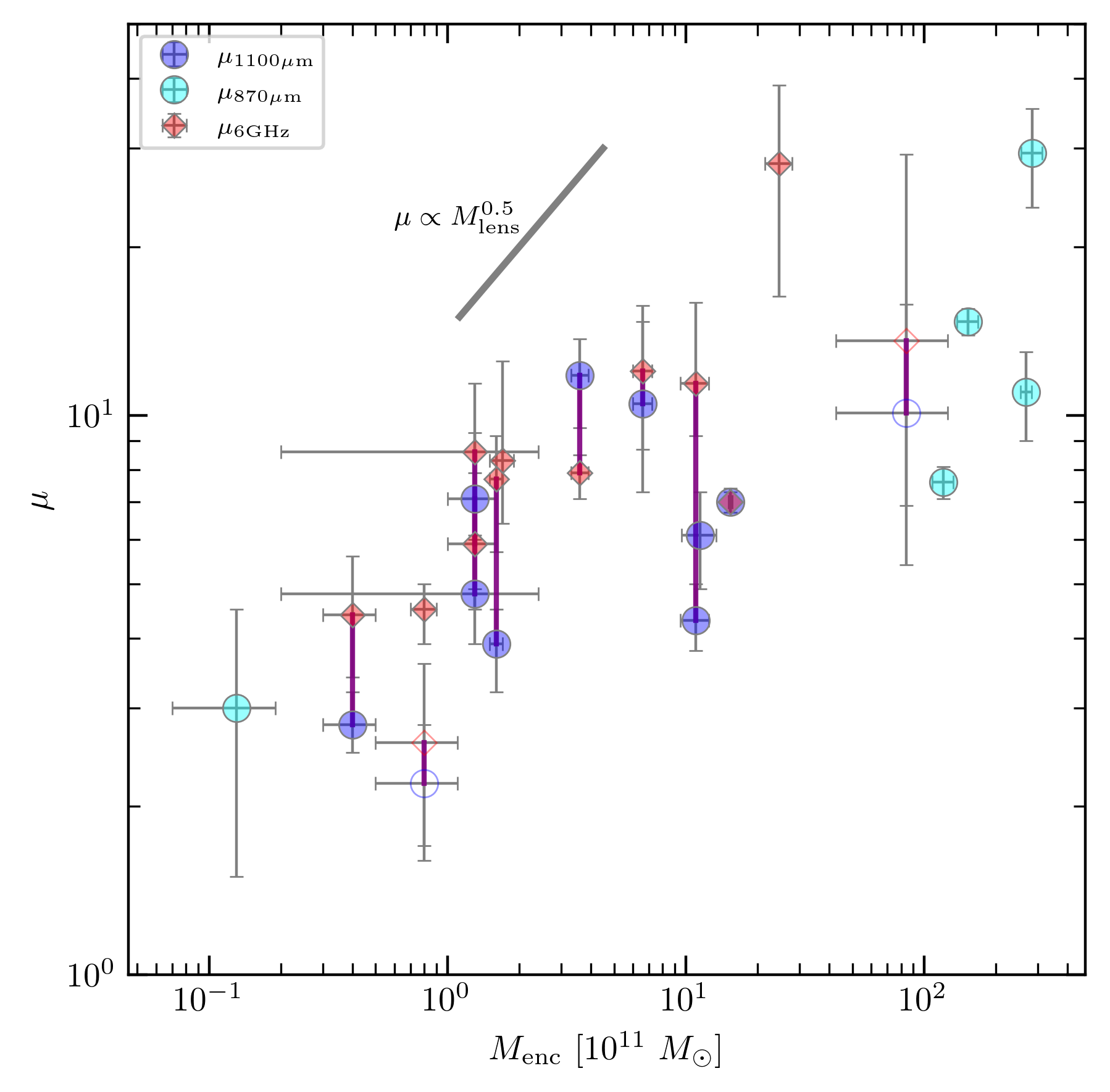}
    \caption{
   Magnification $\mu$ vs. mass enclosed within the Einstein radius (as derived from our lensing models) for the PASSAGES sample (wavelength denoted by color). 
   Open symbols denote 3 objects for which the foreground redshift is preliminary; uncertainties include the wider ranges given in Table~\ref{tab:lensproperties}, which encapsulate some of the effect this has on enclosed mass determination. While shown in this plot, they are not included in our analysis in \S \ref{sec:lens_mass}.
   Purple lines connect radio and far-IR measurements where both exist for the same object. 
   A fiducial line segment shows the slope of $\mu \propto \sqrt{M_{\rm lens}}$, a rough expected scaling relation.
   }
    \label{fig:M_enc}
\end{figure}

\subsection{Connection of lensing halo mass to magnification}
\label{sec:lens_mass}

The distribution of Einstein radii (and masses enclosed within these radii) is summarized in Table~\ref{tab:lensproperties} (also Lowenthal et al.,
%2022, 
in prep.). 
As remarked by \citet{Frye:2019aa} (see their Figure 1), the objects selected via {\it Planck}-{\it Herschel} or {\it Planck}-WISE have a tendency for larger far-IR flux and Einstein radii,
perhaps resulting from the larger-area footprint of {\it Planck}. 
The median Einstein radius is found to be $\sim 1.5\arcsec$, in contrast with the smaller median of $\sim 0.8\arcsec$ for the {\it Herschel} lenses of \citet{Eales:2015ab} and \citet{Amvrosiadis:2018aa}.
While the clear majority are galaxy-scale lenses, the total halo masses enclosed within the Einstein radius for PASSAGES ranges from $1\times10^{10}$ to $\sim 3 \times 10^{13}~M_\odot$, covering galaxy, group, and low-mass cluster scales. 
There is not a systematic effect where the most luminous objects in the sky are all high-magnification cluster-lensed DSFGs; galaxy-scale lenses are just as capable of producing lensed DSFGs with large fluxes.

The source-plane area inside the caustic curves that is capable of producing high magnifications (or lensing cross-section; \citealt{Turner:1984aa, Meneghetti:2003ab}) has been found to correlate tightly with 
%the mass of the foreground lens (along with the Einstein radius). 
the Einstein radius (regardless of definition; \citealt{Meneghetti:2011aa}). 
%In some cases, this is trivial, as some definitions for Einstein radius are based on the area 
%
By extension, as $\theta_{\rm Ein} \propto \sqrt{M}$, this cross-section also should correlate with lensing mass.
This effect might be at play in Fig.~\ref{fig:M_enc}, which shows a positive correlation between magnification and the mass enclosed within the Einstein radius. 
As a general rule-of-thumb, magnification scales approximately as $\mu \propto \sqrt{M_{\rm lens}}$ for a circularly symmetric lens (assuming fixed source position; see e.g. \citealt{Narayan:1996aa,Frye:2019aa}).
Additional factors play a large role in determining magnification\textemdash including perhaps most notably the alignment of background relative to the foreground mass profile\textemdash so there should be large scatter in this relation. We thus do not consider it instructive to parameterize it, but we remark that the distribution of PASSAGES objects in this work is broadly consistent with this trend (shown in Fig.~\ref{fig:M_enc}).

Building on this, 
since
%Seeing as 
we recall that more extended source-plane objects may also be subject to lower overall magnification factors than compact ones (e.g. \citealt{Hezaveh:2012aa}), then a useful parameter might be the ratio of galaxy angular size (effective radius) to the angular Einstein radius of the lens, $R_0 \equiv R_{\rm e} / \theta_{\rm Ein}$. 
If this ratio is close to unity, the size of the background object would be comparable to the areal extent of the caustic network. 
Larger values would necessitate (in general) that the source extends further into lower-magnification regions, resulting in lower, more diluted magnifications. Smaller values, on the other hand, would suggest that the background object can carefully align inside the caustics. 
As remarked in \S \ref{sec:size_bias}, however, smaller source-plane objects have a lower probability of being near to the high-magnification region of the source-plane when positioned at random.
Ideally, one can describe a balance between these two competing effects.

For a general SIS lens, \citet{de-Freitas:2018aa} derived analytic, perturbative solutions for the source-plane cross-sections $\sigma_\mu$\textemdash or the solid angle of the region with magnifications exceeding some threshold $\mu_{\rm thr}$\textemdash as a piecewise function of $R_0$:
\begin{align}
\label{eqn:deF}
\sigma_\mu &= 
	\begin{cases}
	\frac{4\pi}{\mu_{\rm thr}^2} + \frac{\pi}{4}R_0^2, & R_0 <  R_J \\
	\pi (4 R_0^2 - \mu_{\rm thr} R_0^3) \\
	\hspace{1cm}
	- \frac{1}{R_0^2} 
	\big( 
	\frac{3\pi - 8}{(4 - 8/\pi)^2} 
	\big) \\
	\hspace{1cm}
	\times
	(4 R_0^2 - \mu_{\rm thr} R_0^3)^2, & R_0 \ge R_J
	\end{cases}
\end{align}
where the joining point $R_J = 2.15 / \mu_{\rm thr}$ marks a transitional point to the Einstein ring regime. As $\sigma_\mu$ is relative to the Einstein radius, it can be multiplied by $R_0^2$ to obtain an absolute solid angle (e.g. in arcsec$^2$). 
This treatment implies that the peak area for $\mu_{\rm thr} = 5$ is approximately $R_0 = 0.6$ (i.e., source size 60\% of the Einstein radius). 

\begin{figure}
\includegraphics[width=\columnwidth]{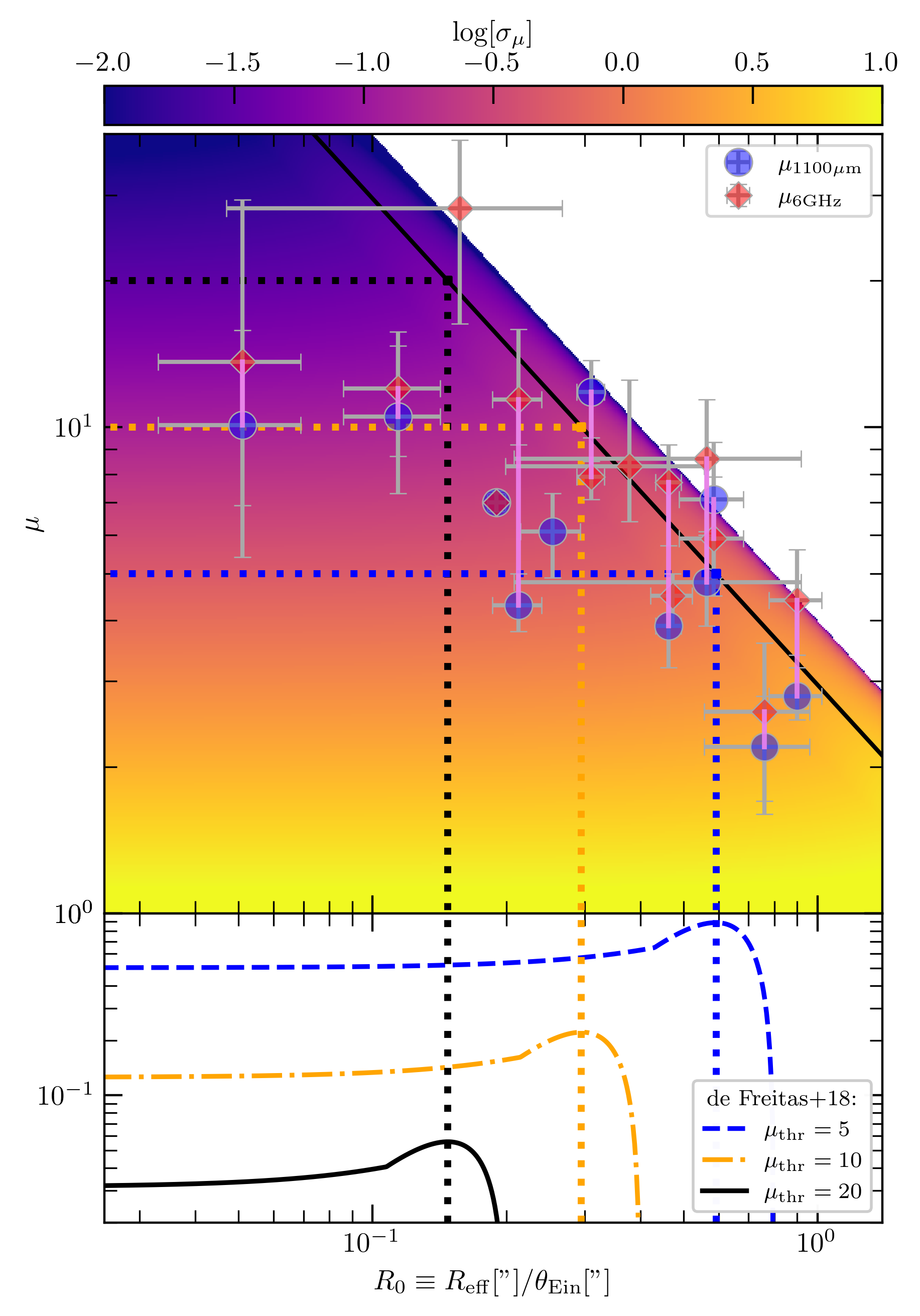}
 \caption{
  {\it Top panel:}
   Magnifications vs. the ratio $R_{\rm e} / \theta_{\rm Ein}$, or source-plane effective radius (in arcsec) to Einstein radius (also in arcsec).
   This ratio can serve as a proxy for the compactness of the background object relative to the foreground lens. There is an apparent negative correlation between this ratio and magnification, suggesting possibly that smaller background objects can align more precisely with the highest-magnification regions in the source plane (which increase in size with Einstein radius). 
   The redshift uncertainty of the foreground for 3 objects does not strongly influence these values, as $\theta_{\rm Ein}$ is computed in angular units.
   {\it Bottom panel:}
   The predicted source-plane cross sections $\sigma_\mu$ as a function of the same ratio, for magnification thresholds of $\mu_{\rm thr} = 5, 10, 20$, using the analytic perturbative solutions calculated by \citet{de-Freitas:2018aa}. The relative solid angle $\sigma_\mu$ can be converted to absolute angular units by multiplying by $\theta_{\rm Ein}^2$.
   Dotted vertical lines indicate the optimal source size for given magnification thresholds (shown as the conjoined horizontal lines in the upper panel) to maximize the cross-sectional area.
   These maxima are located at $\mu \approx 2.9522 / R_0$, which is shown as a black solid line in the top panel.
   The color gradient in the top panel shows the distribution of $\sigma_\mu$ by computing these curves as a function of $R_0$ for a range of continuous $\mu$ values. This indicates that smaller magnifications have larger cross sections, as expected, but there is also a thin diagonal band of elevated cross sections around the black solid line.
    }
    \label{fig:size_ratio_mag}
\end{figure}

These analytic curves are shown in the bottom panel of Fig.~\ref{fig:size_ratio_mag}.
For given magnification $\mu$, the cross-sectional area $\sigma_\mu$ essentially describes the probability of a given $R_0$ ratio. 
We find that the peaks for the selected magnification thresholds ($\mu_{\rm thr} = 5, 10,$ and 20) are approximately concordant with the negative correlation seen in the upper panel of Fig.~\ref{fig:size_ratio_mag} for the PASSAGES sample, as indicated by the conjoined vertical/horizontal dotted lines.
Specifically, the local maxima of $\sigma_\mu$ in Equation \ref{eqn:deF} are found to occur at $\mu(R_0) \approx 2.9522 / R_0$, shown as a solid black line in Fig.~\ref{fig:size_ratio_mag}.
There is perhaps greater deviation from this relation for smaller values of $R_0$, but there are also fewer objects occupying this regime, so the deviation is not a robust conclusion.

The relative probability of lower-magnification lenses is higher, as $\sigma_\mu \sim \mu_{\rm thr}^{-2}$, which would suggest that they would be more numerous. However, this will not be fully reflected for the PASSAGES sample, as the selection 
%effects will lead to a range of preferred magnifications.
is not uniformly sensitive to all magnifications.
This is because selection by apparent flux serves to maximize the multiplicative product of magnification and intrinsic luminosity (the latter of which is itself correlated with source-plane size, as discussed in \S \ref{sec:FIR_size}).
For this reason, the recovered distributions of intrinsic luminosities, magnifications, Einstein radii, and source-plane sizes are all interconnected for PASSAGES and other lensing samples defined by flux.

In future work (Englert at al., in prep.), 
we intend to place the PASSAGES sample in context with other studies that have used magnification and Einstein radius distributions as constraints on the halo mass function and other cosmological parameters
(e.g., \citealt{Eales:2015ab, Amvrosiadis:2018aa}). This is an analysis which benefits immensely from a large sample with consistent selection effects.

\section{Conclusions}
\label{sec:conclusions}

We have presented detailed lens modeling for 15 members of the PASSAGES ({\it Planck} All-Sky Survey to Analyze Gravitationally-lensed Extreme Starbursts) sample of dusty star-forming galaxies ($z=1.1 - 3.3$) that were introduced in \citet{Harrington:2016aa} and \citet{Berman:2022aa}. To do this, we gather complementary information from an extensive set of recent multi-wavelength sub-arcsecond continuum imaging with ALMA, JVLA, {\it HST}, and Gemini-S. We implement parametric lens modeling with \lenstool, where priors on the nature of the foreground mass distribution are established by optical/near-IR imaging, and multiple image positions of background objects (from imaging at all wavelengths) provide modeling constraints. 
In analyzing the results of this modeling and the resultant source-plane reconstructions, we find:

\begin{itemize}

\item The 1 mm and 6 GHz magnification factors of the DSFGs modeled in this sample range from $\mu=2-28$, with more than half at lower magnification, $\mu < 10$. 
Correcting the apparent infrared luminosities from \citet{Harrington:2016aa} and \citet{Berman:2022aa} with these magnification factors, we find intrinsic values of $L_{\rm IR} = 0.2-5.9\times 10^{13}~\Lsun$ (median $1.4\times 10^{13}~\Lsun$), placing them solidly within the regime of hyperluminous infrared galaxies, or HyLIRGs (Fig.~\ref{fig:corrected_luminosities}). 
The corresponding inferred star formation rates (also corrected for lensing) span $200-6300~M_\odot~{\rm yr}^{-1}$ (median $1500~M_\odot~{\rm yr}^{-1}$).
%
%Their extreme rarity 
Their extreme, rare properties are
likely to be a direct result of the all-sky selection method for PASSAGES.
While initially assumed to be starbursts based on their large star formation rates, evidence from their molecular gas depletion times and star formation rate surface densities indicates that\textemdash while nearly all above the center of the star-forming main sequence\textemdash they largely are not predicted to surpass a $\Delta {\rm MS} > 0.6$ dex threshold typically imposed to define starbursts. 
Future work to attempt to constrain stellar masses is thereby warranted in order to directly establish their relation to the main sequence.

\item Despite being among the most IR-luminous galaxies ever discovered in the Universe, owing to their advantageous large-area selection, their submillimeter magnification factors are largely consistent with those of lensed DSFGs identified by the likes of {\it Herschel}, SPT, and ACT \citep{Bussmann:2012aa, Bussmann:2013aa, Bussmann:2015aa, Hezaveh:2013aa, Dye:2014aa, Dye:2018aa, Spilker:2016aa, Enia:2018aa, Rivera:2019aa, Liu:2022ad} and other {\it Planck}-selected lenses not in PASSAGES \citep{Canameras:2015aa, Canameras:2018ab}; see Fig.~\ref{fig:magnifications}.

\item In characterizing their de-lensed, spatial extents in the source plane at rest-frame far-IR and radio,
we find that the PASSAGES DSFGs are generally more extended than typical DSFGs (both lensed and unlensed). Since they appear to be consistent with size-luminosity scaling relations for unlensed submm-bright objects \citep{Fujimoto:2017aa}, we interpret their larger sizes to be a consequence of their higher star formation rates (or vice versa); see Figs.~\ref{fig:LIR_Reff_lensed} and \ref{fig:LIR_Reff}.
They are consistent with a framework of some DSFGs being the result of active growing of and accretion of gas onto a star-forming disk,
with a wider distribution of star formation than low-$z$ ULIRGs, which form stars in compact nuclear regions.

\item We compute the radio-FIR correlation parameter $q_{\rm FIR}$ by assuming a single radio spectral index; the range of values for $q_{\rm FIR}$ is in line with those from \citet{Giulietti:2022aa} for {\it Herschel}-selected lenses, but we find no clear evolution with redshift, as predicted by some recent works. We also find that PASSAGES objects do not show signs of significant contributions from AGN, which might be indicated by a radio excess, $q_{\rm FIR} < 1.8$.

\item In comparing their radio vs. sub-mm sizes, we find some weak dependence on $L_{\rm IR}$, $\Sigma_{\rm SFR}$, and the radio-FIR correlation parameter $q_{\rm FIR}$, with the more luminous and more densely star-forming galaxies having preferentially more compact 1 mm sizes (or more extended 6 GHz halos). Relative to $q_{\rm FIR}$, we find that galaxies closer to being AGN-powered (i.e. showing a radio excess) had generally smaller radio sizes, which might be indicating that deeply-buried AGN are leading to more compact half-light radii. At present, our conclusions are limited by the small number of 10 objects with both radio and FIR sizes measured.

\item The larger sizes of PASSAGES galaxies (relative to the broader population of DSFGs) may also owe to their concentration around Cosmic Noon, $z<3.5$, in contrast to higher-$z$ DSFGs like \citet{Spilker:2016aa}. Nonetheless, we observe no clear trend in size as a function of redshift for the narrow range covered by these PASSAGES objects (Fig.~\ref{fig:z_Reff}).
This agrees with recent results by \citet{Wang:2022ac} for the FIR size of stacked main-sequence galaxies and \citet{Jimenez-Andrade:2019aa} for the radio size of star-forming galaxies, which evolves only shallowly with redshift.

\item There may be some size bias present in the sample (Fig.~\ref{fig:size_bias}), by which more compact objects are capable of producing higher lensing magnifications \citep{Hezaveh:2012aa,Spilker:2016aa}. 
We parameterize this effect by comparing magnifications with the ratio of the angular size of background objects to the Einstein radius of the foreground, and find a modest downward trend as expected (Fig.~\ref{fig:size_ratio_mag}), which is in line with predictions for isothermal lenses (e.g. \citealt{de-Freitas:2018aa}).
That the PASSAGES sample is not fully coincident with the distribution of lensing cross-sections $\sigma_\mu$ as a function of the size ratio and magnification is an indication of the selection function at play, as we are not uniformly sensitive to all magnifications.
Given the selection of these objects by flux, they are likely of an optimal size that is simultaneously sufficiently large to host high intrinsic luminosities and sufficiently small to align with higher-magnification regions of the source plane.

\item Since the spatial extents of their star-forming regions are proportionally larger to match their extreme luminosities, the star-formation surface densities for PASSAGES are significantly sub-Eddington on global scales (in agreement with most DSFGs for which this measurement has been made). 
There is still certainly a possibility that their rapid stellar mass assembly is substantially regulated by radiation pressure, but this effect must be on sub-kpc scales not captured by these current $0.4\arcsec$ (image-plane) resolution ALMA images, perhaps over short timescales that are not contemporaneous between sites of star formation. 
For these observations, ALMA does not significantly resolve the far-IR continuum, and so the peak values of $\Sigma_{\rm SFR}$ in each object are not much larger than the respective globally-averaged values.
In the future, we hope to repeat this analysis with higher angular resolution from ALMA to properly assess the densities of star formation, which we expect may be clumpy and extended beyond just nuclear regions.

\end{itemize}

%% Also note that the akcnowlodgment environment does not support long amounts of text. If you have a lot of people and institutions to acknowledge, do not use this command. Instead, create a new 
\section{Acknowledgments}.
%\begin{acknowledgments}

%\section*{Acknowledgements}

%The Acknowledgements section is not numbered. Here you can thank helpful
%colleagues, acknowledge funding agencies, telescopes and facilities used etc.

PSK would like to thank Bel\'{e}n Alcalde Pampliega and Lilah Mercadante for helpful discussions related to this work, and the scientific staff at the NRAO Array Operations Center for assistance with data reduction.
PSK gratefully acknowledges support from the NRAO Student Observing Support (SOS) award SOSPA7-019 and the Massachusetts Space Grant Consortium.
%
%
%
% HST
This research is based on observations made with the NASA/ESA {\it Hubble Space Telescope} obtained from the Space Telescope Science Institute, which is operated by the Association of Universities for Research in Astronomy, Inc., under NASA contract NAS 5\textemdash 26555. These observations are associated with programs GO-14653 and GO-14223. Support for program GO-14653 was provided by NASA through a grant from the Space Telescope Science Institute.
%
% MAST
Some of the data presented in this paper were obtained from the Multimission Archive at the Space Telescope Science Institute (MAST).
%
% ALMA
This paper makes use of the following ALMA data: ADS/JAO.ALMA 
\#2017.1.01214.S,
\#2015.1.01518.S,
\#2019.1.01636.S. 
%{\red Others?} 
ALMA is a partnership of ESO (representing its member states), NSF (USA) and NINS (Japan), together with NRC (Canada), MOST and ASIAA (Taiwan), and KASI (Republic of Korea), in cooperation with the Republic of Chile. The Joint ALMA Observatory is operated by ESO, AUI/NRAO and NAOJ.
%
% NRAO/JVLA
The National Radio Astronomy Observatory is a facility of the National Science Foundation operated under cooperative agreement by Associated Universities, Inc.
%
% Gemini
This work is based in part on observations obtained at the international Gemini Observatory, a program of NSF's NOIRLab (acquired through the Gemini Observatory Archive and processed using the Gemini IRAF package), which is managed by the Association of Universities for Research in Astronomy (AURA) under a cooperative agreement with the National Science Foundation on behalf of the Gemini Observatory partnership: the National Science Foundation (United States), National Research Council (Canada), Agencia Nacional de Investigaci\'{o}n y Desarrollo (Chile), Ministerio de Ciencia, Tecnolog\'{i}a e Innovaci\'{o}n (Argentina), Minist\'{e}rio da Ci\^{e}ncia, Tecnologia, Inova\c{c}\~{o}es e Comunica\c{c}\~{o}es (Brazil), and Korea Astronomy and Space Science Institute (Republic of Korea).
%
% ADS
This research has made use of NASA's Astrophysics Data System,
%
% APLpy
%This research made use 
of APLpy, an open-source plotting package for Python \citep{Robitaille:2012aa}, 
%Astropy
and of Astropy, a community-developed core Python package for Astronomy \citep{Astropy-Collaboration:2013aa}.
%
% IRAF?

%\end{acknowledgments}

\vspace{5mm}
\facilities{HST(WFC3), LMT, ALMA, JVLA, Gemini}


\software{astropy \citep{Astropy-Collaboration:2013aa,Astropy-Collaboration:2018aa},  
	\lenstool \citep{Kneib:1993aa, Kneib:1996aa,Jullo:2007aa, Jullo:2009aa},
	APLpy \citep{Robitaille:2012aa}, 
	IRAF \citep{Tody:1986aa, Tody:1993aa, Science-Software-Branch-at-STScI:2012aa},
	CASA \citep{McMullin:2007aa},
	Ned Wright's Cosmology Calculator \citep{Wright:2006aa}
          }

%%%%%%%%%%%%%%%%% APPENDICES %%%%%%%%%%%%%%%%%%%%%

\appendix

\section{Notes on individual objects}
\label{sec:object_notes}

Section 5 of \citet{Harrington:2016aa} and Appendix B of \citet{Berman:2022aa} provide notes on each member of the larger PASSAGES sample (in addition to
%\footnote{See also 
Table 1 of \citealt{Harrington:2021aa}).
%}. 
In this section, we review the information necessary for our lens models, including spectroscopic redshifts of the background DSFGs, photometric/spectroscopic redshifts of foreground objects, and multiple image morphologies, summarized in Table \ref{tab:sample}.

{\bf PJ011646.} This galaxy-scale lens system appears to consist of multiple background components lensed by an elliptical galaxy with a 
spectroscopic redshift of $z_{\rm spec} = 0.555$ determined with VLT MUSE observations (Kamieneski et al., in prep.).
There is a smaller, secondary foreground galaxy of unknown redshift to the northeast, but it has minimal impact on the lensing morphology and so we exclude it from our model. The foreground elliptical is modeled with our standard approach: an SIE profile with centroid, ellipticity, orientation, and velocity dispersion all kept as free parameters. The DSFG, with a CO(3-2) detection at $z = 2.125$ by LMT/RSR \citep{Berman:2022aa}, is detected as two primary arcs in ALMA and JVLA imaging and appears to spatially coincide with a bright double image detected by \HST\ (labeled $2ab$ in Figure \ref{fig:postage_stamps}). Another image family (labeled $1abcd$) is a clear quad image, nearly spatially coincident with family $2ab$ but noticeably bluer in the RGB image with Gemini $r'$ and $z'$ (Fig.~\ref{fig:postage_stamps}). It is not entirely clear if this image family is located at the same redshift as $2ab$, but since the two families have very similar Einstein radii 
%({\red be specific?})
and appear to be lensed by virtually the same foreground mass profile, then this should dictate that the background objects lie at similar redshifts. Within the Einstein radius uncertainty for this object (see discussion in Section \ref{sec:Einstein_radius}), $\theta_{\rm Ein,eq} = 2.34 - 2.41\arcsec$, and holding other quantities fixed, this corresponds to a redshift uncertainty of $z\approx 2.0 - 2.4$ (assuming $z_{\rm lens} = 0.555$). 
For the purposes of deriving our lens models, it is sufficient to assume that they lie at the same redshift. Given the proximity of this quad component and the doubly-imaged DSFG component in the source plane ($\lesssim 1\arcsec$, 8.5 kpc at the CO-derived redshift), it seems most likely that these two components may be interacting (or else a very serendipitous alignment of two objects around $z\sim2$).
We also note that these multiple components are aligned with the foreground in such a way as to cover most of the caustic curve region so that there is a very complex set of concentric, partial Einstein ring structures in the image plane, offering excellent constraints on the mass of the foreground elliptical, $(1.54 \pm 0.03) \times 10^{12} M_\odot$ within the Einstein radius at $z_{\rm lens}=0.555$ (see Table~\ref{tab:lensproperties}).

{\bf PJ014341.} 
This compact galaxy-scale lens is one of the lowest-redshift members of our sample, with a CO(2-1) detection from the background DSFG at $z=1.096$. The lensing foreground galaxy is detected spectroscopically at $z_{\rm spec}=0.594$ from SDSS \citep{Berman:2022aa}. Its small Einstein radius of $\approx 0.5\arcsec$ (on the lower end of our sample, see Table \ref{tab:lensproperties}) is due primarily to the small separation in redshift space between the lens and source planes. The enclosed foreground mass is $(1.3 \pm 0.3) \times 10^{11} M_\odot$. The redshift geometry also has the consequence of making lensed image identification difficult, as they are likely heavily blended with foreground light in the optical (Fig.~\ref{fig:postage_stamps}), and only marginally resolved beyond the beam size in the interferometric ALMA/JVLA images. 
Nonetheless, the image morphology is consistent with a classic 4-image fold-caustic geometry, with two isolated images and an additional merging pair (Fig.~\ref{fig:model_SP}).
The radio morphology is similar
(Fig.~\ref{fig:model_SP})
but more difficult to interpret due to its lower signal-to-noise, so it is not included as a constraint in this iteration of the model.
Given the compactness of the lensing configuration, the source-plane area subtended by the caustic curve is likely smaller than or similar in size to the DSFG, so slight offsets between radio and far-IR emitting regions can result in very different image morphologies. 
There appear to be other foreground objects nearby, but given the compactness of the lensing configuration, we only parameterize the primary deflector with our standard SIE model.
The best-fit model finds a high ellipticity, $e=0.77^{+0.15}_{-0.21}$, which suggests these other foreground deflectors may contribute non-trivially. A refined model with more available constraints in the future might allow for a more complex mass distribution.

{\bf PJ020941.}
Also known as {\it 9io9} or the ``Red Radio Ring", PJ020941 has been studied at length by 
\citet{Geach:2015aa, Geach:2018aa},
\citet{Harrington:2016aa, Harrington:2019ab},
\citet{Rivera:2019aa}, 
\citet{Doherty:2020aa},
and
\citet{Liu:2022ad}.
A comparison of our lens model with existing models is provided in Appendix \ref{sec:lens_model_comparison}.
The target was first revealed by the gravitational lens citizen science project {\sc Space Warps} \citep{Geach:2015aa, Marshall:2016aa, More:2016ab} before it was identified independently as a strong sub-millimeter source in surveys by \Planck\ \citep{Harrington:2016aa}, the Atacama Cosmology Telescope \citep{Su:2017aa, Gralla:2020aa}, and \Herschel\ \citep{Viero:2014aa}. Near-infrared $i$-band and $J/K_s$-band imaging from the VISTA-CFHT Stripe 82 survey \citep{Geach:2017ab} and CFHT-MegaCam Stripe 82 survey \citep{Moraes:2014aa} revealed a nearly complete, red Einstein ring \citep{Geach:2015aa}, confirmed with high-resolution $H$-band \HST\ imaging (\citealt{Geach:2018aa}, Lowenthal et al. in prep.).
The 1.4 GHz eMERLIN and 5 GHz JVLA radio images \citep{Geach:2015aa} show a partial Einstein ring that largely coincides spatially with the near-IR emission, which is also borne out by ALMA continuum and spectral line images (\citealt{Geach:2018aa, Doherty:2020aa, Berman:2022aa}; and this work).

PJ020941 has a CO(3-2) detection by LMT/RSR at $z=2.553$ \citep{Harrington:2016aa, Harrington:2021aa}. It is strongly lensed by an elliptical galaxy (SDSS J020941.27+001558.5) at $z_{\rm fg} = 0.202$, and by a secondary galaxy assumed to lie at a similar redshift (which is expected to have a non-negligible influence on the lens, \citealt{Rivera:2019aa}).
Both are included in our model, with a standard SIE approach for the primary elliptical, and an SIS model (with position held fixed to the centroid of the luminous component) for the secondary object.

The overall lens morphology consists of a double image, with an extended arc to the west and a less-magnified counter-image to the northeast. 
Recent Cycle 7 ALMA $\theta \sim 0.15\arcsec$ Band 6 (1.1 mm) continuum imaging (2019.1.01197.S, PI: P. Kamieneski; to be presented in forthcoming work by Kamieneski et al. in prep.) provides further information.
Embedded inside the extended arc are 3 distinct quad-imaged regions, suggestive of a cusp-caustic configuration, which can be identified by matching clumps and using parity information.
This is a commonly-observed lensing configuration where some portion of a galaxy crosses into the inner region of the caustic, which is readily apparent from the morphology in Fig.~\ref{fig:model_SP}.
This quadruply-imaged portion is also evident from subtle surface brightness peaks in the {\it HST} image (Fig.~\ref{fig:postage_stamps}) that coincide with the 1 mm peaks.
%
%}

{\bf PJ022633.} This partial Einstein ring consists of a background object lensed primarily by a group of three foreground objects, assumed to lie at the same redshift.
The DSFG CO(3-2) detection is at $z=3.120$ \citep{Berman:2022aa}, with the brightest lensing galaxy at $z=0.41$ \citep{Wen:2015aa}.
Obvious lensing features include a bright, extended partial Einstein ring to the southeast (images $1bc$) and a counter-image to the northeast (image $1a$), with faint emission connecting the two (more apparent in the optical image). 
There is some evidence that the lensed image expected to be located on the western side of the primary deflector (opposite the semi-ring) is split into two images by the secondary foreground galaxy to the northwest, visible faintly in $H$-band and at 6 GHz (images $1de$, Figs.~\ref{fig:postage_stamps} and \ref{fig:model_SP}). The radio images are nearly spatially coincident with an extended double-lobed radio jet emanating from the primary foreground elliptical, and the {\it HST} image is affected by a bright star nearby, so confirmation of this proposed morphology is still pending. 
However, this interpretation is motivated by the image predictions of a simple zeroth-order lens model including only the mass of the primary and secondary ellipticals. 
The primary is modeled as an SIE profile, with the secondary as an SIS model, with its position allowed to vary to a small amount.

{\bf PJ030510.} This object is a galaxy-scale (or small group-scale) lens, with a CO(3-2) detection of the DSFG at $z=2.263$ \citep{Harrington:2021aa, Berman:2022aa}. 
A preliminary red cluster sequence analysis \citep{Gladders:2000aa} yields a foreground redshift of $z_{\rm phot} = 0.4 - 0.5$. 
Near-IR imaging by \HST\ and Gemini reveal ring-like features surrounding two foreground galaxies in a dense environment. Our model parameterizes the combined mass contribution from both as an SIE model. ALMA 1 mm continuum imaging has a slightly larger beam size of $0.85\times0.59\arcsec$, which leaves the DSFG only marginally resolved. It appears to consist of two lensed images, one of which is notably fainter, on opposite sides of the foreground pair. 
It is possible that higher-resolution imaging might reveal more of a cusp-like quadruple-image morphology, but at present, a double-image configuration is the best explanation.

{\bf PJ105353.} This object was discovered independently by \citealt{Harrington:2016aa} and \citealt{Canameras:2015aa} (also referred to as ``the Ruby" or PLCK\_G244.8+54.9),
and has also been studied by \citet{Canameras:2017aa, Canameras:2017ab, Canameras:2021aa}. 
We compare our modeling results with theirs in Appendix \ref{sec:lens_model_comparison}.
The lensing galaxy is at a higher redshift than most other PASSAGES objects, $z=1.525$, with a source redshift of $z=3.005$. 
It is modeled as an SIE profile, with position, ellipticity, orientation, and velocity dispersion as free parameters.
As we discuss in Appendix \ref{sec:lens_model_comparison}, CO(4-3) imaging presented in \citet{Canameras:2017aa} reveals a complicated image-plane structure, dominated by an apparent quadruply-imaged component 
with no clear counterpart in rest-frame optical (see also \citealt{Canameras:2017aa,Canameras:2017ab}), likely due to blending with the foreground as suggested by \citet{Frye:2019aa}.
High-resolution ($0.07\arcsec$) 1 mm imaging with ALMA (PID 2015.1.01518.S, PI: N. Nesvadba), which was not directly included in the model by \citet{Canameras:2017aa}, leads us to a slightly different conclusion on the quad-image configuration.
For this work, we utilize the Band 7 ($870\mu$m) continuum image to derive the dust magnification, as the observing setup (resolution $\gtrsim 0.15\arcsec$) is closer to that of the other objects.

{\bf PJ112713.} This object has not been imaged by AzTEC or ALMA, but a CO(2-1) line was detected by RSR at $z=1.303$ \citep{Harrington:2021aa, Berman:2022aa}. A photometric redshift for the lens was derived to be $z_{\rm phot} = 0.42$ (Lowenthal et al., in prep.) using SDSS. JVLA imaging indicates two arcs ($1ab$) separated by $\sim1\arcsec$, assumed to be the lensed DSFG, as there are no other bright radio sources within 1$\arcmin$. There is also indication of an incomplete Einstein ring in HST imaging, surrounding a single foreground object, with an Einstein radius that approximately matches the 6 GHz arcs, $\theta_{\rm Ein} = 0.6\arcsec$.
The foreground galaxy is modeled with the standard SIE approach of this work.
Intriguingly, there are two bright peaks in the optical Einstein ring (denoted $2ab$) that roughly correspond to minima in the partial radio ring. We interpret this as possibly the result of radio continuum tracing sites of heavily dust-obscured active star formation, whereas the optical peaks arise from less-extincted sightlines. Both $1ab$ and $2ab$ are utilized as constraints for our lens model.

{\bf PJ113805.} A single CO(2-1) line was detected by RSR and interpreted to lie at $z=2.019$ \citep{Harrington:2021aa, Berman:2022aa}. ALMA imaging reveals a compact galaxy-scale lens that is not resolved into multiple components (due in part to a lower-resolution synthesized beam, $1.03 \times 0.66 \arcsec$). 
The system is however
resolved into two primary components by JVLA 6 GHz imaging (Fig.~\ref{fig:model_SP}). These double images appear to nearly coincide with faint arcs detected by \HST\ ($1ab$), although the latter are heavily contaminated by light from the foreground lensing elliptical, which has a photometric redshift from SDSS of $z_{\rm phot} = 0.52$. Since the radio peaks are slightly offset from the optical images, we add them as independent constraints ($2ab$), although they are likely to provide only minimal new information on the foreground mass.
Given the sparse number of model constraints, this object is modeled only as an SIS profile, with position and velocity dispersion as free parameters.

{\bf PJ113921.} This object was discovered independently by \citet{Canameras:2015aa}, identified as PLCK\_G231.3+72.2, and later reported in \citet{Berman:2022aa}. 
Our model-derived magnification is compared with that of \citep{Canameras:2018ab} in Appendix \ref{sec:lens_model_comparison}.
The RSR CO(3-2) line is interpreted from photometric support to be at $z=2.858$, in agreement with additional spectroscopic coverage \citep{Canameras:2018ab, Nesvadba:2019aa, Harrington:2021aa}. ALMA partially resolves what appears to be a quad-image in 260 GHz continuum ($1abcd$), centered around a foreground elliptical with a photometric redshift of $z=0.57$ \citep{Berman:2022aa},
but likely perturbed by at least one other nearby galaxy. Our model here includes the primary lens as an SIE, and the deflecting galaxy to the west as a simple SIS profile. A faint red feature in \HST\ is possibly affiliated with the background DSFG, as it is close to the ALMA continuum emission, or it may be associated with an apparent foreground spiral galaxy to the southeast. In this iteration of the model, we do not include this feature in constraining our lens model. 

{\bf PJ132630.} This object was discovered independently through our sample \citep{Berman:2022aa} and in the H-ATLAS survey \citep{Bussmann:2013aa,Yang:2017aa}, known also as NAv1.195. A single CO(3--2) line was detected with RSR at $z=2.951$ (confirmed with spectroscopic follow-up by \citealt{Yang:2017aa} and \citealt{Harrington:2021aa}).
\citealt{Bussmann:2013aa} published a lens model for 340 GHz imaging by the SMA ($\mu_{\rm 340 GHz} = 4.1\pm0.3$), with a lens redshift of $z_{\rm fg} = 0.786$. 
We compare their results with our model in Appendix \ref{sec:lens_model_comparison}.
ALMA and JVLA imaging reveals that the DSFG is doubly-imaged, with images separated by 3.7$\arcsec$ in the image plane ($3ab$). An optical counterpart appears to separate into two components that surround the long-wavelength emission; we label these image families $1ab$ and $2ab$ and include them separately in the model. There is evidence for additional background sources lensed by the foreground elliptical, which we model using a standard SIE, but they are at an unknown redshift and are thus not considered as constraints.

{\bf PJ133634.} Two CO lines ($J=3-2$ and $J=4-3$) were detected by RSR at $z=3.254$ \citep{Harrington:2021aa, Berman:2022aa}. A pair of primary lensing galaxies separated by 0.9$\arcsec$ have an SDSS photometric redshift of $z=0.26$. The 6 GHz continuum reveals a clumpy, partial Einstein ring, centered $\approx 0.2\arcsec$ east of the center of the foreground elliptical (likely due to a small perturbation by the secondary). This object is too northern to be visible to ALMA, but the 1.1 mm continuum from AzTEC (8.5$\arcsec$ beam) coincides with the radio continuum, which we thus interpret to originate from the \Planck\ DSFG.
The radio ring consists of multiple clumps larger than the synthesized beam, and we identify 3 doubly-imaged families in the eastern counter-image and western semi-ring to be used in the model ($1ab$, $2ab$, and $3ab$), making use of the expected parity flip between the counter-image and more extended arc.
The foreground mass is modeled as a single SIE profile, as the effect of the secondary perturber is not large enough to break the degeneracy with the primary lens.

{\bf PJ144653.} This lower-redshift member of our sample has a CO(2-1) line detected by RSR at $z=1.084$ \citep{Harrington:2021aa, Berman:2022aa}. The lensing galaxy has an SDSS spectroscopic redshift of $z=0.493$, and has the appearance of a face-on spiral galaxy. However, some of the spiral-like structure may actually be the result of the multiply-imaged background DSFG. At present, it is not feasible to make this distinction without imaging in an additional filter with comparable quality to the \HST\ image. ALMA and JVLA images are consistent with each other and reveal the DSFG to follow a fold-caustic lensing configuration, with two isolated images and an additional merging image pair towards the southeast (labeled $1abc$). The images form a partial Einstein ring with a radius of $\approx 0.9\arcsec$.
The foreground is modeled with a single SIE profile.
The optimized model has a high ellipticity, $e = 0.65^{+0.18}_{-0.25}$, possibly because of the small number of available constraints currently available, which necessitates a perhaps too-simplistic model.

{\bf PJ144958.} This cluster lens has a CO(3-2) line detection at $z=2.153$ \citep{Harrington:2021aa, Berman:2022aa} with a preliminary foreground cluster redshift measured at $z_{\rm phot} =0.4$. Optical emission probed by \HST\ and Gemini appears to be more extended than at 1 mm and 6 GHz, which reveal a set of more point-source-like images. However, there appears at first to be four distinct images all on one side of a foreground galaxy group, with no obvious counter-image. 
We consider it likely that the middle pair of images ($1bc$) is in fact a merging pair on either side of the critical curve (see Fig.~\ref{fig:model_SP}). The northernmost image $1d$ also appears to be split into four images by another intervening cluster member ($4abc$), visible only with \HST's higher resolution and unresolved by ALMA and JVLA. 
As constraints on our model, we include images $1abcd$ (JVLA/ALMA locations), $2abc$ for the peaks of the HST images, and $3abc$ as the fainter tail visible only in optical (adjacent to $2abc$). 
Lastly, we include images $4abc$ for the point-like sources surrounding the perturbing foreground. 
The lensing distribution is parameterized by a cluster-scale NFW 
profile, 3 smaller SIS profiles describing the group of elliptical galaxies interior to the arc, and a fourth SIS profile for the perturbing galaxy at the northern end of the arc.

{\bf PJ160722.} This member of our sample has a CO(2-1) line observed by RSR at $z=1.482$, confirmed with photometric information and subsequent spectroscopic follow-up \citep{Harrington:2016aa, Harrington:2021aa}. \HST\ imaging reveals two apparent foreground lensing objects with a photometric redshift of $z=0.65$, determined from the optical to mid-IR SED in \citep{Harrington:2016aa}. A bright radio AGN is detected in the northern foreground (Fig.~\ref{fig:model_SP}). Both near-IR and radio reveal extended ring-like arcs tightly surrounding the foreground.
Four peaks embedded with the surrounding ring are apparent in the radio and, to a lesser extent, in the optical image, which we interpret as a quadruply-imaged region of the source plane.
The lensing mass distribution is represented by a single SIE model, given the proximity of the two foregrounds (and therefore strong degeneracies in optimal parameters between the two). 

{\bf PJ231356.} This object has a CO(3-2) line observed by RSR at $z=2.215$, confirmed by spectroscopic follow-up \citep{Harrington:2021aa, Berman:2022aa}. \HST/WFC3 1.6\micron\ imaging appears to show a primary lensing galaxy, with a secondary galaxy to the north that appears to have a substantial impact on the lensing morphology, leading to a surprisingly straight arc towards the north. 
Numerous background arcs are visible in \HST, some of which coincide with the arcs in ALMA 260 GHz continuum. 
Most of these multiple images have not been identified definitively, but the quadruple-image family $1abcd$ is a sufficient constraint for the model at present.
The lensing mass distribution is parameterized by the primary elliptical (SIE) and the secondary perturber to the north (SIS). 
While the $x$-position of the primary galaxy is left as a free parameter, we hold fixed its $y$-position to avoid degeneracy with the secondary lens.

\section{Lensed multiple image systems used as constraints}
\label{sec:constraints_table}

In \S \ref{sec:family_id}, we discuss our approach to identifying multiply-imaged features of the background DSFGs using \HST, Gemini, JVLA, and ALMA observations. In Table \ref{tab:lensconstraints}, we summarize these features that were used as constraints for our models, which are also marked on Fig.~\ref{fig:postage_stamps}.

\begin{figure*}
\centering
\begin{minipage}{0.48\textwidth}
\startlongtable
\begin{deluxetable}{ccccc}
	%\tiny
	%\centering
	\tablecolumns{5}
	\tabletypesize{\scriptsize}
	\tablecaption{
	Multiple image positions used as constraints in the lensing models (presented in Table \ref{tab:lensmodels}), as per \S \ref{sec:family_id}. In some rare cases, the location and orientation of caustic curves as inferred by lensing morphologies are used as constraints.
	\label{tab:lensconstraints}
	}
	%\begin{threeparttable}
	%\begin{tabular}{cccccc} % four columns, alignment for each
		%\hline
		%\hline
	\tablehead{
		\colhead{{\bf Field}} & 
		\colhead{Image \#} & 
		\colhead{RA} & 
		\colhead{Dec.} &   
		\colhead{$z$}  \\[-1ex]
		%\colhead{PA} \\[1ex]
		& & \colhead{[deg.]} & \colhead{[deg.]} &  \\[-3ex]
		%\colhead{[\degrees]}
	}
		%\hline
		%\hline\\[-2ex]
	\startdata
		{\bf PJ011646} 	& 1a &19.195512 & -24.617222 & $2.125$  \\
					& 1b &19.194428 & -24.617920	 & \textemdash  \\[1ex]
					& 2a &19.195611 & -24.617580	 & $2.125^a$ \\
		                        	& 2b &19.195158 & -24.616568 & \textemdash \\
		                        	& 2c &19.194320   & -24.617293 & \textemdash  \\		
		                        	& 2d &19.194704 & -24.617876 & \textemdash \\[0.5ex]				                                 
		\hline\\[-2ex]
		{\bf PJ014341} 	& 1a & 25.921762 & -1.790685 & $1.096$ \\
					& 1b & 25.921923 & -1.7905189 & \textemdash \\
					& 1c & 25.921752 & -1.7904362 & \textemdash \\
					& 1d & 25.921651 & -1.7905364 & \textemdash \\[0.5ex]
		\hline\\[-2ex]
		{\bf PJ020941} 	& 1a &32.422486 & 0.26643019 & 2.554 \\
					& 1b &32.421489 & 0.26561695& \textemdash \\[1ex]
					& 2a &32.422507 & 0.26633282 & \textemdash \\
					& 2b &32.421262 & 0.26661786& \textemdash \\[1ex]
					& 3a &32.422500 & 0.26639240& \textemdash \\
					& 3b &32.421370 & 0.26574041& \textemdash \\
					& 3c &32.421216 & 0.26607387& \textemdash \\
					& 3d &32.421181 & 0.26628788& \textemdash \\[1ex]
					& 4a &32.422470 & 0.26642723& \textemdash \\
					& 4b &32.421480 & 0.26561101& \textemdash \\[0.5ex]
		\hline\\[-2ex]
		{\bf PJ022633} 	& 1a & 36.642144 & 23.758811 & $3.120$ \\
					& 1b & 36.642601 & 23.757341 & \textemdash \\
					& 1c & 36.642373 & 23.757083 & \textemdash \\
					& 1d & 36.640842 & 23.75822 & \textemdash \\
					& 1e & 36.640552 & 23.758414 & \textemdash \\[0.5ex]
		\hline\\[-2ex]
		{\bf PJ030510} 	& 1a &46.294327 & -30.608571 & $2.263$ \\
					& 1b &46.294248 & -30.608259	 & \textemdash \\[1ex]
		  (PA $=$ 114\degrees) & critic1 &46.294451 & -30.608419	 & \textemdash  \\
		  (PA $=$ 49\degrees) & critic2 &46.294136 & -30.608439	 & \textemdash  \\[0.5ex]
		\hline\\[-2ex]
		{\bf PJ105353} 	& 1a &163.47142 & 5.9387914 & 3.005\\
					& 1b &163.47121 & 5.9385352	 & \textemdash  \\
					& 1c &163.47158 & 5.9386185	 & \textemdash  \\
					& 1d &163.47144 & 5.9383833	 & \textemdash  \\[1ex]
					& 2a &163.47152 & 5.9386507	 & \textemdash \\
					& 2b &163.47128 & 5.9383962	 & \textemdash \\[0.5ex]
		\hline\\[-2ex]
		{\bf PJ112713} 	& 1a &171.80585 & 46.156730 & 1.303 \\	
					& 1b &171.80619 & 46.156579	 & \textemdash \\[1ex]
					& 2a &171.80595 & 46.156529 & \textemdash \\	
					& 2b &171.80630 & 46.156730	& \textemdash \\[1ex]
					& 3a &171.80619 & 46.156825 & \textemdash \\	
					& 3b &171.80596 & 46.156528	 & \textemdash \\[0.5ex]
		\enddata
	%\multicolumn{5}{p{0.8\columnwidth}}{
	%\end{threeparttable}
\end{deluxetable}
\end{minipage}
\begin{minipage}{0.48\textwidth}
%
%\vspace{0pt}
%
\addtocounter{table}{-1}
\startlongtable
\begin{deluxetable}{ccccc}
	%\tiny
	%\centering
	\tablecolumns{5}
	\tabletypesize{\scriptsize}
	\tablecaption{
	(continued)
	\label{tab:lensconstraints}
	}
	%\begin{threeparttable}
	%\begin{tabular}{cccccc} % four columns, alignment for each
		%\hline
		%\hline
	\tablehead{
		\colhead{{\bf Field}} & 
		\colhead{Image \#} & 
		\colhead{RA} & 
		\colhead{Dec.} &   
		\colhead{$z$}  \\[-1ex]
		%\colhead{PA} \\[1ex]
		& & \colhead{[deg.]} & \colhead{[deg.]} &  \\[-3ex]
		%\colhead{[\degrees]}
	}
		%\hline
		%\hline\\[-2ex]
	\startdata
		{\bf PJ113805} 	& 1a &174.52314 & 32.965895 & 2.019 \\	
					& 1b &174.523     & 32.965711	& \textemdash \\[1ex]
					& 2a &174.52318 & 32.965891	& \textemdash \\
					& 2b &174.52302 & 32.96573 	& \textemdash \\[0.5ex]
		\hline\\[-2ex]
		{\bf PJ113921} 	& 1a &174.84046 & 20.414097 & 2.858 \\
					& 1b &174.84068 & 20.414272 & \textemdash \\
					& 1c &174.84023 & 20.414594 & \textemdash \\
					& 1d &174.84063 & 20.414802 & \textemdash \\[0.5ex]
		\hline\\[-2ex]
		{\bf PJ132630} 	& 1a &201.62551 & 33.735859 & 2.951 \\
					& 1b &201.62637 & 33.735124 & \textemdash \\[1ex]
					& 2a &201.62552 & 33.735968 & \textemdash \\
					& 2b &201.62625 & 33.735165 & \textemdash \\[1ex]
					& 3a &201.62549 & 33.735844 & \textemdash \\
					& 3b &201.62636 & 33.735133 & \textemdash \\[0.5ex]
		\hline\\[-2ex]
		{\bf PJ133634} 	& 1a &204.14616 & 49.220425 & 3.254 \\
					& 1b &204.14534 & 49.220706	 & \textemdash \\[1ex]
					& 2a &204.14607 & 49.220635 & \textemdash \\
					& 2b &204.14515 & 49.220373	 & \textemdash \\[1ex]
					& 3a &204.14586 & 49.220788 & \textemdash \\
					& 3b &204.14545 & 49.220180	 & \textemdash \\[0.5ex]
		\hline\\[-2ex]
		{\bf PJ144653} 	& 1a &221.72176 & 17.876174 & $1.084$ \\
					& 1b &221.72135 & 17.875992	 & \textemdash \\[1ex]
		 (PA $=$ 35\degrees) & critic1 &221.72186 & 17.875843	 & \textemdash  \\[0.5ex]
		\hline\\[-2ex]
		{\bf PJ144958} 	& 1a & 222.49332 & 22.642215 & $2.153$ \\
					& 1b & 222.49347 & 22.643774	 & \textemdash \\
					& 1c & 222.49374 & 22.644154 & \textemdash \\
					& 1d & 222.49416 & 22.644439 & \textemdash \\[1ex]
					& {2a} & 222.49329 & 22.642269 & $2.153$ \\
					& {2b} &222.49346 & 22.643802 & \textemdash  \\
					& {2c} &222.49371 & 22.644152& \textemdash  \\[1ex]
					& 3a &222.49331 & 22.642477& $2.153$  \\
					& 3b &222.49333 & 22.643417& \textemdash \\
					& 3c &222.4939  & 22.644268& \textemdash \\[1ex]
					& 4a &222.49389 & 22.644384& $2.153$ \\
					& 4b &222.49402 & 22.644521& \textemdash  \\
					& 4c &222.49415 & 22.644548& \textemdash  \\[0.5ex]
		\hline\\[-2ex]
		{\bf PJ160722} 	& 1a &241.84517 & 73.783702 & 1.484  \\
					& 1b &241.84558 & 73.784165	 & \textemdash  \\
					& 1c &241.84418 & 73.784227	 & \textemdash  \\
					& 1d &241.84422 & 73.783829	 & \textemdash  \\[0.5ex]
		\hline\\[-2ex]
		{\bf PJ231356} 	& 1a &348.48594 & 1.1542249& 2.217  \\
					& 1b &348.48654 &1.1545733	 & \textemdash  \\
					& 1c &348.48615 &1.1550713	 & \textemdash  \\
					& 1d &348.48519 &1.1548946	 & \textemdash  \\[0.5ex]
		%\hline
		%\hline\\[-0.5ex]
	%\end{tabular}
	\enddata
	\vspace{12pt}
	%\multicolumn{5}{p{0.8\columnwidth}}{
	\tablecomments{
	$^a$ Assumed to be located at the same redshift as the DSFG.
	}
	%\tablecomments{
	%
	%\end{threeparttable}
\end{deluxetable}
\end{minipage}
\end{figure*}

\clearpage

\section{Best-fit lens model parameters}
\label{sec:best_fit_parameters}

In Table~\ref{tab:lensmodels}, we summarize the best-fit lens models for each object. In most cases, we list ``median," ``best," and ``mode" solutions, which represent the set of parameters consisting (respectively) of the median of the posterior distribution, the highest-likelihood solution recovered from the MCMC iterations, and the mode of the posterior distribution. The latter solution is usually employed when the posterior distribution is bimodal, especially in terms of position angle where there may be solutions offset by $90^\circ$.

\startlongtable	
\begin{deluxetable*}{ccccccc}
%\begin{longtable}{lcccccr} % four columns, alignment for each
	%\centering
	\tablecolumns{7}
	%\tabletypesize{\small}
	\tablecaption{
	Optimized gravitational lens model parameters.
	Unless otherwise noted, all potentials are parameterized with singular isothermal ellipsoid (or spheroid) models. For NFW potentials, scale radius in arcsec ($r_s$) replaces velocity dispersion as a parameter.
	While the median of the posterior distribution for each parameter is typically the preferable statistical estimator, multi-modal distributions and highly non-Gaussian distributions may be better summarized by the mode of the posterior or the highest-likelihood (``best") sample. 
	The first solution set listed (median, best, or mode) is taken to be the preferred model, but others are listed for completeness.
	\label{tab:lensmodels}
	}
	\tablehead{
		\colhead{{\bf Field}} & 
		&  
		&
		&
		&    
		& \\[-0.5ex]
		\colhead{{\it Model description}} & 
		\colhead{RA (deg.)} & 
		\colhead{Dec. (deg.)} &  
		\colhead{RMS$_{\rm im} (\arcsec)$} &  
		\colhead{$\sigma_{\rm position} (\arcsec)$} & 
		\colhead{$\nu$} & 
		\colhead{$\chi_\nu^2$} \\[-0.5ex]
		\colhead{potential $1, 2,...,n$} & 
		\colhead{$\Delta \alpha$ ($\arcsec$)} & 
		\colhead{$\Delta \delta$ ($\arcsec$)} & 
		\colhead{$e$} & 
		\colhead{$\theta$ (\degrees)} & 
		\colhead{$\sigma$ (km s$^{-1}$)} & 
		$z$ \\[-3ex]
	}
	\startdata
		%%%%%%%%
		%%%%%%%%
		% PJ011646
		% run 1 (but also run 8, revised z_fg)
		%%%%%%%%
		%%%%%%%%
		{\bf PJ011646} &&&&& \\[0.5ex]
		{\it Median} &19.194989 & -24.617376 & \textemdash & 0.10 & 3 & 2.0\\[0.5ex]
		potential 1 & $-0.12^{+0.03}_{-0.04}$ & $0.29^{+0.03}_{-0.03}$ & $0.37^{+0.04}_{-0.04}$ & $15.8^{+1.2}_{-1.2}$ & $364.0^{+3.0}_{-3.0}$ & $0.555^*$\\[0.5ex]
		\hline\\[-1ex]
		{\it Best} &\textemdash & \textemdash & 0.05 & \textemdash & 3 & 0.5\\[0.5ex]
		potential 1 & $-0.12$ & $0.30$ & $0.36$ & $15.7$ & $363.8$ & $0.555^*$\\
		\hline
		\hline\\
		%%%%%%%%
		%%%%%%%%
		% PJ014341
		% run 11
		%%%%%%%%
		%%%%%%%%
		{\bf PJ014341} &&&&&& \\[0.5ex]
		{\it Median} & 25.921794 & -1.7905377 & \textemdash & 0.17 & 3 & 1.7\\[0.5ex]
		potential 1 & $-0.06^{+0.08}_{-0.07}$ & $-0.03^{+0.06}_{-0.06}$ & {$0.77^{+0.15}_{-0.21}$} & $3.6^{+16.1}_{-11.2}$ & $209.7^{+11.4}_{-10.8}$ & $0.594^*$\\[0.5ex]
		\hline\\[-1ex]
		{\it Best} &\textemdash & \textemdash & 0.04 & \textemdash & 3 & 0.6\\[0.5ex]
		potential 1 & $-0.06$ & $-0.02$ & $0.55$ & $-1.7$ & $202.1$ & $0.594^*$\\
		\hline
		\hline\\
		%%%%%%%%
		%%%%%%%%
		% PJ020941
		% run 5
		%%%%%%%%
		%%%%%%%%
		{\bf PJ020941} &&&&&& \\[0.5ex]
		{\it Median} & 32.42192 & 0.26621843 & \textemdash & 0.15 & 6 & {0.8}\\[0.5ex]
		potential 1 & $-0.28^{+0.13}_{-0.12}$ & $0.12^{+0.04}_{-0.05}$ & $0.36^{+0.08}_{-0.08}$ & $-14.3^{+1.0}_{-0.9}$ & $312.0^{+5.9}_{-8.5}$ & $0.202^*$\\[0.5ex]
		potential 2 (SIS) & $0.22^*$ & $2.59^*$ & \textemdash & \textemdash & $82.6^{+42.2}_{-48.4}$ & $0.202^*$\\[0.5ex]
		\hline\\[-1ex]
		{\it Best} &\textemdash & \textemdash & 0.04 & \textemdash & 6 & 0.1\\[0.5ex]
		potential 1 & $-0.26$ & $0.12$ & $0.33$ & $-14.6$ & $314.7$ & $0.202^*$\\
		potential 2 (SIS) & $0.22^*$ & $2.59^*$ & \textemdash & \textemdash & $67.6$ & $0.202^*$\\
		\hline
		\hline\\
		%%%%%%%%
		%%%%%%%%
		% PJ022633
		% run 27
		%%%%%%%%
		%%%%%%%%
		{\bf PJ022633} &  &&&&& \\[0.5ex]
		%{\blue FINISHED}
		{\it Median} & 36.641654 & 23.757872 & \textemdash & 0.40 & 1 & 5.9\\[0.5ex]
		potential 1 & $-0.08^{+0.23}_{-0.26}$ & $-0.12^*$ & $0.25^{+0.11}_{-0.09}$ & $-9.0^{+6.8}_{-63.9}$ & $376.3^{+10.9}_{-11.5}$ & $0.41^*$\\[0.5ex]
		potential 2 (SIS) & $2.94^{+0.43}_{-0.33}$ & $2.08^{+0.32}_{-0.55}$ & \textemdash & \textemdash & $208.3^{+25.6}_{-26.3}$ & $0.41^*$\\[0.5ex]
		\hline\\[-1ex]
		{\it Best} & \textemdash & \textemdash & 0.10 & \textemdash & 1 & 0.3 \\[0.5ex]
		potential 1 & $-0.08$ & $-0.12^*$ & $0.10$ & $-13.6$ & $383.0$ & $0.41^*$\\
		potential 2 (SIS) & $3.17$ & $1.88$ & \textemdash & \textemdash & $170.7$ & $0.41^*$\\
		\hline
		\hline\\
		%%%%%%%%
		%%%%%%%%
		% PJ030510
		% run 25
		%%%%%%%%
		%%%%%%%%
		{\bf PJ030510} &  &&&&& \\[0.5ex]
		%{\red Not finished?}
		{\it Median} & 46.294291 & -30.608356 & \textemdash & 0.14 & 1 & 2.0  \\[0.5ex]
		potential 1 & $-0.04^{+0.13}_{-0.09}$ & $-0.48^{+0.19}_{-0.14}$ & $0.39^{+0.32}_{-0.28}$ & $3.5^{+56.9}_{45.1}$ & $169.6^{+13.6}_{-11.8}$ & $0.4^*$\\[0.5ex]
		\hline\\[-1ex]
		{\it Best} & \textemdash & \textemdash & $<0.01$ & \textemdash & 3 & 0.002 \\[0.5ex]
		potential 1 & $-0.01$ & $-0.39$ & $0.07$ & $-2.1$ & $165.0$ & $0.4^*$\\
		\hline
		\hline\\
		%%%%%%%%
		%%%%%%%%
		% PJ105353
		% run 16
		%%%%%%%%
		%%%%%%%%
		{\bf PJ105353} &  &&&&& \\[0.5ex]
		%{\blue FINISHED?}
		{\it Median} & 163.4714 & 5.9385728 & \textemdash & 0.10 & 3 & 2.4 \\[0.5ex]
		potential 1 & $0.05^{+0.05}_{-0.04}$ & $0.06^{+0.03}_{-0.02}$ & $0.35^{+0.15}_{-0.11}$ & $-6.8^{+3.9}_{-3.9}$ & $283.0^{+7.2}_{-7.6}$ & $1.525^*$\\[0.5ex]
		\hline\\[-1ex]
		{\it Best} & \textemdash & \textemdash & 0.05 & \textemdash & 3 & 0.5 \\[0.5ex]
		potential 1 & $0.03$ & $0.03$ & $0.20$ & $-5.9$ & $281.7$ & $1.525^*$\\
		\hline
		\hline\\
		%%%%%%%%
		%%%%%%%%
		% PJ112713
		% run 9
		%%%%%%%%
		%%%%%%%%
		{\bf PJ112713} &&&&&& \\[0.5ex]
		{\it Median} &171.80606 & 46.15670 & \textemdash & 0.20 & 3 & 0.8\\[0.5ex]
		potential 1 (SIS) & $0.00^{+0.08}_{-0.09}$ & $-0.32^{+0.08}_{-0.08}$ & \textemdash & \textemdash & $182.7^{+7.4}_{-6.8}$ & $0.415^*$\\[0.5ex]
		\hline\\[-1ex]
		{\it Best} &\textemdash & \textemdash &0.08 & \textemdash & 3 & 0.3\\[0.5ex]
		potential 1 (SIS) & $-0.03$ & $-0.28$ & \textemdash & \textemdash & $179.2$ & $0.415^*$\\[0.5ex]
		\hline
		\hline\\
		%%%%%%%%
		%%%%%%%%
		% PJ113805
		% run 11
		%%%%%%%%
		%%%%%%%%
		{\bf PJ113805} &&&&&& \\[0.5ex]
		{\it Mode} & 174.52305 & 32.965806 & \textemdash & 0.20 & 1 & 0.3\\[0.5ex]
		potential 1 (SIS) & $0.04^{+0.14}_{-0.14}$ & $-0.15^{+0.20}_{-0.17}$ & \textemdash & \textemdash & $145.5^{+11.2}_{-9.6}$ & $0.52^*$\\[0.5ex]
		\hline\\[-1ex]
		{\it Best} &\textemdash & \textemdash & 0.03 & \textemdash & 2 & 0.6\\[0.5ex]
		potential 1 (SIS) & $-0.01$ & $-0.14$ & \textemdash & \textemdash & $146.1$ & $0.52^*$\\[0.5ex]
		\hline
		\hline\\
		%%%%%%%%
		%%%%%%%%
		% PJ113921
		% run 12
		%%%%%%%%
		%%%%%%%%
		{\bf PJ113921} & &&&&& \\[0.5ex]
		%{\blue FINISHED}
		{\it Mode$^a$} & 174.84041 & 20.414479 & \textemdash & 0.39 & 2 & 1.3\\[0.5ex]
		potential 1 & $0.00^*$ & $0.01^*$ & $0.65$ & $28.6$ & $230.5$ & $0.57^*$\\[0.5ex]
		potential 2 (SIS) & $2.56^*$ & $0.58^*$ & \textemdash & \textemdash & $179.0$ & $0.57^*$\\[0.5ex]
		\hline\\[-1ex]
		{\it Median} & \textemdash & \textemdash & \textemdash & \textemdash & 2 & 1.7\\[0.5ex]
		potential 1 & $0.00^*$ & $0.01^*$ & $0.49^{+0.15}_{-0.19}$ & $31.6^{+14.0}_{-12.4}$ & $232.7^{+17.3}_{-18.8}$ & $0.57^*$\\[0.5ex]
		potential 2 (SIS) & $2.56^*$ & $0.58^*$ & \textemdash & \textemdash & $191.5^{+58.9}_{-57.3}$ & $0.57^*$\\[0.5ex]
		\hline\\[-1ex]
		{\it Best} &\textemdash & \textemdash & 0.23 & \textemdash & 2 & 0.7\\[0.5ex]
		potential 1 & $0.00^*$ & $0.01^*$ & $0.69$ & $-74.8$ & $225.5$ & $0.57^*$\\
		potential 2 (SIS) & $2.56^*$ & $0.58^*$ & \textemdash & \textemdash & $274.9$ & $0.57^*$\\
		\hline
		\hline\\
		%%%%%%%%
		%%%%%%%%
		% PJ132630
		% run 11
		%%%%%%%%
		%%%%%%%%
		{\bf PJ132630} &  &&&&& \\[0.5ex]
		%{\blue FINISHED}
		{\it Mode$^a$} & 201.62613 & 33.735272 & \textemdash & 0.38 & 1 & 0.4 \\[0.5ex]
		potential 1 & $-0.03$ & $0.14$ & $0.02$ & $-42.6$ & $331.0$ & $0.786^*$\\[0.5ex]
		\hline\\[-1ex]
		{\it Median} & \textemdash & \textemdash & \textemdash & \textemdash & 1 & 1.3\\[0.5ex]
		potential 1 & $0.01^{+0.14}_{-0.12}$ & $0.15^{+0.13}_{-0.13}$ & $0.19^{+0.18}_{-0.13}$ & $-42.5^{+25.0}_{-29.8}$ & $330.5^{+13.7}_{-12.8}$ & $0.786^*$\\[0.5ex]
		\hline\\[-1ex]
		{\it Best} &\textemdash & \textemdash & 0.02 & \textemdash & 1 & 0.02\\[0.5ex]
		potential 1 & $0.21$ & $0.31$ & $0.04$ & $-1.8$ & $334.5$ & $0.786^*$\\
		\hline
		\hline\\
		%%%%%%%%
		%%%%%%%%
		% PJ133634
		% run 20
		%%%%%%%%
		%%%%%%%%
		{\bf PJ133634} &  &&&&& \\[0.5ex]
		%{\blue FINISHED}
		{\it Mode} & 204.14596 & 49.220455 & \textemdash & 0.30 & 5 & 0.7\\[0.5ex]
		potential 1 & $0.37$ & $0.43$ & $0.01$ & $29.1$ & $218.7$ & $0.26^*$\\[0.5ex]
		\hline\\[-1ex]
		{\it Median} & \textemdash & \textemdash & \textemdash & \textemdash & 5 & 0.8\\[0.5ex]
		potential 1 & $0.37^{+0.26}_{-0.24}$ & $0.40^{+0.14}_{-0.24}$ & $0.20^{+0.28}_{-0.16}$ & $70.3^{+59.5}_{-41.5}$ & $220.2^{+10.7}_{-8.0}$ & $0.26^*$\\[0.5ex]
		\hline\\[-1ex]
		{\it Best} &\textemdash & \textemdash & 0.08 & \textemdash & 5 & 0.1\\[0.5ex]
		potential 1 & $0.71$ & $-0.17$ & $0.12$ & $160.6$ & $227.3$ & $0.26^*$\\
		\hline
		\hline\\
		%%%%%%%%
		%%%%%%%%
		% PJ144653
		% run 11
		%%%%%%%%
		%%%%%%%%
		{\bf PJ144653} & &&&&& \\[0.5ex]
		%{\blue FINISHED? }
		{\it Median} & 221.72172 & 17.875883 & \textemdash & 0.37 & 1 & 3.2\\[0.5ex]
		potential 1 & $0.10^*$ & $0.47^*$ & $0.65^{+0.18}_{-0.25}$ & $95.0^{+14.1}_{-14.0}$ & $262.8^{+18.0}_{-14.7}$ & $0.493^*$\\[0.5ex]
		\hline\\[-1ex]
		{\it Best} &\textemdash & \textemdash & 0.03 & \textemdash & 1 & 0.5\\[0.5ex]
		potential 1 & $0.10^*$ & $0.47^*$ & $0.66$ & $96.4$ & $257.8$ & $0.493^*$\\
		\hline
		\hline\\
		%%%%%%%%
		%%%%%%%%
		% PJ144958
		% run 4
		%%%%%%%%
		%%%%%%%%
		{\bf PJ144958} &&&&&& \\[0.5ex]
		{\it Median} & 222.49514 & 22.642985 & \textemdash & 0.50 & 13 & 0.4\\[0.5ex]
		potential 1 (NFW) & $-1.11^{+2.56}_{-2.87}$ & $-1.55^{+1.40}_{-1.32}$ & $0.64^{+0.12}_{-0.20}$ & $12.2^{+4.2}_{-5.1}$ & $r_s (\arcsec) = 26.1^{+4.0}_{-3.5}$ & $0.4^*$\\[0.5ex]
		potential 2 (SIS) & $3.52^*$ & $5.13^*$ & \textemdash & \textemdash & $114.0^{+10.0}_{-11.3}$ & $0.4^*$\\[0.5ex]
		potential 3 (SIS) & $3.22^*$ & $1.73^*$ & \textemdash & \textemdash & $224.5^{+38.9}_{-35.9}$ & $0.4^*$\\[0.5ex]
		potential 4 (SIS) & $8.44^*$ & $7.02^*$ & \textemdash & \textemdash & $89.6^{+76.4}_{-60.2}$ & $0.4^*$\\[0.5ex]
		potential {5} (SIS) & $-1.63^*$ & $-3.80^*$ & \textemdash & \textemdash & $157.8^{+128.8}_{-111.4}$ & $0.4^*$\\[0.5ex]
		\hline\\[-1ex]
		{\it Best} &\textemdash & \textemdash & 0.12 & \textemdash & 13 & 0.1\\[0.5ex]
		potential 1 (NFW) & $-1.27$ & $-1.96$ & $0.70$ & $10.7$ & $r_s (\arcsec) = 25.2$ & $0.4^*$\\
		potential 2 (SIS) & $3.52^*$ & $5.13^*$ & \textemdash & \textemdash & $107.2$ & $0.4^*$\\
		potential 3 (SIS) & $3.22^*$ & $1.73^*$ & \textemdash & \textemdash & $242.9$ & $0.4^*$\\
		potential 4 (SIS) & $8.44^*$ & $7.02^*$ & \textemdash & \textemdash & $193.4$ & $0.4^*$\\
		potential 5 (SIS) & $-1.63^*$ & $-3.80^*$ & \textemdash & \textemdash& $76.4$ & $0.4^*$\\
		\hline
		\hline\\
		%%%%%%%%
		%%%%%%%%
		% PJ160722
		% run 10
		%%%%%%%%
		%%%%%%%%
		{\bf PJ160722} & &&&&& \\[0.5ex]
		% {\blue FINISHED}
		{\it Best} & 241.84491 & 73.784031 & 0.13  & 0.20 & 1 & 1.6\\[0.5ex]
		potential 1 & $-0.04$ & $-0.10$ & $0.36$ & $-34.0$ & $273.6$ & $0.65^*$\\[0.5ex]
		\hline\\[-1ex]
		{\it Median} & \textemdash & \textemdash & \textemdash & \textemdash & 1 & 5.2\\[0.5ex]
		potential 1 & $-0.03^{+0.15}_{-0.14}$ & $-0.09^{+0.09}_{-0.07}$ & $0.61^{+0.13}_{-0.20}$ & $-32.1^{+82.6}_{-6.6}$ & $278.1^{+12.0}_{-11.6}$ & $0.65^*$\\
		\hline
		\hline\\
		%%%%%%%%
		%%%%%%%%
		% PJ231356
		% run 6
		%%%%%%%%
		%%%%%%%%
		{\bf PJ231356} &&&&&& \\[0.5ex]
		{\it Median} & 348.485940 & 1.154663 & \textemdash & 0.10 & 1 & 4.0 \\[0.5ex]
		potential 1 & $-0.10^{+0.07}_{-0.07}$ & $0.04^*$ & $0.57^{+0.15}_{-0.09}$ & $117.8^{+12.0}_{-4.0}$ & $306.7^{+10.4}_{-12.9}$ & $0.56^*$\\[0.5ex]
		potential 2 (SIS) & $0.25^*$ & $5.26^*$ & \textemdash & \textemdash & $297.8^{+38.5}_{-43.5}$ & $0.56^*$\\[0.5ex]
		\hline\\[-1ex]
		{\it Best} &\textemdash & \textemdash & 0.03 & \textemdash & 1 & 0.4\\[0.5ex]
		potential 1 & $-0.11$ & $0.04^*$ & $0.62$ & $115.6$ & $308.7$ & $0.56^*$\\
		potential 2 (SIS) & $0.25^*$ & $5.26^*$ & \textemdash & \textemdash & $290.4$ & $0.56^*$\\[0.5ex]
	\enddata
	\tablecomments{  %%% This should come before longtable and it is not tablenotes but TableNotes
 	%\small
 	%\item {\bf Notes.} 
	%
  	Fixed parameters are denoted with an asterisk. Asymmetric error bars indicate $1\sigma$ confidence levels and are determined from the resulting MCMC posterior distribution. Ellipticity is defined as $e= (a^2 - b^2) / (a^2 + b^2)$, for semi-major and semi-minor axes $a$ and $b$. The position angle $\theta$ is measured counterclockwise from east. Velocity dispersion $\sigma$ for singular isothermal ellipsoids are functionally equivalent to physical velocity dispersions.
	The Right ascension (RA) and Declination (Dec.) given are reference positions (to which $\Delta\alpha$ and $\Delta\delta$ are relative), and are typically chosen as the centroid of the brightest foreground. 
	The positional uncertainty $\sigma_{\rm position}$ is supplied for each object, and is equal to the size of the synthesized beam for ALMA or JVLA (whichever is lowest resolution). 
	RMS$_{\rm im}$ is the RMS deviation between observed and modeled multiple image positions for each field, calculated in the image plane for the lowest-$\chi^2$ solution.
	%
	%}
	The reduced $\chi_\nu^2$ is reported for the given number of degrees of freedom $\nu$.
  %
 	 %\newline{}
  %
  $^a$ In some cases, the mode solution of the optimization is used instead of the median, which is less useful when the posterior distribution is bimodal or if one parameter's optimization is sensitive to the choice of parameter bounds (e.g., when ellipticity is poorly constrained by the multiple images, the posterior tends to follow more of a uniform distribution, so the median tends towards the midpoint of the parameter bounds). For these solutions, the uncertainty in the median is taken to be representative of the uncertainty in the mode. 
  }
  \end{deluxetable*}

\section{Comparison with existing lens models}

\label{sec:lens_model_comparison}

Several of the objects included in this work have already been the subject of gravitational lens modeling, but we independently derive models here to both incorporate information from new observations and to ensure consistency between models for this subsample.
In particular, we examine discrepancies in derived magnifications as a benchmark.

First, PJ020941 (or {\it 9io9}) is perhaps the most well-studied DSFG included, with independent modeling efforts by \citet{Geach:2015aa, Geach:2018aa, Rivera:2019aa}, and \citet{Liu:2022ad}.
The initial model by \citet{Geach:2015aa} was based on 4 different radio observations and near-IR ground-based imaging (seeing $\approx 0.8\arcsec$), and was parameterized by an isothermal ellipsoid and a satellite isothermal spheroid for the secondary foreground, with added external shear. With the {\it gravlens} software \citep{Keeton:2011aa}, they modeled the near-IR as a single S\'ersic profile and the radio as two Gaussian components, finding a magnification of $\mu \approx 10$, in agreement with what we find, $\mu_{\rm 1mm} =10.5 \pm 3.7$ and $\mu_{\rm 6 GHz} = 12.0 \pm 3.5$ (Table~\ref{tab:lensproperties}). 
Revising this model using CO(4-3) ALMA spectro-imaging and a semi-linear inversion approach \citep{Warren:2003aa,Dye:2018aa}, \citet{Geach:2018aa} found a slightly higher magnification of $\mu_{\rm CO4-3} = 14.7 \pm 0.3$.
\citet{Rivera:2019aa} incorporated CO(3-2) imaging with NOEMA, with 32 independent velocity channels as constraints, finding a smooth velocity gradient consistent with the structure found by \citet{Geach:2018aa}. Also using a 2-component lens model with external shear, they found that the magnification varied strongly with velocity, ranging from $\mu \approx 7 - 22$, but with a luminosity-weighted mean of $\langle \mu \rangle \approx 13$.
Most recently, \citet{Liu:2022ad} used newly-collected adaptive optics $H$ and $K_s$-band imaging, alongside \HST/WFC3 F125W, SMA 870 $\mu$m, and multi-J CO observations with ALMA in order to build their own lens model.
As with our work, \citeauthor{Liu:2022ad} used \lenstool\ to fit the primary and secondary foreground with an SIE and SIS profile, constrained by the F125W image. 
This approach is the one most similar to our own, so an equivalent comparison is possible. The authors found optimized parameters that agreed with all in our model (Table \ref{tab:lensmodels}) within uncertainties, except for only a small disagreement for position angle (PA $ = -8 \pm 2$\degrees\ vs. our $-14 \pm 1$\degrees), and their source-plane reconstruction is visually very similar to ours (Fig.~\ref{fig:model_SP}).
As a result, \citeauthor{Liu:2022ad} found a dust continuum magnification of $\mu_{\rm dust} = 12.8 \pm 0.3$ and a stellar component magnification of $\mu_{\rm star} = 13.6 \pm 0.4$, consistent with our value of $\mu_{\rm dust} \approx 10.5$, especially given the different interferometric configurations of ALMA vs. SMA.

PJ105353 (or G244.8+54.9, the ``Ruby") was previously discussed and modeled by \citet{Canameras:2017aa,Canameras:2017ab, Frye:2019aa}. 
\citeauthor{Canameras:2017aa} used \lenstool\ to model the $z=1.525$ foreground with a pseudo-isothermal elliptical mass density (PIEMD) profile (with the poorly-constrained core and cut radii held fixed at 0.15 kpc and 100 kpc), based on image locations and parities from CO(4-3) observations (angular resolution 0.1$\arcsec$).
However, the authors acknowledged that these image identifications were not completely unambiguous, but settled ultimately on a pair of source-plane objects imaged 2 and 4 times, respectively.
An additional 0.07$\arcsec$-resolution Band 6 (1 mm) observation with ALMA (Program 2015.1.01518.S, PI: N. Nesvadba) was not included as a constraint, but our interpretation of the image configuration differs slightly. Whereas \citeauthor{Canameras:2017aa} identify a cusp-like configuration, the 1 mm image appears to instead show more of an Einstein cross, with 4 well-separated images (nearly aligned north, south, east, and west). However, the second, doubly-imaged system is in accord with \citeauthor{Canameras:2017aa}. 
Unfortunately, the non-detection of the background object with \HST\ (as it is blended with the foreground; \citealt{Frye:2019aa}) makes confirmation of this morphology difficult.
\citeauthor{Canameras:2017aa} find a dust magnification of $\mu_{\rm 3mm} = 21.8 \pm 0.6$ (reported in \citealt{Canameras:2018ab}), with magnifications for individual clumps ranging from $\mu \approx 7 - 38$.
In our case, we find a significantly different total value of $\mu_{870 \mu {\rm m}} = 7.6 \pm 0.7$ (or $\mu_{\rm 1 mm} \approx 12$).
Given the difficulty in deriving robust constraints on the model, a discrepancy is not particularly unexpected, especially as both interpretations agree that the source-plane galaxy lies near to a caustic, where the magnification gradient is steep and highly sensitive to the model. On the other hand, values like the Einstein radius are not as sensitive, and our value is in close agreement with that of \citet{Canameras:2017aa}.

While a lens model for PJ113921 (G231.3+72.2) has not yet been published, \citet{Canameras:2018ab} provide magnifications of $\mu_{\rm dust} = 7.9 \pm 0.3$ and $\mu_{\rm gas} = 6.0 \pm 0.5$, based on SMA 880 $\mu$m continuum and multi-J CO imaging with the IRAM Plateau de Bure Interferometer (PdBI), which are broadly consistent with our measurements of $\mu \approx 5 - 9$ from ALMA and JVLA.

Lastly, PJ132630 was modeled by \citet{Bussmann:2013aa} based on HST F110W near-IR and SMA 880 $\mu$m imaging observations, using an interferometric visibility-based approach. They found an Einstein radius of $\theta_E = 1.80 \pm 0.02$ (consistent with our $\theta_{\rm Ein} = 1.78 \pm 0.18$; Table \ref{tab:lensproperties}), a lens ellipticity of $\epsilon = 0.26 \pm 0.04$ (similar to our median solution, $e = 0.19 \pm 0.16$; Table~\ref{tab:lensmodels}), and a dust continuum magnification of $\mu_{880\mu{\rm m}} = 4.1 \pm 0.3$ (in close agreement with our $\mu_{\rm 1mm} = 4.3 \pm 0.6$). 
While magnifications are generally robust between various lens modeling approaches, the non-trivial differences between this work and others should not be discounted.
\citet{Spilker:2016aa} computes magnification as flux-weighted average over the elliptical model components describing the source plane, whereas \citet{Bussmann:2013aa, Bussmann:2015aa} use the ratio of image-plane to reconstructed source-plane flux.
On the other hand, \citet{Dye:2018aa} explore the sensitivity of magnification to imaging resolution and depth by calculating magnification as a function of fraction of source-plane flux (i.e. surface brightness threshold). 
The authors found that, while variation can be small, in the vicinity of a caustic in particular, magnification can vary by up to 25\% as a function of interferometric configuration.
Our work does not make explicit assumptions of source-plane structure, but also does not take advantage of the entirety of information present in the interferometric images like the visibility-modeling approach of these other works.
Future work to apply a visibility-based modeling approach to these objects will allow for a more direct comparison.
An additional promising method that we hope to explore in the future involves Regularized Semi-linear Inversion \citep{Warren:2003aa,Nightingale:2015aa,Enia:2018aa}, by which the source plane is tessellated but not assumed to follow a specific parametric form.

In summary, the general agreement for most objects with prior models (derived from independent constraints) helps to lend credence to their robustness. We interpret any tensions in results\textemdash in particular, for PJ105353\textemdash to be more influenced by different observing setups than by different lens modeling approaches.

%
%

%%%%%%%%%%%%%%%%%%%%%%%%%%%%%%%%%%%%%%%%%%%%%%%%%%



%\bibliographystyle{mnras}
\bibliographystyle{aasjournal}
%\bibliography{/Users/chatham64/Documents/UMass-Amherst/RESEARCH_YEAR_2/RESEARCH_2017} % if your bibtex file is called example.bib

% COPIED FROM Kamieneski2023.bbl



%\begin{thebibliography}{99}
%\end{thebibliography}

%%%%%%%%%%%%%%%%%%%%%%%%%%%%%%%%%%%%%%%%%%%%%%%%%%


%%%%%%%%%%%%%%%%%%%%%%%%%%%%%%%%%%%%%%%%%%%%%%%%%%

% Don't change these lines
%\bsp	% typesetting comment
%\label{lastpage}
\end{document}